\documentclass[preprint,12pt,nopreprintline]{elsarticle}

\usepackage{setspace}
\usepackage{chngcntr}
\counterwithout{figure}{section}

\usepackage{bm}
\usepackage{bbm}
\usepackage[figuresleft]{rotating}
\usepackage{graphicx,epstopdf}
\usepackage{multirow}
\usepackage[caption=false]{subfig}
\usepackage{algorithm}

\usepackage{algorithmicx}
\usepackage{algpseudocode}
\usepackage{adjustbox}
\usepackage{mathtools}
\usepackage{array}
\usepackage{placeins}
\usepackage{xcolor}
\usepackage{tikz}
\usetikzlibrary{arrows.meta,positioning,fit,backgrounds}

\usepackage{amsmath,amssymb,amsfonts,amsthm}
\usepackage{textcomp}
\usepackage{hyperref}
\usepackage{etoolbox}
\AtBeginEnvironment{table}{\setlength{\belowcaptionskip}{6pt}}
\AtBeginEnvironment{table*}{\setlength{\belowcaptionskip}{6pt}}

\theoremstyle{definition}

\newcommand{\arxivdir}{}
\IfFileExists{arxiv/pausct-arxiv.tex}{\renewcommand{\arxivdir}{arxiv/}}{}
\newcommand{\pausct}{\texttt{PAUSCT}}

\renewcommand{\vec}[1]{\mathbf{#1}}

\newcommand{\p}{\partial}
\newcommand{\pd}[2]{\frac{\partial #1}{\partial #2}}
\newcommand{\dd}[2]{\frac{\mathrm{d} #1}{\mathrm{d} #2}}

\DeclareMathOperator{\sinc}{sinc}

\newcommand{\had}{\mathbin{\!\odot\!}}
\newcommand{\hdiv}{\mathbin{\!\oslash\!}}

\newcommand{\slp}{\scriptstyle (\ell\!+\!\frac{1}{2})}
\newcommand{\slm}{\scriptstyle (\ell\!-\!\frac{1}{2})}
\newcommand{\slpm}{\scriptstyle (\ell\!\pm\!\frac{1}{2})}
\newcommand{\skp}{\scriptstyle (k\!+\!\frac{1}{2})}
\newcommand{\skm}{\scriptstyle (k\!-\!\frac{1}{2})}

\newcommand{\Nr}{N_{\mathrm{rx}}}
\newcommand{\Ntx}{N_{\mathrm{tx}}}
\newcommand{\dx}{d_{x_m}}
\newcommand{\dt}{d_t}
\newcommand{\br}{\bm{r}}
\newcommand{\hbr}{\hat{\bm{r}}}
\newcommand{\brt}{\br,t}

\newcommand{\bT}{\boldsymbol{\Theta}}
\newcommand{\bt}{\boldsymbol{\theta}}

\newcommand{\Dt}{\Delta_t}

\newcommand{\sspj}{j^{\mkern-2mu{\scriptscriptstyle +}}}
\newcommand{\ssp}{m^{\mkern-2mu{\scriptscriptstyle +}}}

\newcommand{\pml}{\boldsymbol{\Xi}_{m}}
\newcommand{\pmlp}{\boldsymbol{\Xi}_{\ssp}}
\newcommand{\Kp}{\boldsymbol{\kappa}}

\newcommand{\sfFr}{\mathsf{F}_{\br}}
\newcommand{\sfFri}{\mathsf{F}^{-1}_{\br}}

\newcommand{\sfFt}{\mathsf{F}_{t}}
\newcommand{\sfFti}{\mathsf{F}^{-1}_{t}}

\newcommand{\sfIp}{\mathsf{I}_m^+}
\newcommand{\sfIm}{\mathsf{I}_m^-}

\newcommand{\sfD}{\mathsf{D}}
\newcommand{\sfDp}{\mathsf{D}_{m}^+}
\newcommand{\sfDm}{\mathsf{D}_{m}^-}
\newcommand{\sfDj}{\mathsf{D}_{j}^-}
\newcommand{\sfDpm}{\mathsf{D}_{m}^{\pm}}

\newcommand{\divg}{\mathsf{D}_{\Sigma}}

\newcommand{\sfH}{\mathsf{H}}
\newcommand{\sfB}{\mathsf{B}}
\newcommand{\sfS}{\mathsf{S}}
\newcommand{\sfP}{\mathsf{P}}
\newcommand{\sfG}{\mathsf{G}}
\newcommand{\sfN}{\mathsf{N}_{B\!/\!A}}
\newcommand{\sfX}{\mathsf{X}}
\newcommand{\sfJ}{\mathsf{J}}
\newcommand{\sfR}{\mathsf{R}}

\newcommand{\sfT}{\mathsf{T}}
\newcommand{\sfQ}{\mathsf{Q}}
\newcommand{\sfKsrc}{\mathsf{K}_{\mathrm{src}}}

\newcommand{\sftS}{\tilde{\mathsf{S}}}

\newcommand{\sfhG}{\hat{\mathsf{G}}}

\newcommand{\cS}{\mathcal{S}}
\newcommand{\cG}{\mathcal{G}}
\newcommand{\cR}{\mathcal{R}}
\newcommand{\cL}{\mathcal{L}}
\newcommand{\cN}{\mathcal{N}}
\newcommand{\cQ}{\mathcal{Q}}

\newcommand{\lu}{\boldsymbol{\Phi}_{\ssp}}
\newcommand{\lr}{\boldsymbol{\Psi}_{m}}
\newcommand{\lrj}{\boldsymbol{\Psi}_{j}}
\newcommand{\lrs}{\boldsymbol{\Psi}_{\Sigma}}

\newcommand{\lp}{\boldsymbol{\Lambda}}

\newcommand{\vn}{\vec{n}}

\newcommand{\vK}{\vec{K}}

\newcommand{\vtFm}{\widetilde{\vec{F}}_{\ssp}}

\newcommand{\vV}{\vec{V}}
\newcommand{\vW}{\vec{W}}
\newcommand{\vU}{\vec{U}}
\newcommand{\vR}{\vec{R}}
\newcommand{\vH}{\vec{H}}
\newcommand{\vG}{\vec{G}}

\newcommand{\vCo}{\vec{C}_0}
\newcommand{\vBo}{\vec{B}_0}
\newcommand{\vAo}{\vec{A}_0}
\newcommand{\vRo}{\vec{R}_0}
\newcommand{\vTo}{\vec{T}_0}
\newcommand{\vEo}{\vec{E}_0}
\newcommand{\vRs}{\vec{R}_{\Sigma}}

\newcommand{\vX}{\vec{X}}
\newcommand{\vY}{\vec{Y}}

\newcommand{\vP}{\vec{P}}
\newcommand{\vPo}{\vec{P}_0}
\newcommand{\vUm}{\vec{U}_{\ssp}}
\newcommand{\vUj}{\vec{U}_{\sspj}}
\newcommand{\vRm}{\vec{R}_{m}}
\newcommand{\vtMm}{\widetilde{\vec{M}}_m}

\newcommand{\vhA}{\vec{\hat{A}}}
\newcommand{\vhUm}{\vec{\hat{U}}_{\ssp}}
\newcommand{\vhU}{\vec{\hat{U}}}
\newcommand{\vhR}{\vec{\hat{R}}}
\newcommand{\vhRm}{\vec{\hat{R}}_{m}}
\newcommand{\vhP}{\vec{\hat{P}}}
\newcommand{\vhRs}{\vec{\hat{R}}_{\Sigma}}
\newcommand{\vhbR}{\vec{\hat{R}}_{\mathrm{Eff.}}}

\newcommand{\vhX}{\vec{\hat{X}}}
\newcommand{\vhY}{\vec{\hat{Y}}}

\newcommand{\vbR}{\vec{R}_{\mathrm{Eff.}}}

\newcommand{\vdP}{\dot{\vP}}
\newcommand{\vdU}{\dot{\vU}}
\newcommand{\vdR}{\dot{\vR}}
\newcommand{\vdUm}{\vdU_{\ssp}}
\newcommand{\vdRm}{\vdR_m}
\newcommand{\vdRs}{\vdR_{\Sigma}}
\newcommand{\vdbR}{\vdR_{\mathrm{Eff.}}}

\newcommand{\adsrc}{\boldsymbol{\xi}}

\DeclareMathOperator*{\argmin}{arg\,min}

\begin{document}

\begin{frontmatter}

\author[1]{Gangwon Jeong}
\ead{gangwon@wustl.edu}
\affiliation[1]{organization={Mallinckrodt Institute of Radiology, Washington University in St. Louis},
            addressline={510 South Kingshighway Blvd},
            city={St. Louis},
            state={MO},
            postcode={63110},
            country={USA}}

\author[1]{Mark A. Anastasio\corref{cor1}}
\cortext[cor1]{Corresponding author}
\ead{anastasio@wustl.edu}
\title{\pausct: A Discrete-Adjoint Framework for Iterative Full-Wave Image Reconstruction in Ultrasound and Photoacoustic Computed Tomography
}

\begin{abstract}
Optimization-based image reconstruction in ultrasound computed tomography (USCT) and photoacoustic computed tomography (PACT) requires accurate signal predictions and parameter gradients.
We present \pausct{}, an open-source Python/JAX package that provides a unified discrete-adjoint framework for 2D and 3D reconstruction on GPUs.
The forward model applies the \textit{k}-space pseudospectral time-domain method to established first-order acoustic equations with heterogeneous speed of sound (SOS) and density, power-law acoustic attenuation and dispersion, and cumulative nonlinearity.
The model represents finite-aperture transmitters and receivers with spatial or spatiotemporal responses.
Explicit gradients of the data-fidelity term with respect to medium parameters, initial pressure, transmit waveforms, and receive-response coefficients are algebraically consistent with this model.
Adjoint-based gradients share one adjoint solution per acquisition and objective, with forward quantities stored or recomputed according to the selected parameters.
The same adjoint formulation supports a range of differentiable data-fidelity functions.
In 2D PACT and USCT tests, selected spatial and temporal gradient profiles agreed with forward-mode automatic differentiation (AD) applied one parameter entry at a time.
Relative root-mean-square (RMS) differences were below 0.006\%.
For 2D SOS gradients of the same model under matched checkpointing schedules, reverse-mode AD took approximately 6\% to 29\% longer, and the discrete adjoint used less GPU memory for the lossy models.
In a 3D virtual imaging study with a numerical breast phantom, noise, and model mismatch, PACT reconstruction yielded a relative RMS error in initial pressure of 29.06\% with USCT-derived SOS and 60.30\% with homogeneous SOS.

\end{abstract}

\end{frontmatter}

\section{Introduction}
Ultrasound computed tomography (USCT) produces estimates of spatially varying tissue properties from measurements of transmitted and scattered ultrasound wavefields.
Such tissue properties include the speed of sound (SOS), acoustic attenuation (AA), and acoustic impedance, which have been investigated for tissue differentiation, lesion characterization, breast-density assessment, and cancer-risk estimation~\cite{wuOptimalTransportMultiparameter2025,liInVivoBreastSoundSpeed2009,duricBreastDensityMeasurements2013,duricWholeBreastRisk2020}.
Full-waveform inversion (FWI) has been used to estimate these distributions at high resolution by iteratively minimizing the discrepancy between simulated and measured signals, using a wave-propagation model that accounts for diffraction and multiple scattering~\cite{wuOptimalTransportMultiparameter2025,wangWaveformInversionSource2015,matthewsRegularizedDualAveraging2017,perez-livaTimeDomainReconstruction2017,luckaHighResolution3d2022,liuThreeDimensionalTimeDomain2025}.

Photoacoustic computed tomography (PACT) combines optical excitation with ultrasound detection, providing optical-absorption contrast for vascular imaging, cancer detection, and, with multispectral measurements, blood-oxygenation assessment~\cite{linSingleBreathHoldPhotoacoustic2018,tzoumasEigenspectraOptoacoustic2016}.
Its acoustic inverse problem is to recover the initial pressure (IP) distribution generated by optical absorption of pulsed illumination.
Analytic methods offer rapid IP estimation under simplifying assumptions such as acoustic homogeneity~\cite{poudelSurveyComputationalFrameworks2019}, but neglecting heterogeneous acoustic properties, particularly SOS, can introduce artifacts and reduce reconstruction accuracy~\cite{huangFullWaveIterative2013,matthewsParameterizedJointReconstruction2018}.
This has motivated reconstruction methods that incorporate SOS maps estimated by USCT, which are available, for example, in hybrid photoacoustic and ultrasound breast tomography~\cite{dantumaHybridThreeDimensional2026}. Joint estimation of IP and SOS from PACT measurements alone or from combined PACT and USCT measurements has been proposed~\cite{huangJointReconstructionAbsorbed2016,matthewsJointReconstructionInitial2017,matthewsParameterizedJointReconstruction2018,jeongRevisitingJointEstimation2025}.

These reconstruction methods require numerical forward models of the measured signals.
The \texttt{k-Wave} toolbox is widely used for biomedical ultrasound propagation modeling~\cite{treebyWaveMatlabToolbox2010}.
Its wave model consists of coupled first-order acoustic equations that account for heterogeneous SOS and density, power-law absorption and dispersion, and cumulative nonlinearity~\cite{treebyModelingPowerLaw2010,treebyModelingNonlinearUltrasound2012}.
These equations are solved by \textit{k}-space pseudospectral time-domain (\textit{k}-PSTD) time marching, which combines Fourier collocation for spatial differentiation with a \textit{k}-space correction for temporal numerical dispersion~\cite{treebyWaveMatlabToolbox2010,treebyModelingNonlinearUltrasound2012}.
It has been used in iterative PACT~\cite{huangFullWaveIterative2013} and USCT~\cite{perez-livaTimeDomainReconstruction2017,luckaHighResolution3d2022} reconstruction, including three-dimensional (3D) studies.
Transducer properties also influence reconstruction accuracy in PACT~\cite{wangImagingModelIncorporating2011,mitsuhashiInvestigationFarfieldApproximation2014} and USCT~\cite{yuanFullWaveformInversionBreast2023,liForwardModelIncorporating2023}.
Finite-aperture effects in transmission and reception can be modeled using band-limited interpolation and numerical aperture integration, avoiding staircasing artifacts in the aperture representation~\cite{wiseRepresentingArbitraryAcoustic2019}.
Under a linear time-invariant assumption, the temporal receive response can be modeled by convolution~\cite{wangImagingModelIncorporating2011}.

First-order optimization methods require gradients of the objective function with respect to the parameters being estimated.
The adjoint-state method evaluates these gradients without explicitly forming the Jacobian of the predicted data with respect to those parameters~\cite{plessixReviewAdjointState2006}.
Adjoint formulations have been developed for IP reconstruction in heterogeneous media~\cite{huangFullWaveIterative2013,arridgeAdjointOperatorPhotoacoustic2016,javaherianMultiGridIterative2017}, joint IP and SOS estimation in PACT~\cite{matthewsParameterizedJointReconstruction2018}, and acoustic-property estimation in USCT~\cite{perez-livaTimeDomainReconstruction2017,luckaHighResolution3d2022}.
These formulations target particular combinations of unknown parameters and modeled propagation and transducer effects; extending their scope can require additional gradient derivation and implementation.
A unified formulation covering these effects would make the corresponding gradients available for different reconstruction problems without separate derivations for each configuration.
Because an iterative method minimizes the discrete objective, the relevant gradients are those of the complete discrete computation, including the transducer operations.
A discretize-then-adjoint (DTA) formulation derives these gradients directly from that computation~\cite{huangFullWaveIterative2013}, ensuring algebraic consistency with the discrete forward model; independently discretizing a continuous adjoint~\cite{arridgeAdjointOperatorPhotoacoustic2016} does not, in general, guarantee this consistency.

Several open-source packages support gradient-based acoustic inversion.
Stride~\cite{cuetoStrideFlexibleSoftware2022} for USCT and JUDI~\cite{witteJUDI2019} for seismic imaging provide adjoint operators for finite-difference discretizations.
The JAX-based \texttt{j-Wave}~\cite{stanziolaJWaveOpensourceDifferentiable2023,stanziolaResearchFrameworkWriting2021} provides a differentiable \textit{k}-PSTD solver whose gradients are obtained by automatic differentiation (AD), without a separately derived adjoint.
Its time-domain solver does not include power-law AA or cumulative nonlinearity.
Line apertures are represented by grid-aligned masks, and receive-response convolution and temporal filtering are not provided.
Reverse-mode AD can use segment-based checkpointing to reduce memory, but parameter-specific storage reductions may require additional changes to the forward implementation.

This work presents \pausct{}, an open-source Python/JAX package for optimization-based reconstruction in 2D and 3D PACT and USCT.\footnote{Source code and examples are available at \url{https://github.com/gangwonee/pausct}.}
It combines \textit{k}-PSTD acoustic forward simulation with gradient computation, accounting for transducer responses.
The main contributions are:
\begin{itemize}
    \item Explicit gradients for SOS, ambient density, the AA prefactor and spatially constant power-law exponent, the nonlinearity parameter $B/A$, IP, transmit waveforms, and receive-response coefficients, algebraically consistent with the complete discrete forward model.
    \item An interface for finite-aperture transmission and reception with user-prescribed spatial or spatiotemporal responses, using numerical aperture integration and matched adjoints to support a range of element designs.
    \item {\emergencystretch=1em Memory-efficient evaluation of the requested adjoint-based gradients from one adjoint solution per acquisition and objective, using parameter-specific history storage, spatial restriction of selected histories, and checkpointing.\par}
    \item Support for a range of differentiable data-fidelity functions using the same discrete-adjoint formulation for acoustic propagation and transducer modeling.
    Derivatives with respect to predicted data can be supplied explicitly or computed by reverse-mode AD.
\end{itemize}
Numerical tests demonstrate consistency between the derived gradients and independently computed reference derivatives.
The efficiency of spatially restricted history storage and checkpointing is assessed through time and memory comparisons with reverse-mode AD in JAX, using the same discrete model, gradient regions, and checkpointing schedules.
Prescribed spatiotemporal aperture responses for transmission and reception are illustrated with a concave-lens model, and the corresponding discrete adjoints are verified by inner-product tests.
A 3D virtual imaging study provides an example of USCT-augmented PACT reconstruction using \pausct{}.
Data were generated using an anatomically realistic numerical breast phantom~\cite{liStochasticNumericalBreast2022,parkStochasticThreeDimensional2023,parkVirtualImaging2026} and a hemispherical array of finite-aperture elements shared by USCT and PACT.
With added noise and prescribed model mismatch, SOS is first estimated from USCT data and then used in PACT IP reconstruction to account for SOS heterogeneity.

The remainder of this paper is organized as follows.
Section~\ref{back} presents the continuous and discrete forward models, and Section~\ref{sec:dta} derives the adjoint recursions and parameter-gradient formulas.
Section~\ref{sec:implementation} outlines the software, presents numerical verification, and evaluates memory-efficient gradient computation and an example of spatiotemporal transducer modeling.
Section~\ref{sec:applications} presents the 3D virtual imaging study of USCT-augmented PACT reconstruction.
Sections~\ref{sec:discussion} and~\ref{sec:conclusion} provide the discussion and conclusions.

\section{Continuous and discrete forward models for PACT and USCT}\label{back}
This section describes the continuous forward model for PACT and USCT data acquisition and its discretization.
The wave equation and its \textit{k}-PSTD discretization follow Treeby \textit{et al.}~\cite{treebyModelingNonlinearUltrasound2012} and \texttt{k-Wave}~\cite{treebyWaveMatlabToolbox2010}.
The apodized finite-aperture source and receive models, the effective temporal impulse response of each receiving element, and the temporal receive operator are specified for the discrete-adjoint derivation in Section~\ref{sec:dta}.

\subsection{Continuous-to-discrete forward model}\label{back:cd}
\subsubsection{Continuous wave equation}\label{back:cd:wave}
The reduced first-order acoustic model includes power-law AA, Kramers--Kronig-consistent dispersion, and the cumulative effects of convective and material nonlinearities~\cite{treebyModelingNonlinearUltrasound2012,hamiltonNonlinearAcoustics2024}.
For $\br \in \mathbb{R}^{d}$, $d\in\{2,3\}$, and $t\in(0,\infty)$, the coupled system is
\begin{align}
\label{eq:momentum}
\pd{\bm{u}}{t}(\brt) &= -\frac{1}{\rho_0(\br)}\nabla p(\brt) + \bm{f}(\brt), \\
\label{eq:mass}
\pd{\rho}{t}(\brt) &= -\bigl(2\rho(\brt)+\rho_0(\br)\bigr)\nabla\cdot\bm{u}(\brt) -\bm{u}(\brt)\cdot\nabla\rho_0(\br) +m(\brt),\\
\label{eq:eos}
p(\brt) &= c_0(\br)^2 \bigg[\rho(\brt) +\bm{d}(\brt)\cdot\nabla\rho_0(\br)
-\cQ[\rho](\brt)+\cN[\rho](\brt)
\bigg].
\end{align}
Here, $c_0(\br)$ in $\mathrm{m\,s^{-1}}$, $\rho_0(\br)$ in $\mathrm{kg\,m^{-3}}$, $\alpha_0(\br)$ in $\mathrm{Np}\,(\mathrm{rad\,s^{-1}})^{-\gamma}\mathrm{m}^{-1}$, and the dimensionless, spatially constant $\gamma$ denote the SOS, ambient density, power-law AA prefactor, and power-law exponent, respectively, with $\alpha(\br,\omega)=\alpha_0(\br)|\omega|^\gamma$.
A prefactor $\alpha_0^{\mathrm{dB}}$ given in the conventional units $\mathrm{dB}\,\mathrm{MHz}^{-\gamma}\mathrm{cm}^{-1}$ is converted as $\alpha_0(\br)=\alpha_0^{\mathrm{dB}}(\br)\,(100/(20\log_{10}e))\,(2\pi\times10^{6})^{-\gamma}$.
Throughout this work, AA-prefactor gradients are taken with respect to the converted SI coefficient $\alpha_0$.
The fields $\bm{u}(\br,t)$, $\rho(\br,t)$, and $p(\br,t)$ denote particle velocity in $\mathrm{m\,s^{-1}}$, the deviation of density from $\rho_0$ in $\mathrm{kg\,m^{-3}}$, and pressure in $\mathrm{Pa}$, respectively, and $\bm{d}(\br,t)=\int_0^t \bm{u}(\br,\tau)\,d\tau$ is the displacement in $\mathrm{m}$.

The operator $\cQ$ models power-law absorption and dispersion for $0<\gamma<3$, $\gamma\neq 1$~\cite{treebyModelingPowerLaw2010,treebyModelingNonlinearUltrasound2012}:
\begin{equation}
\label{eq:loss_operator}
\cQ[\rho](\br,t) = \tau_0(\br) \frac{\p}{\p t} \left(-\nabla^2\right)^{\frac{\gamma}{2}-1} \rho(\br,t) + \eta_0(\br) \left(-\nabla^2\right)^{\frac{\gamma-1}{2}} \rho(\br,t),
\end{equation}
where $\tau_0(\br)$ and $\eta_0(\br)$ are the absorption and dispersion coefficients:
\begin{equation*}
\tau_0(\br) = -2\alpha_0(\br)c_0(\br)^{\gamma-1}, \qquad \eta_0(\br) = 2\tan\left(\frac{\pi\gamma}{2}\right)\alpha_0(\br)c_0(\br)^{\gamma}.
\end{equation*}
The operator $\cN$ accounts for finite-amplitude effects in the pressure--density relation, with dimensionless nonlinearity parameter $(B/A)(\br)$:
\begin{equation}
\label{eq:finite_amplitude_operator}
\cN[\rho](\br,t)
= \frac{(B/A)(\br)\rho(\br,t)^2}{2\rho_0(\br)},
\end{equation}
where $A(\br)$ and $B(\br)$ are the coefficients of $\rho/\rho_0$ and $(\rho/\rho_0)^2/2$, respectively, in the isentropic Taylor expansion of pressure about the ambient state~\cite[Eq.~(2.2)]{hamiltonNonlinearAcoustics2024}.

The source terms $\bm{f}(\br,t)$ and $m(\br,t)$ denote force per unit mass in $\mathrm{m\,s^{-2}}$ and mass-injection rate per unit volume in $\mathrm{kg\,m^{-3}\,s^{-1}}$, respectively.
For the $\Ntx$ transmitting elements indexed by $i_{\mathrm{tx}}=0,\dots,\Ntx-1$, the prescribed pressure and normal particle-velocity waveforms are denoted by $s_p^{[i_{\mathrm{tx}}]}(t)$ in $\mathrm{Pa}$ and $s_u^{[i_{\mathrm{tx}}]}(t)$ in $\mathrm{m\,s^{-1}}$, respectively.
These prescribed waveforms are causal and bounded, and vanish outside $0<t\le T_s$, where $T_s$ is the source duration.
Each element has aperture $\Gamma^{[i_{\mathrm{tx}}]}\subset\mathbb{R}^d$ of finite length in 2D or finite area in 3D.
With time-independent spatial apodization~\cite{jensenCalculationPressureFields1992}, the excitation is the prescribed waveform multiplied by a real-valued factor $a_{\mathrm{tx}}^{[i_{\mathrm{tx}}]}(\br')$.
The resulting source terms follow the additive convention~\cite{treebyWaveMatlabToolbox2010}:
\begin{equation}
\label{eq:continuous-source-waveforms}
\begin{aligned}
m(\br,t)
&=2\sum_{i_{\mathrm{tx}}=0}^{\Ntx-1}w_{\mathrm{tx}}^{[i_{\mathrm{tx}}]}s_p^{[i_{\mathrm{tx}}]}(t)\int_{\Gamma^{[i_{\mathrm{tx}}]}}
\frac{a_{\mathrm{tx}}^{[i_{\mathrm{tx}}]}(\br')}{c_0(\br')}
\delta(\br-\br')\,d\Gamma^{[i_{\mathrm{tx}}]}(\br'),\\
\bm{f}(\br,t)
&=2\sum_{i_{\mathrm{tx}}=0}^{\Ntx-1}w_{\mathrm{tx}}^{[i_{\mathrm{tx}}]}s_u^{[i_{\mathrm{tx}}]}(t)\int_{\Gamma^{[i_{\mathrm{tx}}]}}
c_0(\br')a_{\mathrm{tx}}^{[i_{\mathrm{tx}}]}(\br')\bm{\nu}^{[i_{\mathrm{tx}}]}
\delta(\br-\br')\,d\Gamma^{[i_{\mathrm{tx}}]}(\br').
\end{aligned}
\end{equation}
Here, $\delta$ denotes the $d$-dimensional Dirac delta.
All transmitting apertures are assumed to lie in water with a prescribed constant SOS $c_{\mathrm{water}}$, so that $c_0(\br')=c_{\mathrm{water}}$ over each aperture.
For particle-velocity sources, each aperture is assumed to be a straight line segment in 2D or a planar surface in 3D, with a constant unit normal $\bm{\nu}^{[i_{\mathrm{tx}}]}$.
The dimensionless coefficients $w_{\mathrm{tx}}^{[i_{\mathrm{tx}}]}$ select and weight the elements for transmit beamforming.

Equation~\eqref{eq:continuous-source-waveforms} assumes separability between spatial apodization and the temporal waveform within each transmitting aperture.
More generally, a spatiotemporal aperture response allows waveform shape and timing to vary across the aperture~\cite{cuetoSpatialResponseIdentification2022}.
For convenience, temporal effects common to all points on $\Gamma^{[i_{\mathrm{tx}}]}$ are retained in the prescribed waveform, while the remaining position-dependent effects are represented by convolving this waveform with the aperture response.
This transmit model is detailed in~\ref{app:aperture-temporal}.

The initial density is prescribed using the linear pressure--density relation~\cite{huangFullWaveIterative2013}, with absorption, dispersion, and nonlinear corrections neglected in the initial conditions:
\begin{equation}
\label{eq:ic}
p(\br,0)=p_0(\br), \qquad
\bm{u}(\br,0)=\bm{0}, \qquad \rho(\br,0) = \frac{p_0(\br)}{c_0(\br)^2}.
\end{equation}
Here, $p_0(\br)$ denotes the IP distribution induced by optical absorption in PACT; for USCT, $p_0(\br)\equiv0$.

\subsubsection{Continuous-to-discrete receive model}\label{back:cd:receive}
Let $i_{\mathrm{rx}}=0,\dots,\Nr-1$ index the $\Nr$ receiving elements. The receiver and transmitter numberings are independent, even when physical elements are shared.
With time-independent spatial apodization, the effect of the spatial impulse response (SIR) is represented by the weighted average of the incident pressure over the receiving aperture $\Gamma^{[i_{\mathrm{rx}}]}\subset\mathbb{R}^d$~\cite{wangImagingModelIncorporating2011,wiseRepresentingArbitraryAcoustic2019}:
\begin{equation}\label{eq:sir-continuous}
    \bar{p}^{[i_{\mathrm{rx}}]}(t)
    =
    \frac{1}{\lvert\Gamma^{[i_{\mathrm{rx}}]}\rvert}
    \int_{\Gamma^{[i_{\mathrm{rx}}]}} a_{\mathrm{rx}}^{[i_{\mathrm{rx}}]}(\br')\,p(\br',t)\,d\Gamma^{[i_{\mathrm{rx}}]}(\br').
\end{equation}
Here, $\lvert\Gamma^{[i_{\mathrm{rx}}]}\rvert=\int_{\Gamma^{[i_{\mathrm{rx}}]}}d\Gamma^{[i_{\mathrm{rx}}]}(\br')$ is the aperture measure.
This normalization makes $\bar{p}^{[i_{\mathrm{rx}}]}$ an aperture-averaged pressure in $\mathrm{Pa}$, whereas the transmit source in Eq.~\eqref{eq:continuous-source-waveforms} retains the aperture measure.
The apodization $a_{\mathrm{rx}}^{[i_{\mathrm{rx}}]}(\br')$ describes the relative spatial sensitivity and is assumed bounded, dimensionless, real-valued, and independent of time and frequency.

Let $e^{[i_{\mathrm{rx}}]}(t)$ denote the impulse response of the linear time-invariant receive chain, including its electroacoustic response and element-specific gain.
When the aperture average of $a_{\mathrm{rx}}^{[i_{\mathrm{rx}}]}(\br')$ is nonzero, it can be normalized to one, with the overall scale absorbed into $e^{[i_{\mathrm{rx}}]}$.
For an acquisition duration $T>0$ and sampling interval $\Dt>0$, let $S=\lfloor T/\Dt\rfloor$ and define $t^{(s)}=s\Dt$, $s=0,\dots,S$. The sampling aperture centered at $t^{(s)}$ is denoted by $g^{[i_{\mathrm{rx}}]}(t-t^{(s)})$.
Combining the receive-chain impulse response $e^{[i_{\mathrm{rx}}]}$ with the finite temporal sampling aperture defines the effective temporal impulse response (ETIR) $h^{[i_{\mathrm{rx}}]}$ of receiving element $i_{\mathrm{rx}}$:
\begin{equation}
\label{eq:effective_temporal_response}
h^{[i_{\mathrm{rx}}]}(\tau)
=
\int_{\mathbb{R}}
g^{[i_{\mathrm{rx}}]}(\tau')\,e^{[i_{\mathrm{rx}}]}(\tau'+\tau)\,d\tau' .
\end{equation}
The ETIR is common to all points within a receiving aperture but may differ between elements.
These effective responses are assumed causal relative to the chosen acquisition time origin, with a common support interval $[0,T_r]$, where $T_r<T$.
The noise-free data matrix $\vG\in\mathbb{R}^{(S+1)\times\Nr}$ is modeled as
\begin{equation}
\label{eq:meas}
[\vG]_{s,i_{\mathrm{rx}}}
=
\int_{0}^{T_r} h^{[i_{\mathrm{rx}}]}(\tau)\,\bar{p}^{[i_{\mathrm{rx}}]}(t^{(s)}-\tau)\,d\tau,
\end{equation}
for $s=0,\dots,S$ and $i_{\mathrm{rx}}=0,\dots,\Nr-1$, where $\bar{p}^{[i_{\mathrm{rx}}]}(t)=0$ for $t<0$.
Measured data additionally contain noise; in subsequent sections, $\vG$ denotes the measured data.

As for transmission, the spatial apodization in Eq.~\eqref{eq:sir-continuous} can be replaced by a spatiotemporal aperture response.
For convenience, the common receive-chain and sampling-aperture response is retained in the ETIR, while the remaining position-dependent effects are represented by convolving the incident pressure at each aperture point with the aperture response.
This receive model is detailed in~\ref{app:aperture-temporal}.

\subsection{Discrete-to-discrete forward model} \label{back:dd}
Discretizing the wave and receive models in Section~\ref{back:cd} yields the forward map used to derive the discrete adjoint in Section~\ref{sec:dta}.
At each \textit{k}-PSTD time step, the pressure is averaged over the receiving apertures to account for the SIR.
The resulting pressure traces are interpolated to the measurement time grid and convolved with the ETIR.

\subsubsection{Discrete wave model using \textit{k}-PSTD}\label{back:dd:wave}
A bounded open domain $\Omega\subset\mathbb{R}^d$ is discretized on a Cartesian grid with spacings $d_{x_m}$, $m=1,\dots,d$, and index set $\mathbb{N}_{\Omega}\coloneqq\prod_{m=1}^d\{0,\dots,N_m-1\}$.
The total number of grid points is $N=\prod_{m=1}^{d}N_m$.
The domain includes a perfectly matched layer (PML) at its outer boundaries.
The grid coordinates satisfy $x_m^{[\vn]}=x_m^{[\vec{0}]}+d_{x_m}n_m$ for $\vn=(n_1,\dots,n_d)\in\mathbb{N}_{\Omega}$, with $\br^{[\vn]}=(x_1^{[\vn]},\dots,x_d^{[\vn]})$.
The points $\{\br^{[\vn]}:\vn\in\mathbb{N}_{\Omega}\}$ form the primary grid.
The $m^+$-staggered grid is obtained by shifting these points by $d_{x_m}/2$ in the positive $x_m$ direction.
The time grid is $t^{(\ell)}=\ell\dt$, $\ell=0,\dots,L$, where $L=\lceil S\Dt/\dt\rceil$ ensures coverage of the measurement interval even when $\dt\neq\Dt$.
The time step is set through the Courant--Friedrichs--Lewy (CFL) number $\dt\,\max_{\vn}[\vCo]_{\vn}/\min_m d_{x_m}$.
The sampled source waveforms have indices $\ell=0,\dots,L_s$, where $L_s=\lfloor T_s/\dt\rfloor<L$.

\begin{table}[!htb]
\centering
\caption{Discrete representations of quantities and spatial operators for \textit{k}-PSTD.}
\label{tab:wave-discretization}
\small
\renewcommand{\arraystretch}{1.25}
\begin{tabular}{@{}p{0.25\linewidth}p{\dimexpr0.75\linewidth-2\tabcolsep\relax}@{}}
\hline\hline
 & Discrete representation \\
\hline
Acoustic properties
& $[\vCo]_{\vn}=c_0(\br^{[\vn]})$, $[\vRo]_{\vn}=\rho_0(\br^{[\vn]})$,
  $[\vAo]_{\vn}=\alpha_0(\br^{[\vn]})$,\newline $[\vBo]_{\vn}=(B/A)(\br^{[\vn]})$ \\[3pt]
\hline
Sources
& $[\vPo]_{\vn}=p_0(\br^{[\vn]})$,
  $[\vec{s}_p^{[i_{\mathrm{tx}}]}]_{\ell}=s_p^{[i_{\mathrm{tx}}]}(t^{\slp})$,
  $[\vec{s}_u^{[i_{\mathrm{tx}}]}]_{\ell}=s_u^{[i_{\mathrm{tx}}]}(t^{(\ell)})$ \\[3pt]
\hline
Acoustic fields
& $[\vP^{(\ell)}]_{\vn}=p(\br^{[\vn]},t^{(\ell)})$ \\[3pt]
& $\vRs^{(\ell)}=\sum_{m=1}^{d}\vRm^{(\ell)}$ \\[3pt]
& $[\vUm^{\slpm}]_{\vn}=u_m\!\big(\br^{[\vn]}\!+\!\tfrac{d_{x_m}}{2}\bm{e}_m,t^{(\ell)}\pm\tfrac{\dt}{2}\big)$, $\vU^{\slpm}=(\vUm^{\slpm})_{m=1}^{d}$ \\
\hline
$\p/\p x_m$
\newline $(\sfDpm:\cR\to\cR)$
& $\sfDpm(\vV)=\sfFri\!\big(\imath\vK_m\had\Kp\had\exp(\pm\tfrac{\imath d_{x_m}}{2}\vK_m)\had\sfFr(\vV)\big)$ \\[3pt]
$\nabla\cdot$
\newline $(\divg:\cR^d\to\cR)$
& $\displaystyle\divg\big((\vV_{\ssp})_{m=1}^{d}\big)=\sum_{m=1}^{d}\sfDm(\vV_{\ssp})$ \\[3pt]
$(-\nabla^2)^a$
\newline $(\sfD^a:\cR\to\cR)$
& $\sfD^a(\vV)=\sfFri\!\big((\vK^2)^a\had\sfFr(\vV)\big)$ \\[3pt]
Staggering
\newline $(\sfIp:\cR\to\cR)$
& $[\sfIp(\vV)]_{\vn}=\tfrac{1}{2}\big([\vV]_{\vn}+[\vV]_{\vn+s\bm{e}_m}\big)$, $s=\min\{1,N_m-1-n_m\}$ \\[3pt]
PML
& $\pml$ (primary grid) and $\pmlp$ ($m^+$-staggered grid) along axis $m$\\
\hline\hline
\end{tabular}
\par\smallskip
{\footnotesize\raggedright
Here, $\bm{e}_m$ is the unit vector along axis $m$; $\sfFr$ and $\sfFri$ denote the spatial discrete Fourier transform and its inverse; $\imath=\sqrt{-1}$; $\had$ and $\hdiv$ denote elementwise multiplication and division, respectively.
The wavenumber arrays satisfy $\vK=\sqrt{\sum_{m=1}^{d}\vK_m^2}$, where $\vK_m$ corresponds to axis $m$.
For absorption and dispersion, the multiplier of $\sfD^a$ at zero wavenumber is set to zero.
The \textit{k}-space correction is $\Kp=\sinc(\dt c_{\mathrm{ref}}\vK/2)$ for reference SOS $c_{\mathrm{ref}}$, where $\sinc(x)=\sin(x)/x$ and $\sinc(0)=1$.
\par}
\end{table}

Table~\ref{tab:wave-discretization} summarizes the discrete representations used in the \textit{k}-PSTD updates.
Scalar spatial fields are represented by arrays in $\cR\coloneqq\mathbb{R}^{N_1\times\dots\times N_d}$.
As in \texttt{k-Wave}, the acoustic density is split into $d$ Cartesian components $\vRm^{(\ell)}$, $m=1,\dots,d$, so that directional PML damping can be applied to each~\cite{treebyModelingNonlinearUltrasound2012}.
Pressure and these split-density variables are represented on the primary grid, whereas velocity is staggered in space and time.
The discrete operators $\sfDpm$ approximate $\p/\p x_m$: $\sfDp$ maps primary-grid values to derivatives on the $m^+$-staggered grid, whereas $\sfDm$ maps $m^+$-staggered values to derivatives on the primary grid.
The input and output of $\sfD^a$ are both defined on the primary grid.
The divergence $\divg$ sums the derivatives of the staggered component arrays on the primary grid.

The \textit{k}-PSTD updates follow Treeby \textit{et al.}~\cite{treebyModelingNonlinearUltrasound2012}.
The terms $-\bm{u}\cdot\nabla\rho_0$ in Eq.~\eqref{eq:mass} and $\bm{d}\cdot\nabla\rho_0$ in Eq.~\eqref{eq:eos} cancel in the coupled system to the order of approximation retained there and are omitted from the discrete updates.
Consequently, $\vRs^{(\ell)}$ is an auxiliary density variable rather than a direct sampling of $\rho$ in Eqs.~\eqref{eq:mass} and~\eqref{eq:eos}.
At time index $\ell$, the discrete acoustic state is defined as
\begin{equation} \label{eq:discrete_state}
\vX^{(\ell)} \coloneqq \Big( \vU_{1^+}^{\slm},\dots,\vU_{d^+}^{\slm}, \vR_1^{(\ell)},\dots,\vR_d^{(\ell)}, \vP^{(\ell)} \Big) \in\cR^{2d\!+\!1}.
\end{equation}

The initial conditions in Eq.~\eqref{eq:ic} are discretized as follows:
\begin{equation}\label{eq:fwd_ic}
 \vP^{(0)} = \vPo, \qquad \vUm^{\scriptstyle (\!-\!\frac{1}{2})}=\tfrac{\dt}{2}\sfDp\big(\vPo\big)\hdiv\sfIp\big(\vRo\big),\qquad\vRm^{(0)}=\tfrac{1}{d}\vPo\hdiv\vCo^2.
\end{equation}
For $\ell=0,\dots,L-1$, the \textit{k}-PSTD time-marching updates are given by
\begin{align}
\vUm^{\slp} &= \pmlp\had\Big[\pmlp\had\vUm^{\slm}
  \!-\!\dt\,\sfDp\big(\vP^{(\ell)}\big)\hdiv\sfIp\big(\vRo\big)\Big]\!+\!\vtFm^{(\ell)},\label{eq:fwd_u}\\
\vRm^{(\ell\!+\!1)} &= \pml\had\Big[\pml\had\vRm^{(\ell)}
  \!-\!\dt\,\vbR^{(\ell)}\had\sfDm\big(\vUm^{\slp}\big)\Big]\!+\!\vtMm^{\slp},\label{eq:fwd_r}\\
\vP^{(\ell\!+\!1)}&=\vCo^{2}\had\Big[\vRs^{(\ell\!+\!1)}\!+\!\vTo\had\sfQ_{\mathrm{A}}\big(\vU^{\slp}\big)\nonumber\\
&\qquad\quad-\vEo\had\sfQ_{\mathrm{D}}\big(\vRs^{(\ell\!+\!1)}\big)\!+\!\sfN\big(\big(\vRs^{(\ell\!+\!1)}\big)^2\big)\Big]. \label{eq:fwd_p}
\end{align}
Here, $\vtMm^{\slp}$ and $\vtFm^{(\ell)}$ are the additive split-density and particle-velocity source increments, respectively.
The source construction for the separable excitation in Eq.~\eqref{eq:continuous-source-waveforms} is detailed in~\ref{app:source-convention}; its spatiotemporal extension is given in~\ref{app:aperture-temporal-discrete}.
The operator $\sfIp$ in Table~\ref{tab:wave-discretization} maps primary-grid quantities to the staggered grid along axis $m$.
The coefficients $\vTo=-2\vAo\had\vCo^{\gamma-1}$ and $\vEo=2\tan(\pi\gamma/2)\vAo\had\vCo^{\gamma}$ correspond to the discrete representation of $\tau_0(\br)$ and $\eta_0(\br)$, respectively, with powers evaluated elementwise.
The operators $\sfQ_{\mathrm{A}}:\cR^d\to\cR$, $\sfQ_{\mathrm{D}}:\cR\to\cR$, and $\sfN:\cR\to\cR$ represent absorption, dispersion, and material nonlinearity, respectively:
$\sfQ_{\mathrm{A}}(\vV)\coloneqq\sfD^{\frac{\gamma-2}{2}}\big(\vRo\had\divg(\vV)\big)$,
$\sfQ_{\mathrm{D}}(\vV)\coloneqq\sfD^{\frac{\gamma-1}{2}}(\vV)$, and
$\sfN(\vV)\coloneqq\tfrac{1}{2}\vBo\had\vV\hdiv\vRo$.
The factor $\vbR^{(\ell)}\coloneqq2\vRs^{(\ell)}+\vRo$ accounts for convective nonlinearity in the mass-conservation update~\eqref{eq:fwd_r}.
It is evaluated explicitly at the current time level~\cite{treebyNonlinearUltrasoundSimulation2020}.
For the linear acoustic model, $\vbR^{(\ell)}=\vRo$ and $\sfN(\cdot)\equiv\vec{0}_{\cR}$.
The arrays $\pml$ and $\pmlp$ specify the PML damping factors along axis $m$ on the primary and $m^+$-staggered grids, respectively.
Following \texttt{k-Wave}~\cite{treebyWaveMatlabToolbox2010}, their entries are $\exp\!\big(-\tfrac{\dt}{2}a_m\big)$ with $a_m=\alpha_{\mathrm{PML}}\,(c_{\mathrm{ref}}/d_{x_m})\,(\delta_m/N_{\mathrm{PML},m})^4$.
Here, $\delta_m$ is the depth into the PML along axis $m$ in grid points, evaluated on the corresponding grid, and $N_{\mathrm{PML},m}$ is the PML thickness along axis $m$ in grid points.
The absorption strength $\alpha_{\mathrm{PML}}$ is expressed in nepers per grid point.

\subsubsection{Discrete-to-discrete receive model}\label{back:dd:receive}
Let $\bar{\vP}\in\mathbb{R}^{(L+1)\times\Nr}$ denote the discrete representation of the aperture-averaged pressure, with $[\bar{\vP}]_{\ell,i_{\mathrm{rx}}}$ corresponding to $\bar{p}^{[i_{\mathrm{rx}}]}(t^{(\ell)})$. At each time step, the discrete spatial receive operator $\sftS$ is defined by
\begin{equation}
\label{eq:dd-sir-step}
    [\bar{\vP}]_{\ell,\cdot} \coloneqq \sftS(\vP^{(\ell)})
    =
    \bigl(\langle \vec{S}_{\mathrm{rx}}^{[i_{\mathrm{rx}}]},\vP^{(\ell)}\rangle_{\cR}\bigr)_{i_{\mathrm{rx}}=0}^{\Nr-1},
\end{equation}
where $\vec{S}_{\mathrm{rx}}^{[i_{\mathrm{rx}}]}\in\cR$ denotes the discrete spatial receive kernel for receiving element $i_{\mathrm{rx}}$~\cite{wiseRepresentingArbitraryAcoustic2019}; see~\ref{app:transducer-operators} for its expression.
This time-independent receive action corresponds to the separable model in Eq.~\eqref{eq:sir-continuous}; its spatiotemporal generalization is given in~\ref{app:aperture-temporal-discrete}.
Collecting these receiver vectors as rows for successive time steps defines the discrete spatial receive operator $\sfS$:
\begin{equation}
\label{eq:dd-sir-full}
    \bar{\vP}
    =
    \sfS(\vP)
    =
   \bigl(\sftS(\vP^{(\ell)})\bigr)_{\ell=0}^{L},
\end{equation}
where $\vP=\bigl(\vP^{(\ell)}\bigr)_{\ell=0}^{L}$.

The data sampling interval $\Dt$ is assumed small enough to resolve the bandwidth retained in the ETIR-convolved measurements and to satisfy $\Dt\leq\dt$.
Let $\cG\coloneqq\mathbb{R}^{(S+1)\times\Nr}$ denote the discrete data space, and let $\mathsf{Z}:\mathbb{R}^{(L+1)\times\Nr}\to\cG$ denote temporal interpolation from the \textit{k}-PSTD time grid $(\ell\dt)_{\ell=0}^{L}$ to the data time grid $(s\Dt)_{s=0}^{S}$.
Thus, $\bar{\vP}_{\Dt}=\mathsf{Z}(\bar{\vP})$, with $\mathsf{Z}$ performing temporal upsampling when $\Dt<\dt$ and reducing to the identity when $\Dt=\dt$.

With $S_r=\lfloor T_r/\Dt\rfloor$, the discrete ETIR coefficients are collected in $\vH\in\mathbb{R}^{(S_r+1)\times\Nr}$ with entries
\begin{equation}
    [\vH]_{r,i_{\mathrm{rx}}}\coloneqq\Dt\,h^{[i_{\mathrm{rx}}]}(r\Dt),
    \qquad r=0,\dots,S_r.
\end{equation}
For $\vec{A}\in\cG$, define the causal channelwise convolution $\mathsf{C}_{\vH}:\cG\to\cG$ by
\begin{equation*}
    [\mathsf{C}_{\vH}(\vec{A})]_{s,i_{\mathrm{rx}}}
    \coloneqq
    \sum_{r=0}^{\min(s,S_r)}[\vH]_{r,i_{\mathrm{rx}}}[\vec{A}]_{s-r,i_{\mathrm{rx}}},
    \qquad s=0,\dots,S.
\end{equation*}
An optional temporal filter $\sfB$ can attenuate frequencies for which the spatial grid has fewer than a prescribed number $n_{\mathrm{PPW}}$ of points per wavelength (PPW).
The same filter is applied to the measured data before comparison.
It limits the bandwidth used for data comparison but does not correct propagation errors already introduced by the discretization.
The total temporal receive operator is
\begin{equation}
\label{eq:dd-eir-full}
    \sfH\coloneqq\sfB\circ\mathsf{C}_{\vH}\circ\mathsf{Z},
\end{equation}
with $\sfB$ set to the identity when filtering is omitted.
The temporal receive operators implemented in this work are detailed in~\ref{app:temporal-operators}.

For the discrete-adjoint derivation in Section~\ref{sec:dta}, the complete forward model is written compactly as
\begin{equation}
\label{eq:dd-full}
\sfG(\cS,\vH)
\coloneqq
\sfhG_{\vH}\big(\sfX(\cS)\big)
\in\cG.
\end{equation}
Here, $\cS$ collects the medium and source parameters required by the chosen propagation model and excitation type.
For a fixed acquisition configuration, $\sfX(\cS)=\vX$ is the state sequence obtained by \textit{k}-PSTD time marching.
The receive map is $\sfhG_{\vH}\coloneqq\sfH\circ\sfS\circ\sfP$, where $\sfP$ extracts the pressure sequence.

\section{Discrete adjoint method}\label{sec:dta}
Let $\bt$ collect the unknown entries selected from $\cS$ and the ETIR coefficients $\vH$.
It may be a scalar, such as $\gamma$, a single parameter array, or a collection of parameter arrays and scalars.
For one acquisition event with measured data $\vG$, parameter estimation is formulated as
\begin{equation}\label{eq:inv_prob}
    \min_{\bt\in\bT}
    \left\{
    \sfJ_{\sfR}(\bt)
    \coloneqq
    \sfJ\big(\sfG(\cS,\vH);\vG\big)
    +
    \sfR(\bt)
    \right\},
\end{equation}
where $\bT$ denotes the parameter space and $\sfR:\bT\to\mathbb{R}\cup\{+\infty\}$ is a regularization functional.
The data-fidelity functional $\sfJ:\cG\times\cG\to\mathbb{R}$ measures the discrepancy between predicted and measured data and is differentiable with respect to the predicted data.
The regularization term $\sfR$ is handled separately, using its derivative when differentiable or a proximal method for nonsmooth penalties.
We therefore focus on deriving the total derivative $\mathrm{d}\sfJ/\mathrm{d}\bt$ from the discrete forward model in Section~\ref{back:dd} using DTA.

\subsection{Derivation of the discrete-adjoint system}
Introduce an auxiliary state sequence $\vhX = (\vhX^{(\ell)})_{\ell=0}^{L}$, treated as independent of $\bt$, with the same component ordering as the forward state in Eq.~\eqref{eq:discrete_state}:
\begin{equation*}
\vhX^{(\ell)}\coloneqq(\vhX_i^{(\ell)})_{i=1}^{2d\!+\!1}\coloneqq\left(\vhU_{1^+}^{\slm},\dots,\vhU_{d^+}^{\slm}, \vhR^{(\ell)}_1,\dots,\vhR^{(\ell)}_d, \vhP^{(\ell)}\right) \in \cR^{2d\!+\!1}.
\end{equation*}
The \textit{k}-PSTD updates in
Eqs.~\eqref{eq:fwd_u}--\eqref{eq:fwd_p} are expressed through the residual sequences $\big(\big(\vhA_j^{(\ell)}\big)_{j=1}^{2d+1}\big)_{\ell=0}^{L}$.
The nonzero residual scalings below simplify the subsequent adjoint equations without changing the forward solution.
For $m=1,\dots,d$ and $\ell=1,\dots,L$, define
\begin{equation}\label{eq:fwd_u_res}
\vhA_m^{(\ell)} \coloneqq \tfrac{1}{\dt}\Big[\pmlp\had\vhUm^{\scriptstyle (\ell\!-\!\frac{3}{2})}\!-\!\big(\vhUm^{\slm}\!-\!\vtFm^{(\ell\!-\!1)}\big)\hdiv \pmlp\Big]\had\sfIp\big(\vRo\big) \!-\!\sfDp\big(\vhP^{(\ell\!-\!1)}\big),
\end{equation}
\begin{equation}\label{eq:fwd_r_res}
\vhA_{m\!+\!d}^{(\ell)} \coloneqq \tfrac{1}{\dt}\Big[
\pml\had\vhRm^{(\ell\!-\!1)} \!-\!\big(\vhRm^{(\ell)}\!-\!\vtMm^{\slm}\big)\hdiv\pml\Big]
  \!-\!\vhbR^{(\ell\!-\!1)}\had\sfDm\big(\vhUm^{\slm}\big),
\end{equation}
\begin{equation}\label{eq:fwd_p_res}
\vhA_{2d\!+\!1}^{(\ell)}\coloneqq \vhRs^{(\ell)}\!+\!\vTo\had\sfQ_{\mathrm{A}}\big(\vhU^{\slm}\big)\!-\!\vEo\had\sfQ_{\mathrm{D}}\big(\vhRs^{(\ell)}\big)\!+\! \sfN\big(\big(\vhRs^{(\ell)}\big)^2\big)\!-\!\vhP^{(\ell)}\hdiv \vCo^{2},
\end{equation}
and the initial conditions in Eq.~\eqref{eq:fwd_ic} give
\begin{align}
    \vhA_m^{(0)} \coloneqq& \tfrac12\sfDp(\vPo) - \tfrac{1}{\dt}\vhUm^{\scriptstyle (\!-\!\frac{1}{2})}\had\sfIp\big(\vRo\big),\\
    \vhA_{m+d}^{(0)} \coloneqq& \tfrac{1}{d\,\dt}\vPo\hdiv\vCo^2\!-\!\tfrac{1}{\dt}\vhRm^{(0)},\\
    \label{eq:res-p}
    \vhA_{2d+1}^{(0)} \coloneqq& \vPo - \vhP^{(0)}.
\end{align}
Here, $\vhRs^{(\ell)}\!\coloneqq\!\sum_{m=1}^d\vhRm^{(\ell)}$, $\vhbR^{(\ell\!-\!1)}\!\coloneqq\! 2\vhRs^{(\ell\!-\!1)}+\vRo$, and $\vhU^{\slm}\!\coloneqq\! \big(\vhUm^{\slm}\big)_{m=1}^d$.
For linear acoustics, $\vhbR^{(\ell\!-\!1)}=\vRo$ and $\sfN(\cdot)\equiv\vec{0}_{\cR}$, as in Section~\ref{back:dd:wave}.
Setting all residuals to zero recovers both the \textit{k}-PSTD updates and their initial conditions.

Minimizing the data-fidelity term alone is therefore equivalent to the constrained problem
\begin{equation}
    \min_{\bt\in\bT,\,\vhX}\sfJ(\sfhG_{\vH}(\vhX);\vG)  \quad\text{subject to } \vhA^{(\ell)}_j=\vec{0}_{\cR},
\end{equation}
for $j=1,\dots,2d+1$ and $\ell=0,\dots,L$.
Let $\vhY \coloneqq((\vhY^{(\ell)}_j)_{j=1}^{2d\!+\!1})_{\ell=0}^L$, with $\vhY_j^{(\ell)}\in\cR$, denote the Lagrange multipliers for these constraints.
The Lagrangian is
\begin{equation}
\cL(\bt, \vhX,\vhY) \coloneqq \sfJ(\sfhG_{\vH}(\vhX);\vG) + \sum_{\ell=0}^{L}\sum_{j=1}^{2d\!+\!1} \Big\langle\vhA^{(\ell)}_j,\vhY^{(\ell)}_j \Big\rangle_{\cR}.
\end{equation}
Here, $\langle\cdot,\cdot\rangle_{\cR}$ is the standard inner product on $\cR$.
When the constraints are satisfied, $\vhX$ coincides with the forward-state sequence $\vX$ and the Lagrangian equals the data-fidelity term.
For a given $\bt$ and its forward solution $\vX$, choose the multipliers $\vhY=\vY \coloneqq((\vY^{(\ell)}_j)_{j=1}^{2d\!+\!1})_{\ell=0}^L$ so that the Lagrangian is stationary with respect to the auxiliary states:
\begin{equation}\label{eq:adjoint-stationarity}
\left.\nabla_{\vhX_i^{(k)}}\cL(\bt,\vhX,\vY)\right|_{\vhX=\vX}=\vec{0}_{\cR},\qquad
i=1,\dots,2d+1,\quad k=0,\dots,L.
\end{equation}
For an entry $\theta\in\mathbb{R}$ of $\bt$, the vanishing residuals and this stationarity condition eliminate the chain-rule terms involving derivatives of $\vY$ and $\vX$, respectively.
The total derivative is therefore
\begin{align}\label{eq:grad}
    \dd{\sfJ\big(\sfG(\cS,\vH);\vG\big)}{\theta} &=
    \dd{\cL(\bt,\vX,\vY)}{\theta}
    \nonumber\\&=\left.\pd{\sfJ\big(\sfhG_{\vH}(\vhX);\vG\big)}{\theta}\right|_{\vhX=\vX}
    +\sum_{\ell=0}^{L}\sum_{j=1}^{2d+1}
    \bigg\langle \pd{\vhA^{(\ell)}_{j}}{\theta}\bigg|_{\vhX=\vX},\vY_j^{(\ell)}\bigg\rangle_{\cR}.
\end{align}
The partial derivatives in Eq.~\eqref{eq:grad} hold the auxiliary states fixed.
The first term accounts for the ETIR coefficients $\vH$ in the receive operator $\sfhG_{\vH}$.
It vanishes for the parameters in $\cS$, which affect the data only through $\vX$.
Expanding the stationarity condition in Eq.~\eqref{eq:adjoint-stationarity} yields the discrete-adjoint system, for
$i=1,\dots,2d+1$ and $k=0,\dots,L$:
\begin{equation}\label{eq:AS1}
    \sum_{\ell=0}^{L}\sum_{j=1}^{2d\!+\!1}\left(\frac{\p\vhA^{(\ell)}_j}{\p\vhX^{(k)}_i}\bigg|_{\vhX=\vX}\right)^*\vY^{(\ell)}_j  = -\left.\nabla_{\vhX_i^{(k)}}\sfJ\big(\sfhG_{\vH}(\vhX);\vG\big)\right|_{\vhX=\vX}.
\end{equation}
Here, $^*$ denotes the adjoint of each residual Jacobian with respect to the inner product on $\cR$.
The spatial operators satisfy $(\sfDp)^*=-\sfDm$ and $(\sfDm)^*=-\sfDp$, including the Nyquist bins; real even Fourier multipliers and real elementwise multipliers are self-adjoint.

Expanding this system componentwise gives a backward time-marching scheme for the discrete adjoint of the \textit{k}-PSTD updates.
The adjoint states are written in the corresponding primary- and staggered-grid notation as
$\lu^{\skm}=\vY_{m}^{(k)}$, $\lr^{(k)}= \vY_{m\!+\!d}^{(k)}$, and $\lp^{(k)} = \vY_{2d\!+\!1}^{(k)}$ for $k=0,\dots,L$.
With the terminal conditions
$\lu^{\scriptstyle (L\!+\!\frac{1}{2})}=\vec{0}_{\cR}$ and
$\lr^{(L\!+\!1)}=\vec{0}_{\cR}$ for $m=1,\dots,d$, the resulting adjoint recursion for $k=L,\dots,1$ is given by
\begin{equation}\label{eq:AS2-lp}
    \lp^{(k)} = \vCo^2\had\Big[\divg\big(\boldsymbol{\Phi}^{\scriptstyle(k\!+\!\frac12)}\big)+ \adsrc^{(k)}\Big],
\end{equation}
\begin{equation}
    \label{eq:AS2-lr}
    \lr^{(k)} = \pml\had\Big[\pml\had\lr^{(k\!+\!1)} \!+\! \dt \vec{I}\big(\boldsymbol{\Psi}^{(k\!+\!1)},\lp^{(k)}\big)\Big],
\end{equation}
\begin{equation}\label{eq:AS2-lu}
    \lu^{\skm} = \pmlp\had\Big[\pmlp\had\lu^{\skp}\!+\! \dt\vec{J}_{m}\big(\lr^{(k)},\lp^{(k)}\big)\Big],
\end{equation}
where $\boldsymbol{\Phi}^{\scriptstyle(k\!+\!\frac12)}\!\coloneqq\!\big(\lu^{\scriptstyle(k\!+\!\frac12)}\big)_{m=1}^d$, $\boldsymbol{\Psi}^{(k)}\!\coloneqq\!\big(\lr^{(k)}\big)_{m=1}^d$, and $\lrs^{(k)}\coloneqq\sum_{m=1}^d\lr^{(k)}$.
The divergence $\divg$ is the operator defined in Table~\ref{tab:wave-discretization}.
The increment terms $\vec{I}$ and $\vec{J}_{m}$ are given by
\begin{align}\label{eq:inc_1}
    \vec{I}\big(\boldsymbol{\Psi}^{(k\!+\!1)},\lp^{(k)}\big)\!=&\lp^{(k)}\underbrace{\!-\sfD^{\frac{\gamma-1}{2}}\big(\vEo\had\lp^{(k)}\big)}_{\text{Dispersion}} \underbrace{\!-2\sum_{j=1}^{d}\lrj^{(k\!+\!1)}\had\sfDj\big(\vUj^{\skp}\big)}_{\text{Convective nonlinearity}}\nonumber\\ &\!\underbrace{+2\sfN\big(\vRs^{(k)}\had\lp^{(k)}\big)}_{\text{Material nonlinearity}},
\end{align}
and
\begin{equation}\label{eq:inc_2}
    \vec{J}_{m}\big(\lr^{(k)},\lp^{(k)}\big) \!=\!\sfDp\big(\big(\underbrace{2\vRs^{(k\!-\!1)}}_{\text{\shortstack{Convective \\ nonlinearity}}} \!+ \vRo\big)\had\lr^{(k)}\underbrace{\!-\vRo\had\sfD^{\frac{\gamma-2}{2}}\big(\vTo\had\lp^{(k)}\big)}_{\text{Absorption}}\big)\hdiv\sfIp(\vRo).
\end{equation}
For linear acoustics, the marked convective- and material-nonlinearity terms are omitted, consistent with Section~\ref{back:dd:wave}.
At $k=L$, the terms multiplied by $\boldsymbol{\Psi}^{(L+1)}$ vanish, so no forward state beyond the final time index is needed.
At $k=0$,
\begin{align}
    \lp^{(0)} =& \divg\big(\boldsymbol{\Phi}^{\scriptstyle(\frac12)}\big)+ \adsrc^{(0)}, \\
    \lr^{(0)} =& \pml\had\lr^{(1)}\!\underbrace{-2\dt\sum_{j=1}^{d}\lrj^{(1)}\had\sfDj\big(\vUj^{\scriptstyle (\frac{1}{2})}\big)}_{\text{Convective nonlinearity}}, \\
    \lu^{\scriptstyle (\!-\!\frac{1}{2})} =& \pmlp\had\lu^{\scriptstyle (\frac{1}{2})}.
\end{align}
The different scaling of $\lp$ arises because the pressure residual contains $-\vhP^{(k)}\hdiv\vCo^2$ for $k\geq1$, but $-\vhP^{(0)}$ at $k=0$.

The adjoint source term $\adsrc^{(k)}=\nabla_{\vhP^{(k)}} \sfJ(\sfhG_{\vH}(\vhX);\vG)\big|_{\vhX=\vX}$ in Eq.~\eqref{eq:AS2-lp} can be computed as
\begin{equation}
\label{eq:adjoint-source}
  \adsrc^{(k)} = \sftS^{*}\big(\big[\vW^{\sfT}\big]_{\cdot,k}\big), \qquad \vW = \sfH^*\big(\nabla_{\vV}\sfJ(\vV;\vG)\big|_{\vV = \sfhG_{\vH}(\vX)}\big).
\end{equation}
Here, $\vW\in\mathbb{R}^{(L+1)\times\Nr}$ is defined on the propagation time grid.
The temporal adjoint $\sfH^*=\mathsf{Z}^*\circ\mathsf{C}_{\vH}^*\circ\sfB^*$ applies the adjoints of the operations in Eq.~\eqref{eq:dd-eir-full} in reverse order.
The adjoint of $\sftS$, $\sftS^{*}$, maps $\vec{w}\in\mathbb{R}^{\Nr}$ to a spatial array through a weighted sum of the receive kernels:
\begin{equation}
\label{eq:dd-sir-adjoint}
\sftS^{*}(\vec{w})
=
\sum_{i_{\mathrm{rx}}=0}^{\Nr-1}[\vec{w}]_{i_{\mathrm{rx}}}\,\vec{S}_{\mathrm{rx}}^{[i_{\mathrm{rx}}]} .
\end{equation}

\subsection{Gradient formulas}\label{sec:parameter-gradients}
Substituting the discrete-adjoint states $\vY$ into the second term of Eq.~\eqref{eq:grad} gives the gradients with respect to the parameters in $\cS$.
All requested gradients evaluated this way share the same adjoint solution.
The ETIR gradient follows from the first term and does not require adjoint recursion.
The derivative with respect to the scalar power-law exponent $\gamma$ can alternatively be evaluated by direct differentiation (DD) of the discrete time-marching equations, as detailed below.

For linear acoustics, the adjoint recursion in Eqs.~\eqref{eq:AS2-lp}--\eqref{eq:AS2-lu} requires no forward time histories.
For nonlinear acoustics, the increments in Eqs.~\eqref{eq:inc_1} and~\eqref{eq:inc_2} require $\big(\sfDm(\vUm^{\slm})\big)_{m=1}^{d}$ for $\ell=1,\dots,L$ and $\vRs^{(\ell)}$ for $\ell=0,\dots,L$ over $\mathbb{N}_{\Omega}$, where $\vRs^{(0)}=\vPo\hdiv\vCo^2$ follows from the initial condition in Eq.~\eqref{eq:fwd_ic} and vanishes for USCT.
Additional histories needed for each parameter gradient are specified below; histories shared by several gradients are stored only once.

\subsubsection{Spatial distributions}\label{sec:grad-spatial}
To reduce forward-history storage, the gradients with respect to $\vCo$, $\vAo$, $\vRo$, and $\vBo$ are evaluated over a prescribed region of interest (ROI) rather than the full spatial grid.
This ROI is defined as a grid-aligned rectangle in 2D or rectangular cuboid in 3D, lying entirely within the non-PML region.
Let $\mathbb{N}_{\mathrm{ROI}}\subset\mathbb{N}_{\Omega}$ denote its grid indices.

\paragraph{SOS}
For $\vCo$, the gradient is
\begin{equation}
\dd{\sfJ}{[\vCo]_{\vn}}
=\frac{2}{[\vCo]_{\vn}}
\sum_{\ell=1}^{L}[\lp^{(\ell)}]_{\vn}[\vV_c^{(\ell)}]_{\vn}
-\frac{2[\vPo]_{\vn}[\lrs^{(0)}]_{\vn}}{d\,\dt[\vCo]_{\vn}^{3}},\label{eq:grad-sos}
\end{equation}
where
\begin{equation*}
\vV_c^{(\ell)}=\vP^{(\ell)}\hdiv\vCo^2+
\frac{\gamma-1}{2}\vTo\had\sfQ_{\mathrm{A}}\big(\vU^{\slm}\big)
-\frac{\gamma}{2}\vEo\had\sfQ_{\mathrm{D}}\big(\vRs^{(\ell)}\big).
\end{equation*}
The reference SOS $c_{\mathrm{ref}}$ and the prescribed water SOS $c_{\mathrm{water}}$ are held fixed in this derivative.
Here, $\sfQ_{\mathrm{A}}\big(\vU^{\slm}\big)$ and $\sfQ_{\mathrm{D}}\big(\vRs^{(\ell)}\big)$ are already computed during forward \textit{k}-PSTD time marching.
Only the history of $\vV_c^{(\ell)}$ within the ROI is retained for this gradient.
The IP term vanishes for USCT.

\paragraph{AA prefactor}
For $\vAo$ in $\mathrm{Np}\,(\mathrm{rad\,s^{-1}})^{-\gamma}\mathrm{m}^{-1}$, the gradient is
\begin{equation}
\dd{\sfJ}{[\vAo]_{\vn}}
=-2[\vCo]_{\vn}^{\gamma-1}
\sum_{\ell=1}^{L}[\lp^{(\ell)}]_{\vn}[\vV_{\alpha}^{(\ell)}]_{\vn},
\label{eq:grad-aa}
\end{equation}
where
\begin{equation*}
\vV_{\alpha}^{(\ell)}
=\sfQ_{\mathrm{A}}\big(\vU^{\slm}\big)
+\tan\left(\frac{\pi\gamma}{2}\right)\vCo\had\sfQ_{\mathrm{D}}\big(\vRs^{(\ell)}\big).
\end{equation*}
Likewise, only the history of $\vV_{\alpha}^{(\ell)}$ within the ROI is retained.

\paragraph{Ambient density}
Within the ROI, $[\pml]_{\vn}=1$ for every direction $m$.
Since the increment in Eq.~\eqref{eq:AS2-lr} is independent of $m$ and the split-density adjoints have zero terminal values,
\begin{equation*}
[\lr^{(\ell)}]_{\vn}
=\frac{1}{d}[\lrs^{(\ell)}]_{\vn}, \qquad \vn \in \mathbb{N}_{\mathrm{ROI}},
\quad m=1,\dots,d,\quad \ell=0,\dots,L.
\end{equation*}
Using this equation, the gradient with respect to $\vRo$ is
\begin{align}
\dd{\sfJ}{[\vRo]_{\vn}}
={}&\sum_{\ell=1}^{L}\Bigg\{
\sum_{m=1}^{d}\left[\sfIm\!\left(
\lu^{\slm}\had\sfDp\big(\vP^{(\ell-1)}\big)\hdiv\sfIp(\vRo)
\right)\right]_{\vn}\nonumber\\
&\quad+\left(
\left[\sfD^{\frac{\gamma-2}{2}}\big(\vTo\had\lp^{(\ell)}\big)\right]_{\vn}
-\frac{[\lrs^{(\ell)}]_{\vn}}{d}
\right)\left[\divg\big(\vU^{\slm}\big)\right]_{\vn}\nonumber\\
&\quad-\frac{[\vBo]_{\vn}[\lp^{(\ell)}]_{\vn}}{2}
\left(\frac{[\vRs^{(\ell)}]_{\vn}}{[\vRo]_{\vn}}\right)^2
\Bigg\}\nonumber\\
&-\frac{1}{\dt}\sum_{m=1}^{d}\left[\sfIm\!\left(
\lu^{\scriptstyle(-\frac12)}\had\vUm^{\scriptstyle(-\frac12)}
\right)\right]_{\vn},
\label{eq:grad-den}
\end{align}
where $\divg(\vU^{\slm})=\sum_{m=1}^{d}\sfDm(\vUm^{\slm})$.
Within the ROI, which excludes the outermost computational grid nodes, $\sfIm$ represents the adjoint of $\sfIp$ by averaging its input at $\vn$ and $\vn-\bm e_m$.
The histories of $\sfDp(\vP^{(\ell-1)})$ are therefore stored on $\mathbb{N}_{\mathrm{ROI}}^{-}$, the rectangular ROI extended by one grid point on the negative side of each axis.

For linear acoustics, the term proportional to $\vBo$ in Eq.~\eqref{eq:grad-den} is omitted.
The computed $\sfDm(\vUm^{\slm})$ are summed over $m$ and stored within the ROI during forward time marching.
For nonlinear acoustics, $\vRs^{(\ell)}$ and $\sfDm(\vUm^{\slm})$, $m=1,\dots,d$, are already stored over the full grid for the adjoint recursion in Eqs.~\eqref{eq:inc_1} and~\eqref{eq:inc_2}.
Their ROI values are reused in Eq.~\eqref{eq:grad-den}, with the divergence formed by summing the stored $\sfDm(\vUm^{\slm})$ over $m$, so no separate divergence or density-sum history is needed.

\paragraph{Nonlinearity}
For the nonlinear model, the gradient with respect to $\vBo$ is
\begin{equation}
\dd{\sfJ}{[\vBo]_{\vn}}
=\frac{1}{2[\vRo]_{\vn}}
\sum_{\ell=1}^{L}[\lp^{(\ell)}]_{\vn}[\vRs^{(\ell)}]_{\vn}^{2}.
\label{eq:grad-BonA}
\end{equation}
The full-grid history of $\vRs^{(\ell)}$ required by the increment term in Eq.~\eqref{eq:inc_1} is reused on $\mathbb{N}_{\mathrm{ROI}}$, so no additional forward history is required.

\paragraph{IP}
Let $\mathbb{N}_{\mathrm{PML}}\subset\mathbb{N}_{\Omega}$ denote the PML indices.
The IP distribution can be nonzero and spatially varying outside the object being imaged, so its gradient is returned over the entire non-PML region:
\begin{equation}
\dd{\sfJ}{[\vPo]_{\vn}}
=[\lp^{(0)}]_{\vn}
-\frac12\left[\sfD_{\Sigma}\big(\boldsymbol\Phi^{\scriptstyle(-\frac12)}\big)\right]_{\vn}
+\frac{[\lrs^{(0)}]_{\vn}}{d\,\dt[\vCo]_{\vn}^{2}},
\label{eq:grad-p0}
\end{equation}
where $\vn\in\mathbb{N}_{\Omega}\setminus\mathbb{N}_{\mathrm{PML}}$.
This gradient requires no forward time histories beyond those needed by the adjoint recursion.

\subsubsection{Temporal waveforms and responses}\label{sec:grad-temporal}
\paragraph{Transmit waveforms}
The transmit-waveform gradients require no forward time histories beyond those needed by the adjoint recursion.
The geometric-mean grid spacing $\bar d_x$ and aperture grid measure $\Delta_{\Gamma}^{[i_{\mathrm{tx}}]}$ are defined in~\ref{app:source-convention}.
The spatial transmit coefficient array $\vec{S}_{\mathrm{tx}}^{[i_{\mathrm{tx}}]}$ is defined in~\ref{app:transducer-operators}, and $\sfKsrc$ is the source \textit{k}-space correction operator in Eq.~\eqref{eq:source-kspace-correction}.
Differentiating the source increments in Eqs.~\eqref{eq:scaled-pressure-source} and~\eqref{eq:scaled-velocity-source} with the excitation coefficients $w_{\mathrm{tx}}^{[i_{\mathrm{tx}}]}$ held fixed gives
\begin{align}
\dd{\sfJ}{[\vec{s}_p^{[i_{\mathrm{tx}}]}]_{\ell}}
&=
\frac{2w_{\mathrm{tx}}^{[i_{\mathrm{tx}}]}}{d\,c_{\mathrm{water}}\bar d_x}
\frac{\lvert\Gamma^{[i_{\mathrm{tx}}]}\rvert}{\Delta_{\Gamma}^{[i_{\mathrm{tx}}]}}
\left\langle
\sfKsrc(\vec{S}_{\mathrm{tx}}^{[i_{\mathrm{tx}}]}),
\sum_{m=1}^{d}\big(\lr^{(\ell\!+\!1)}\hdiv\pml\big)
\right\rangle_{\cR},
\label{eq:grad-source-pressure}\\
\dd{\sfJ}{[\vec{s}_u^{[i_{\mathrm{tx}}]}]_{\ell}}
&=
2c_{\mathrm{water}}w_{\mathrm{tx}}^{[i_{\mathrm{tx}}]}\frac{\lvert\Gamma^{[i_{\mathrm{tx}}]}\rvert}{\Delta_{\Gamma}^{[i_{\mathrm{tx}}]}}
\sum_{m=1}^{d}\frac{\nu_m^{[i_{\mathrm{tx}}]}}{\bar d_x}
\left\langle
\sfKsrc(\vec{S}_{\mathrm{tx}}^{[i_{\mathrm{tx}}]}),
\lu^{\slp}\had\sfIp(\vRo)\hdiv\pmlp
\right\rangle_{\cR}.
\label{eq:grad-source-velocity}
\end{align}
These expressions apply for $\ell=0,\dots,L_s$.

\paragraph{ETIR}\label{sec:grad-response}
Using the temporal receive operators in~\ref{app:temporal-operators}, define the adjoint quantity at the output of the ETIR convolution as
\begin{equation*}
\vec{Q}
\coloneqq
\sfB^{*}
\left[
\nabla_{\vV}\sfJ\big(\vV;\vG\big)
\big|_{\vV=\sfH(\bar{\vP})}
\right].
\end{equation*}
With $\bar{\vP}_{\Dt}=\mathsf{Z}(\bar{\vP})$, the gradient with respect to the discrete ETIR coefficients is
\begin{equation}
\label{eq:grad-response}
\dd{\sfJ}{[\vH]_{r,i_{\mathrm{rx}}}}
=
\sum_{s=r}^{S}
[\vec{Q}]_{s,i_{\mathrm{rx}}}[\bar{\vP}_{\Dt}]_{s-r,i_{\mathrm{rx}}},
\qquad r=0,\dots,S_r.
\end{equation}
When one ETIR is shared by all receiving elements, the corresponding gradient is obtained by summing Eq.~\eqref{eq:grad-response} over $i_{\mathrm{rx}}$.

\subsubsection{Power-law exponent}\label{sec:grad-gamma}
For the derivative with respect to $\gamma$, the converted SI coefficient $\vAo$, rather than the coefficient in dB units, is held fixed.
The DTA expression is\footnote{The dispersion coefficient and its derivative are singular at $\gamma=1$; an optimization that updates $\gamma$ must keep it away from this value.}:
\begin{equation}
\dd{\sfJ}{\gamma}
=\sum_{\ell=1}^{L}\sum_{\vn\in\mathbb{N}_{\Omega}}
[\lp^{(\ell)}]_{\vn}[\vV_{\gamma}^{(\ell)}]_{\vn},
\label{eq:grad-gamma}
\end{equation}
where the forward quantity retained over $\mathbb{N}_{\Omega}$ is
\begin{align*}
\vV_{\gamma}^{(\ell)}
&=\vTo\had\Big(\ln(\vCo)\had\sfQ_{\mathrm{A}}\big(\vU^{\slm}\big)
+\frac{\p\sfQ_{\mathrm{A}}}{\p\gamma}\big(\vU^{\slm}\big)\Big)\\
&\quad-\vEo\had\Big(\ln(\vCo)\had\sfQ_{\mathrm{D}}\big(\vRs^{(\ell)}\big)
+\frac{\p\sfQ_{\mathrm{D}}}{\p\gamma}\big(\vRs^{(\ell)}\big)\Big)\\
&\quad-\pi\sec^2\left(\frac{\pi\gamma}{2}\right)\vAo\had\vCo^{\gamma}\had\sfQ_{\mathrm{D}}\big(\vRs^{(\ell)}\big).
\end{align*}
The partial operator derivatives, with their arguments held fixed, are
\begin{align*}
\frac{\p\sfQ_{\mathrm{A}}}{\p\gamma}\big(\vU^{\slm}\big)
&=\sfFri\!\left(\ln(|\vK|)\had|\vK|^{\gamma-2}\had
\sfFr\!\left(\vRo\had\sum_m\sfDm\big(\vUm^{\slm}\big)\right)\right),\\
\frac{\p\sfQ_{\mathrm{D}}}{\p\gamma}\big(\vRs^{(\ell)}\big)
&=\sfFri\!\left(\ln(|\vK|)\had|\vK|^{\gamma-1}\had
\sfFr\big(\vRs^{(\ell)}\big)\right).
\end{align*}
The zero-wavenumber entries of these derivative multipliers are set to zero.
Evaluating Eq.~\eqref{eq:grad-gamma} requires a full-grid history of $\vV_{\gamma}$, with $NL$ scalar values.

In contrast, DD avoids storing this history by computing the derivative of the pressure sequence $\vP=\sfP(\vX)$ during forward time marching.
The receive operators are linear and independent of $\gamma$, so the chain rule gives
\begin{equation}
\label{eq:grad-gamma-forward}
\dd{\sfJ}{\gamma}
=\left\langle
\left.\nabla_{\vV}\sfJ(\vV;\vG)\right|_{\vV=\sfhG_{\vH}(\vX)},
\sfH\!\left(\sfS\!\left(\dd{\vP}{\gamma}\right)\right)
\right\rangle_{\cG}.
\end{equation}
The pressure derivatives $\dd{\vP^{(\ell)}}{\gamma}\in\cR$ are advanced using the recursion in Eq.~\eqref{eq:gamma-recursion} of~\ref{app:gamma-recursion}.
At each step, $\sftS(\dd{\vP^{(\ell)}}{\gamma})\in\mathbb{R}^{\Nr}$ is evaluated and stored.
After time marching, $\sfH$ is applied to these derivative traces to evaluate Eq.~\eqref{eq:grad-gamma-forward}.
In addition to the current derivative states, this approach stores $\Nr L$ scalar values in the derivative traces for $\ell=1,\dots,L$; the known zero samples at $\ell=0$ are excluded from this count.

With DD, no adjoint recursion is required when only the $\gamma$ derivative is requested.
If an adjoint solution is already required for other gradients, the $\gamma$ derivative can instead be accumulated during the same recursion using Eq.~\eqref{eq:grad-gamma}, provided the full-grid history of $\vV_{\gamma}$ fits in memory.

\section{Software implementation and numerical evaluation}\label{sec:implementation}
This section describes the implementation, verifies the parameter gradients in 2D PACT and USCT simulations, and evaluates the memory savings from spatially restricted storage and checkpointing.
An example with concave acoustic lenses illustrates spatiotemporal transmission and reception.

\subsection{Software overview}\label{sec:software}

\pausct{} is implemented in Python using JAX~\cite{jax2018github}; JAX and JAXlib 0.4.38 were used for the computations reported here.
Forward and gradient computations are performed using just-in-time (JIT) compilation for GPU execution.
The software was built and tested with CUDA 12.9 on an NVIDIA RTX 6000 Ada GPU (48~GiB).

The main software features are:
\begin{itemize}
    \item \textit{Interface.}
    Users specify the grid, time axis, medium, excitation, and spatial and temporal transducer operators.\footnote{During early development, selected domain, field, and medium representations were adapted from \texttt{j-Wave} and \texttt{JaxDF}~\cite{stanziolaJWaveOpensourceDifferentiable2023,stanziolaResearchFrameworkWriting2021} and subsequently modified for \pausct{}.}
    The solver returns simulated data or the data-fidelity value and requested parameter gradients for use in a user-defined optimization method.
    \item \textit{Memory-efficient gradient evaluation.}
    The solver stores only the forward histories needed for the requested gradients, using ROI-restricted storage where permitted by the gradient formulas.
    With checkpointing, histories are retained for one time segment at a time and earlier segments are recomputed during backward time marching (Section~\ref{sec:check}).
    \item \textit{User-defined data fidelity.}
    Users can choose a differentiable data-fidelity function while reusing the same discrete-adjoint formulation.
    Its derivative $\nabla_{\vV}\sfJ$ with respect to predicted data is used to form the adjoint source in Eq.~\eqref{eq:adjoint-source}.
    This derivative can be supplied explicitly or computed by reverse-mode AD applied only to the data-fidelity function.
    \item {\emergencystretch=1em \textit{Sparse and batched spatial operators.}
    For time-independent apodization, each element pressure in Eq.~\eqref{eq:dd-sir-step} is evaluated as a weighted sum of grid pressures.
    These weights are computed before time marching from the quadrature weights, apodization, and spatial interpolation.
    The weights and grid indices are stored only over the union of supports in Eq.~\eqref{eq:R-def-total} and reused for the adjoint operations.
    Elements with similar numbers of stored weights are evaluated in GPU batches at each time step, with the weight arrays zero-padded to a common length within each batch.
    Spatiotemporal responses use separate spatial interpolation weights for each aperture quadrature point, combined with the pointwise temporal operators in~\ref{app:aperture-temporal-discrete}.\par}
\end{itemize}

\subsection{Gradient verification}\label{sec:verification}
The parameter gradients were verified for the half-squared-sum data-fidelity function
\begin{equation}
\label{eq:verification-fidelity}
    \sfJ\big(\sfG(\cS,\vH);\vG\big)
    =\tfrac{1}{2}\big\|\sfG(\cS,\vH)-\sfB(\vG)\big\|_{\mathrm{F}}^2,
\end{equation}
where $\|\cdot\|_{\mathrm{F}}$ denotes the Frobenius norm.

\begin{figure*}[!htbp]
\centering
\includegraphics[width=0.8\textwidth]{\arxivdir 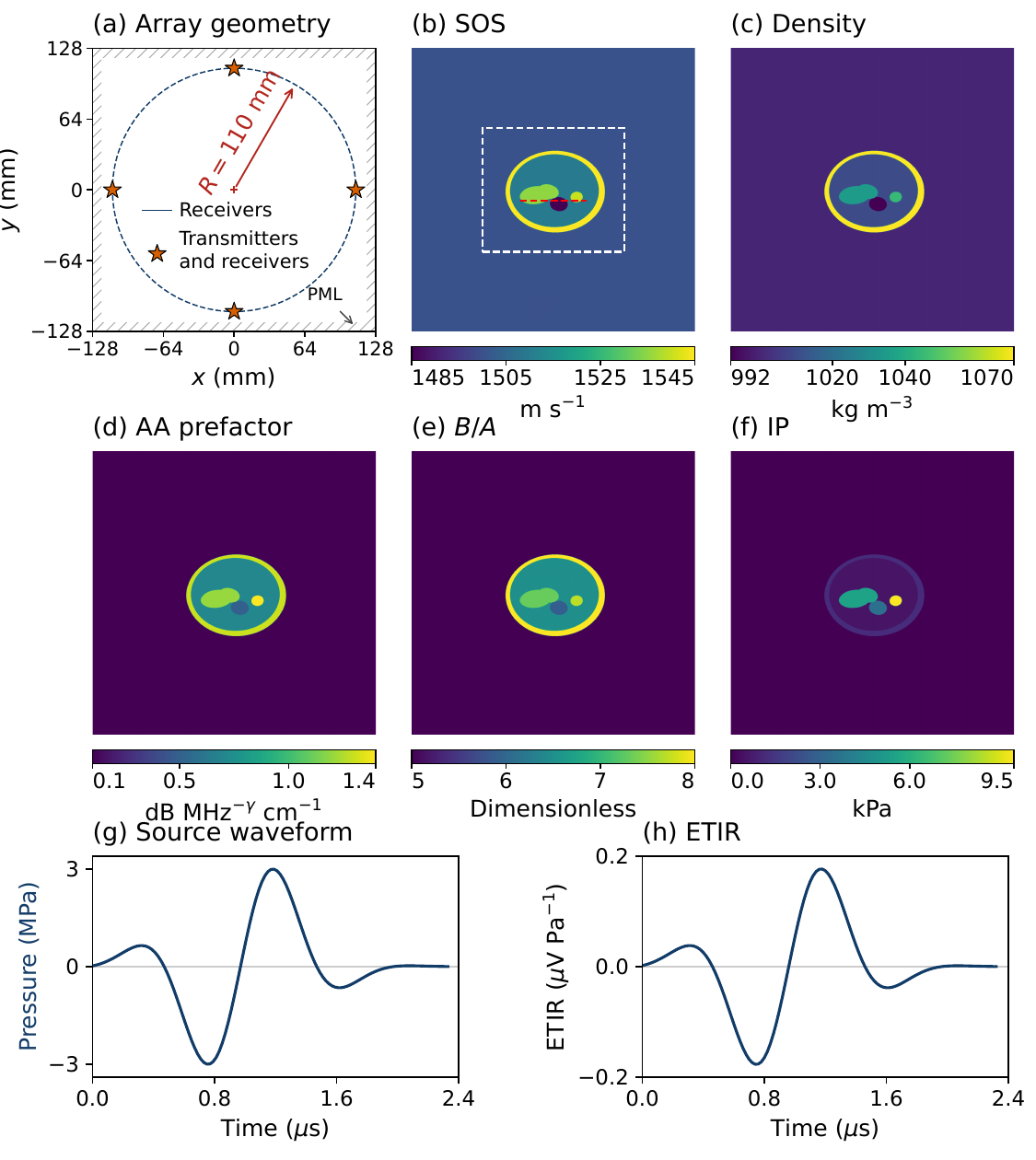}
\caption{Numerical phantom and transducer array for the 2D PACT and USCT experiments. (a) Line apertures drawn to scale. All elements receive; the four orange stars mark the elements that also transmit in USCT. (b)--(f) SOS, density, AA prefactor, $B/A$, and IP on the data-generation grid. Panels (a)--(f) span $256\times256$~mm, with the 8-mm PML hatched in (a). The white dashed rectangle in (b) marks the $128\times112$~mm region of interest (ROI) used in Section~\ref{sec:verification} and \ref{app:ad-dta-comparison}. The red dashed segment marks the gradient-comparison line, $-30\leq x\leq30$~mm at $y=-10$~mm. The AA prefactor is shown in $\mathrm{dB}\,\mathrm{MHz}^{-\gamma}\mathrm{cm}^{-1}$, with $\gamma=1.5$. (g) USCT pressure waveform. (h) Discrete ETIR coefficients in $\mu\mathrm{V}\,\mathrm{Pa}^{-1}$. Both are sampled at $0.02~\mu\mathrm{s}$.}
\label{fig:verification-setup}
\end{figure*}

\paragraph{Data generation} Figure~\ref{fig:verification-setup} shows the 2D numerical phantom and circular array used to generate the synthetic PACT and USCT data.
The SOS, density, and AA prefactor were spatially heterogeneous, while the power-law exponent was fixed at $\gamma=1.5$.
The $256\times256$~mm domain was discretized on a $2560\times2560$ grid with $\dx=0.1$~mm, including an 80-point (8-mm) PML on each side.
The reference SOS $c_{\mathrm{ref}}$ was set to the maximum SOS over the full grid.
The array had 96 equally spaced receiving elements with 5-mm line apertures tangent to a circle of radius 110~mm and normals directed toward its center.
Quadrature points were placed along each aperture using the midpoint rule with an upsampling rate of four relative to the spatial grid.
The receive apodization was constant over the central 80\% of each line aperture and tapered to zero with a raised-cosine profile over each outer 10\%, then normalized to have an aperture average of one.
The truncation parameter was $\mu=0.05$ in Eq.~\eqref{eq:R-def}.

For PACT, linear propagation was simulated with the IP distribution in Fig.~\ref{fig:verification-setup}(f) and no time-varying source.
For USCT, nonlinear propagation was simulated with the heterogeneous $B/A$ map in Fig.~\ref{fig:verification-setup}(e) and zero IP.
The source \textit{k}-space correction in Eq.~\eqref{eq:source-kspace-correction} was enabled for all USCT simulations unless stated otherwise.
The four transmitting elements, marked by orange stars in Fig.~\ref{fig:verification-setup}(a), emitted the same 1-MHz pressure waveform in Fig.~\ref{fig:verification-setup}(g) simultaneously.
Their aperture, quadrature, apodization, and truncation parameter were identical to those used for reception.
In both PACT and USCT, $L=8640$ time steps of $\dt=0.02~\mu\mathrm{s}$ were used over an acquisition duration of $172.8~\mu\mathrm{s}$.
Pressure was recorded at $\Dt=\dt$, giving $S+1=8641$ samples per receiver.
The pressure traces were convolved with the ETIR in Fig.~\ref{fig:verification-setup}(h), shared by all receiving elements, and then filtered with a 7-MHz zero-phase low-pass filter.\footnote{All temporal low-pass filters used in this work were designed by the windowed-sinc method with a Hamming window~\cite{harrisUseWindows1978}, using SciPy~\cite{virtanenSciPy2020} to compute real, symmetric kernels of 401 coefficients.
The integer group delay was removed, and the specified cutoffs refer to the nominal half-amplitude ($-6$-dB) points.}
This cutoff corresponded to two PPW on the data-generation grid at a prescribed minimum SOS of $1400~\mathrm{m\,s^{-1}}$.
The resulting filtered traces were taken as the synthetic measured data $\vG$.
A comparison of forward simulations with \texttt{k-Wave} is provided in~\ref{app:kwave-validation}.

\paragraph{Gradient evaluation} Gradients were computed on a coarser $640\times640$ grid with 0.4-mm spacing.
The PML had 20 grid points (8~mm) on each side.
The acoustic-property maps and IP were subsampled onto this grid and smoothed with a Gaussian kernel of standard deviation 4~mm.
The constant background was restored outside the ROI for the acoustic-property maps and within the PML for the IP.
The power-law exponent was set to $\gamma=1.3$, rather than the value of 1.5 used for data generation.
The propagation models, array geometry, quadrature, and apodization were unchanged.
The grid index sets and spatial interpolation coefficients were recomputed on the coarser grid using the same $\mu$.

The \textit{k}-PSTD time step was $\dt=0.08~\mu\mathrm{s}$, with $L=2160$ updates over the same acquisition duration.
The USCT pressure waveform was subsampled directly to $\dt=0.08~\mu\mathrm{s}$, since its $-30$-dB amplitude band (relative to the spectral peak) extended only to approximately 2.11~MHz, well below the corresponding 6.25-MHz Nyquist frequency.
The upsampling operator $\mathsf{Z}$ was configured to map $L+1=2161$ pressure samples at $\dt=0.08~\mu\mathrm{s}$ to $S+1=8641$ samples at $\Dt=0.02~\mu\mathrm{s}$.
The convolution operator $\mathsf{C}_{\vH}$ used the same ETIR as for data generation.
The zero-phase filter $\sfB$ in Eq.~\eqref{eq:verification-fidelity} was designed for the coarser evaluation grid, at sampling interval $\Dt$ with a 1.75-MHz cutoff corresponding to two PPW.
These components define the temporal receive operator $\sfH$ in Eq.~\eqref{eq:dd-eir-full}, with its discrete adjoint $\sfH^*$ used in DTA.
The DTA gradients and reference derivatives were evaluated for the same objective in Eq.~\eqref{eq:verification-fidelity}, with the synthetic data held fixed.

\begin{itemize}
    \item \textbf{DTA}: The SOS, density, and AA-prefactor gradients\footnote{All AA-prefactor gradients reported in this work are taken with respect to the converted SI coefficient $\vAo$ in $\mathrm{Np}\,(\mathrm{rad\,s^{-1}})^{-\gamma}\mathrm{m}^{-1}$, as in Eq.~\eqref{eq:grad-aa}.} were evaluated for PACT and USCT within the $128\times112$~mm ROI ($320\times280$ grid points) in Fig.~\ref{fig:verification-setup}(b), spanning $-64\leq x<64$~mm and $-56\leq y<56$~mm.
    The $B/A$ gradient was also evaluated within this ROI for USCT.
    The IP gradient for PACT was evaluated throughout the non-PML region, and the USCT pressure-waveform gradient was evaluated for all samples of the shared pressure waveform.
    All these gradients were computed together in one forward--adjoint evaluation per modality.
    \item \textbf{Reference derivatives}: The derivatives $\mathrm{d}\sfJ/\mathrm{d}[\bt]_j$ for selected entries $j$ were computed as Jacobian--vector products (JVPs) using forward-mode AD through \texttt{jax.jvp}, independently of the adjoint formulas.
    Finite differences were not used because, in single precision, round-off errors can dominate the difference between objective values for small perturbations.
    Larger perturbations reduce this problem but increase the error of the derivative approximation.
    Each JVP was evaluated as $(\partial f/\partial\bt)(\bt)\,\delta\bt_j$.
    Here, $f(\bt)$ denotes Eq.~\eqref{eq:verification-fidelity} with all other inputs held fixed, and $\delta\bt_j$ has value one at entry $j$ and zero elsewhere; for the scalar $\gamma$, $\delta\gamma=1$.
    For each spatial parameter, these reference derivatives were evaluated at the 151 grid points along the marked segment in Fig.~\ref{fig:verification-setup}(b).
    For the shared USCT pressure waveform, JVP was evaluated separately for each of its $L_s+1=30$ samples on the evaluation grid ($\ell=0,\dots,29$, corresponding to $0.04$--$2.36~\mu\mathrm{s}$).
    Each direction selected the same temporal entry at all four transmitting elements.
\end{itemize}

For both PACT and USCT, the explicitly derived ETIR gradient in Eq.~\eqref{eq:grad-response} was also verified by evaluating JVP separately for each of the $S_r+1=117$ shared coefficients ($r=0,\dots,116$).
The receiver-pressure traces computed on the gradient-evaluation grid were reused because they are independent of the ETIR.

\begin{figure*}[!htbp]
\centering
\includegraphics[width=\textwidth]{\arxivdir 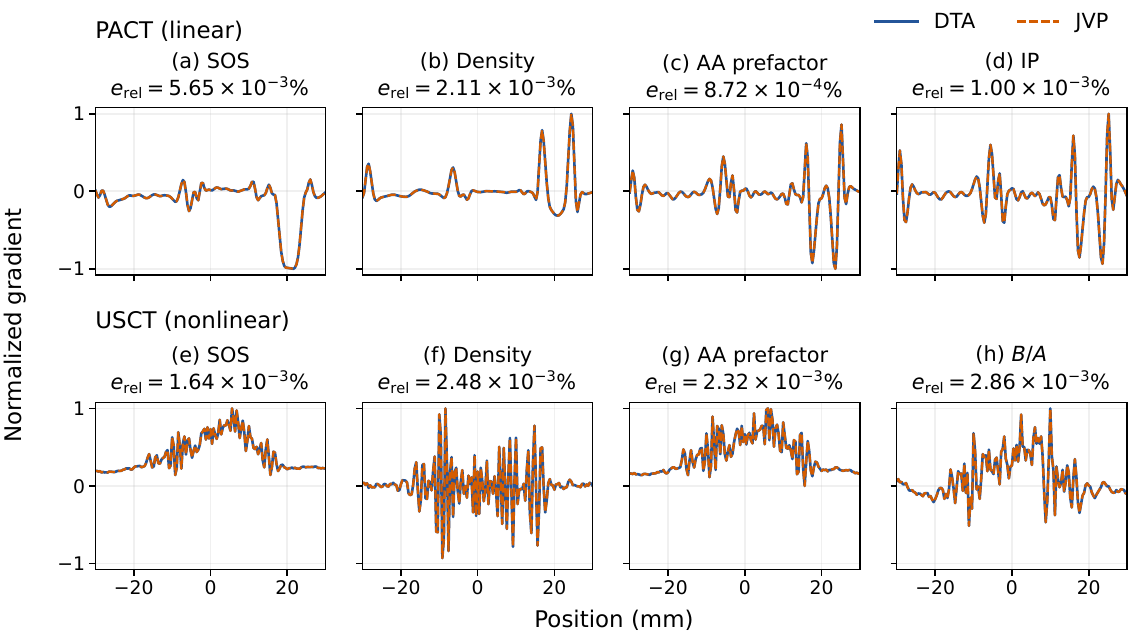}
\caption{Comparison of gradients computed by DTA and JVP along the segment in Fig.~\ref{fig:verification-setup}(b). The top row shows linear PACT results, and the bottom row shows nonlinear USCT results.
Columns 1--3 show SOS, density, and AA-prefactor gradients; the last column shows IP for PACT and $B/A$ for USCT. Here, $e_{\mathrm{rel}}$ denotes the relative RMS difference defined in the text. For display, both profiles in each panel are divided by the maximum absolute value of the DTA profile.
}
\label{fig:gradient-check}
\end{figure*}

\paragraph{Results} Figure~\ref{fig:gradient-check} compares the spatial profiles obtained with DTA and JVP.
The relative root-mean-square (RMS) difference, $e_{\mathrm{rel}}$, is the RMS of the pointwise differences divided by the RMS of the JVP values, expressed as a percentage and evaluated before display normalization.
Table~\ref{tab:verification-summary} summarizes the results.
The relative RMS differences were below $5.7\times10^{-3}\%$ for all eight spatial profiles and below $5\times10^{-5}\%$ for the pressure-waveform and ETIR gradients, which are not plotted.
In single precision, the unit round-off is about $6\times10^{-8}$ ($6\times10^{-6}\%$); the observed differences are consistent with round-off accumulated differently by the forward-mode and adjoint computations over 2160 time steps.
Separate dot-product adjoint checks also passed for spatial reception, temporal interpolation, ETIR convolution, and zero-phase filtering on small test arrays.
The particle-velocity waveform gradient in Eq.~\eqref{eq:grad-source-velocity} was not included in these tests.

\begin{table}[!htb]
\centering
\caption{Summary of gradient verification against forward-mode AD (JVP) on the 2D evaluation grid. DD denotes direct differentiation. For array-valued parameters, $e_{\mathrm{rel}}$ is the relative RMS difference over the tested entries; for $\gamma$, it is the relative absolute difference. PACT used the linear lossy model and USCT the nonlinear lossy model.}
\label{tab:verification-summary}
\small
\begin{tabular}{@{}llll@{}}
\hline
Parameter & Modality & Tested entries & $e_{\mathrm{rel}}$ (\%) \\
\hline
\multirow{2}{*}{SOS} & PACT & \multirow{8}{*}{151 grid points} & $5.65\times10^{-3}$ \\
 & USCT & & $1.64\times10^{-3}$ \\
\multirow{2}{*}{Density} & PACT & & $2.11\times10^{-3}$ \\
 & USCT & & $2.48\times10^{-3}$ \\
\multirow{2}{*}{AA prefactor} & PACT & & $8.72\times10^{-4}$ \\
 & USCT & & $2.32\times10^{-3}$ \\
IP & PACT & & $1.00\times10^{-3}$ \\
$B/A$ & USCT & & $2.86\times10^{-3}$ \\
\hline
Pressure waveform & USCT & 30 samples & $4.51\times10^{-5}$ \\
\hline
\multirow{2}{*}{ETIR coefficients} & PACT & \multirow{2}{*}{117 coefficients} & $3.22\times10^{-5}$ \\
 & USCT & & $1.86\times10^{-5}$ \\
\hline
\multirow{2}{*}{$\gamma$ (DTA)} & PACT & \multirow{4}{*}{scalar} & $7.15\times10^{-6}$ \\
 & USCT & & $1.43\times10^{-3}$ \\
\multirow{2}{*}{$\gamma$ (DD)} & PACT & & $1.14\times10^{-4}$ \\
 & USCT & & $4.69\times10^{-5}$ \\
\hline
\end{tabular}
\end{table}

\paragraph{Power-law exponent}
For the scalar $\gamma$, three methods were compared for both PACT and USCT: DTA (Eq.~\eqref{eq:grad-gamma}), DD (Eq.~\eqref{eq:grad-gamma-forward}) using the explicit recursion in~\ref{app:gamma-recursion}, and JVP of the objective.
In each case, the converted SI AA coefficient was held fixed, as specified in Section~\ref{sec:grad-gamma}.
Here, relative differences are the absolute differences divided by the absolute JVP value, expressed as percentages.
Table~\ref{tab:verification-summary} reports the relative differences.

Both DD and JVP advance the forward fields together with their derivatives with respect to $\gamma$.
DD computes these derivatives using the explicit recursion in Eq.~\eqref{eq:gamma-recursion}, whereas JVP computes them through JAX forward-mode AD.
With only the $\gamma$ derivative requested, peak memory with non-checkpointed DTA was 3.353~GiB for PACT and 13.242~GiB for USCT. Both DD and JVP used 0.124~GiB for PACT and 0.119~GiB for USCT.\footnote{Peak memory denotes the maximum memory in use recorded by the JAX allocator over the lifetime of each process.
Memory preallocation was disabled, and the allocator limit was set to 95\% of GPU memory. The \texttt{peak\_bytes\_in\_use} entry of \texttt{jax.devices()[0].memory\_stats()} was read after synchronizing GPU execution.}
DTA can share an adjoint solution already required for other gradients, but also requires the full-grid history described in Section~\ref{sec:grad-gamma}.

\subsection{Memory-efficient DTA with checkpointing}\label{sec:check}
The memory required for gradient evaluation can be reduced by recomputing forward fields instead of storing their complete time histories~\cite{griewankAlgorithm799Revolve2000}.
The $L$ updates are divided into $N_c+1$ time segments, with full-grid states $\vX$ in Eq.~\eqref{eq:discrete_state} saved at $N_c$ internal checkpoints during the forward run.
The histories needed for the adjoint recursion and requested gradients are retained only for the final segment (Section~\ref{sec:parameter-gradients}).
During backward time marching, earlier segments are recomputed from their saved starting states, retaining only one segment's histories at a time.
Algorithm~\ref{alg:checkpoint} gives the detailed procedure, performed entirely on the GPU.
Increasing the number of checkpoints shortens the retained histories but increases the storage required for checkpoint states.
With uniform segments, recomputing only the earlier segments adds approximately $N_c/(N_c+1)$ of a forward solve per gradient evaluation.

\begin{algorithm}[tbp]
\caption{Checkpointed DTA gradient evaluation}
\label{alg:checkpoint}
\footnotesize
\begin{algorithmic}[1]
\Require Model parameters $(\cS,\vH)$, measured data $\vG$, requested DTA gradients, and internal checkpoint count $N_c$
\State Choose boundaries $0=\ell_0<\cdots<\ell_{N_c+1}=L$ with segment lengths differing by at most one
\State Initialize and retain $\vX^{(0)}$; record $\sftS(\vP^{(0)})$ in $\bar{\vP}$
\For{$k=0,\dots,N_c$}
    \State Save the checkpoint state $\vX^{(\ell_k)}$ if $k>0$
    \State Advance updates $\ell_k+1,\ldots,\ell_{k+1}$ and record $\sftS(\vP^{(\ell)})$ in $\bar{\vP}$
    \State Retain the required forward histories during these updates only if $k=N_c$
\EndFor
\State Compute $\vG_{\mathrm{sim}}=\sfH(\bar{\vP})$
\State Compute $\vW=\sfH^*\!\left(\nabla_{\vG_{\mathrm{sim}}}\sfJ(\vG_{\mathrm{sim}};\vG)\right)$
\State Initialize the terminal adjoint states and set the gradient accumulators to zero
\For{$k=N_c,\dots,0$}
    \If{$k<N_c$}
        \State Recompute updates $\ell_k+1,\ldots,\ell_{k+1}$ from $\vX^{(\ell_k)}$, retaining the required histories without recording receiver pressures
    \EndIf
    \State Run the discrete-adjoint recursion backward over the segment with sources from $\vW$ (Eq.~\eqref{eq:adjoint-source}), accumulating the requested gradients
    \State Release the segment histories, retaining the boundary-step values needed when the recursion continues into the preceding segment (e.g., $\sfDm(\vUm^{\scriptstyle(\ell_k+\frac12)})$ for nonlinear models)
\EndFor
\State Evaluate the time-zero adjoint and gradient terms and return the requested gradients
\end{algorithmic}
\end{algorithm}

For SOS gradients, the history of $\vV_c$ in Eq.~\eqref{eq:grad-sos} can be restricted to the ROI, whereas nonlinear adjoint increments in Eqs.~\eqref{eq:inc_1} and~\eqref{eq:inc_2} require additional histories over the full grid.

All reported \pausct{} forward and gradient evaluations used single precision.
Unless stated otherwise, forward and gradient evaluation times in Sections~\ref{sec:implementation} and~\ref{sec:applications} were averaged over ten repetitions after JIT compilation and one warm-up evaluation.
GPU execution was synchronized for each measurement.
For the SOS-gradient problems detailed in~\ref{app:ad-dta-comparison}, checkpointing enabled nonlinear evaluations that exceeded GPU memory without it, while ROI restriction further reduced the storage needed for the SOS history.
The DTA gradients agreed across checkpoint settings, including no checkpointing where feasible, with relative RMS differences below $8.5\times10^{-5}\%$.

\begin{figure*}[!htbp]
\centering
\includegraphics[width=\textwidth]{\arxivdir 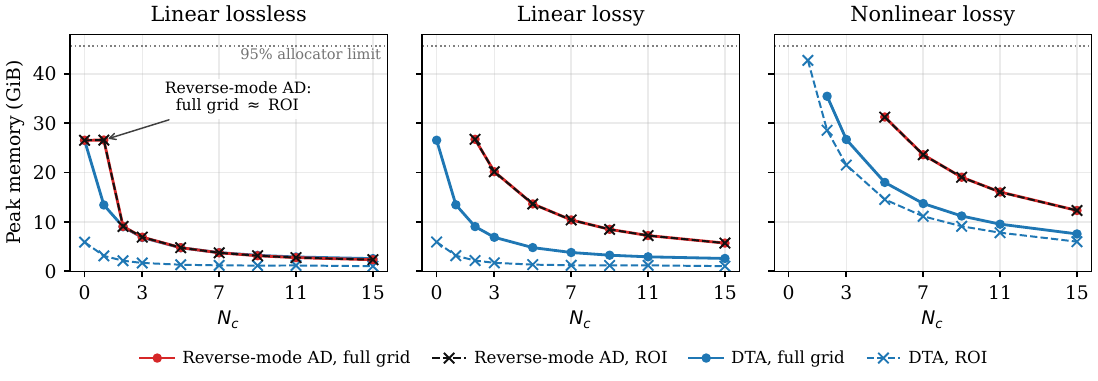}
\caption{Peak memory for evaluating the SOS gradient and data-fidelity value with reverse-mode AD and DTA as the number of internal checkpoints $N_c$ varies. The grid is $1280\times1280$, with 4320 updates. Reverse-mode AD uses red for the full grid and black for the ROI; DTA uses blue. Solid lines with circles denote gradients over the full grid including the PML; dashed lines with crosses denote gradients within the ROI. The reverse-mode AD curves nearly coincide. Missing points indicate failed GPU allocation. The dotted line marks the allocator limit (95\% of the 48-GiB GPU memory). Values are from Table~\ref{tab:ad-dta-checkpoint}.}
\label{fig:ad-dta-memory}
\end{figure*}

For comparison, the SOS gradients were evaluated using JAX reverse-mode AD, with segment-based checkpointing through \texttt{jax.\allowbreak checkpoint}.
For the same discrete problem, reverse-mode AD and DTA are mathematically equivalent~\cite{zhu2021general}.
Their implementations, however, need not retain the same intermediate quantities.
The explicit DTA formulas allow forward quantities to be combined before storage; segment-based checkpointing alone does not prescribe such parameter-specific storage choices for reverse-mode AD.
Figure~\ref{fig:ad-dta-memory} summarizes the peak-memory comparison for the SOS gradient in~\ref{app:ad-dta-comparison}, using the same time-segment boundaries and recomputed updates for both methods.
Within \pausct{}, DTA required less memory for the lossy models in every configuration completed by both methods.
ROI restriction reduced memory use for DTA but left that of reverse-mode AD nearly unchanged.
Reverse-mode AD took approximately 6--29\% longer than DTA in all 42 configurations in which both methods completed.
Full timings and memory values are given in Table~\ref{tab:ad-dta-checkpoint}.

\FloatBarrier
\subsection{Example with spatiotemporal responses: concave acoustic lenses}\label{sec:lens-example}
{
\subsubsection{Concave acoustic lens model}\label{sec:lens-model}
Spatiotemporal transmission and reception in \pausct{} are illustrated with the concave-lens model used by Li \textit{et al.}~\cite{liForwardModelIncorporating2023} for ring-array USCT.
Under the geometrical-acoustics approximation, the lens is represented by spatially varying propagation times across the aperture.
Let $\br'=(x',y',z')$ denote an aperture point, with $z'$ measured along the element height, $z_e^{[i]}$ the center coordinate along this direction, and $F$ the nominal focal distance.
For uniform spatial sensitivity and a parabolic lens profile along $z$, the transmit and receive kernels in~\ref{app:aperture-temporal} are
\begin{equation}\label{eq:lens-spatiotemporal-kernel}
 k_{\sigma}^{[i]}(\br',\tau)
 =\delta\!\left(\tau+\frac{(z'-z_e^{[i]})^2}{2c_{\mathrm{water}}F}\right),
\end{equation}
for $\sigma\in\{\mathrm{tx},\mathrm{rx}\}$.
At quadrature point $\hbr^{[i,j]}$, the Fourier multiplier in~\ref{app:aperture-temporal-discrete} has components
\begin{equation}\label{eq:lens-temporal-operator}
 \left[\widehat{\vec{k}}_{\sigma}^{[i,j]}\right]_q
 =\exp\!\left\{\imath2\pi f_q
 \frac{(\hat z^{[i,j]}-z_e^{[i]})^2}{2c_{\mathrm{water}}F}\right\}.
\end{equation}
Here, the nonnegative FFT frequencies are $f_q=q/(n_{\mathrm{FFT}}\dt)$ for transmission and $f_q=q/(n_{\mathrm{FFT}}\Dt)$ for reception, with $q=0,\dots,\lfloor n_{\mathrm{FFT}}/2\rfloor$.
The transform length $n_{\mathrm{FFT}}$ includes zero padding.

The grid-aligned line-aperture representation used by Li \textit{et al.}~\cite{liForwardModelIncorporating2023} ties aperture sampling to the \textit{k}-PSTD primary grid and neglects the horizontal width.
Here, Eq.~\eqref{eq:sensitivity-point-off} enables off-grid aperture quadrature to be refined independently of the \textit{k}-PSTD Cartesian grid, while surface integration retains both the horizontal width and height of a rectangular aperture.

\subsubsection{Forward-model validation and adjoint tests}\label{sec:lens-water}
Two facing rectangular elements were placed 200~mm apart along the $x$-axis in homogeneous water with $c_0(\br)=1500~\mathrm{m\,s^{-1}}$ and $\rho_0(\br)=1000~\mathrm{kg\,m^{-3}}$.
Propagation was linear, with absorption and dispersion omitted.
Each element had a 2-mm width along $y$, an 18-mm height along $z$, and a lens with $F=100$~mm (Fig.~\ref{fig:lens-example}(b)).
The computational grid had $1280\times240\times320$ points with 0.2-mm spacing and a 40-point (8-mm) PML on each face ($\alpha_{\mathrm{PML}}=2$).
The time step and receiver sampling interval were $\dt=\Dt=0.04~\mu\mathrm{s}$, with $L=S=3600$ updates over $144~\mu\mathrm{s}$.
The transmit pressure waveform was a Gaussian-modulated 2-MHz pulse with peak amplitude 500~kPa, envelope center $5~\mu\mathrm{s}$, and envelope standard deviation $0.6~\mu\mathrm{s}$.
An aperture-quadrature upsampling rate of four gave 3600 triangle-centroid points per element.
Spatial interpolation used the sinc-support truncation parameter $\mu=0.01$.
The ETIR and filtering operators were the identity.
Figure~\ref{fig:lens-example}(a) shows the combined focusing region, represented by the normalized product of the two time-harmonic pressure-field magnitudes at 2~MHz.

For comparison, the received pressure was computed by numerically evaluating the Rayleigh--Sommerfeld surface integral~\cite{oneilTheoryFocusingRadiators1949,liForwardModelIncorporating2023}, with normal velocity $\frac{s_p(t)}{\rho_0(\br)c_0(\br)}$, followed by aperture averaging at the receiver.
The lens responses in Eq.~\eqref{eq:lens-spatiotemporal-kernel} were applied on both apertures, each integrated using 8 Gauss--Legendre points along its horizontal width and 96 along its height.\footnote{Increasing these counts to 12 and 144, respectively, changed the reference by less than $10^{-8}\%$ in relative RMS.}
Figure~\ref{fig:lens-example}(c) compares the received pressure traces; their relative RMS difference, normalized by the reference RMS, was $0.0572\%$.

The form of the discrete adjoint recursion in Eqs.~\eqref{eq:AS2-lp}--\eqref{eq:AS2-lu} is independent of the transducer response model.
The additional adjoint tests therefore focused on the spatiotemporal pressure-source and receive maps in Eqs.~\eqref{eq:pointwise-aperture-transmit} and~\eqref{eq:pointwise-aperture-receive}, with the receive adjoint given by Eq.~\eqref{eq:spatiotemporal-adjoint-source}.
For each map, 100 pairs of independent random input and test arrays were generated.
Transmission used a 250-sample source waveform as input and a $250\times N$ test array for one density-source component.
Reception used a $3601\times N$ pressure array as input and a 3601-sample test trace for the single receiver, where $N=1280\times240\times320$.
The transmit test arrays and receive inputs were represented by random spatial arrays scaled by random temporal coefficients, with these factors stored separately.
The inner product of the forward output with the test array was compared with that of the input with the adjoint applied to the same test array.
The absolute difference between the two inner products was divided by the larger of their absolute values.
The mean relative differences were $1.32\times10^{-6}$ for transmission and $6.60\times10^{-6}$ for reception.

\begin{figure}[!htbp]
\centering
\includegraphics[width=\textwidth]{\arxivdir 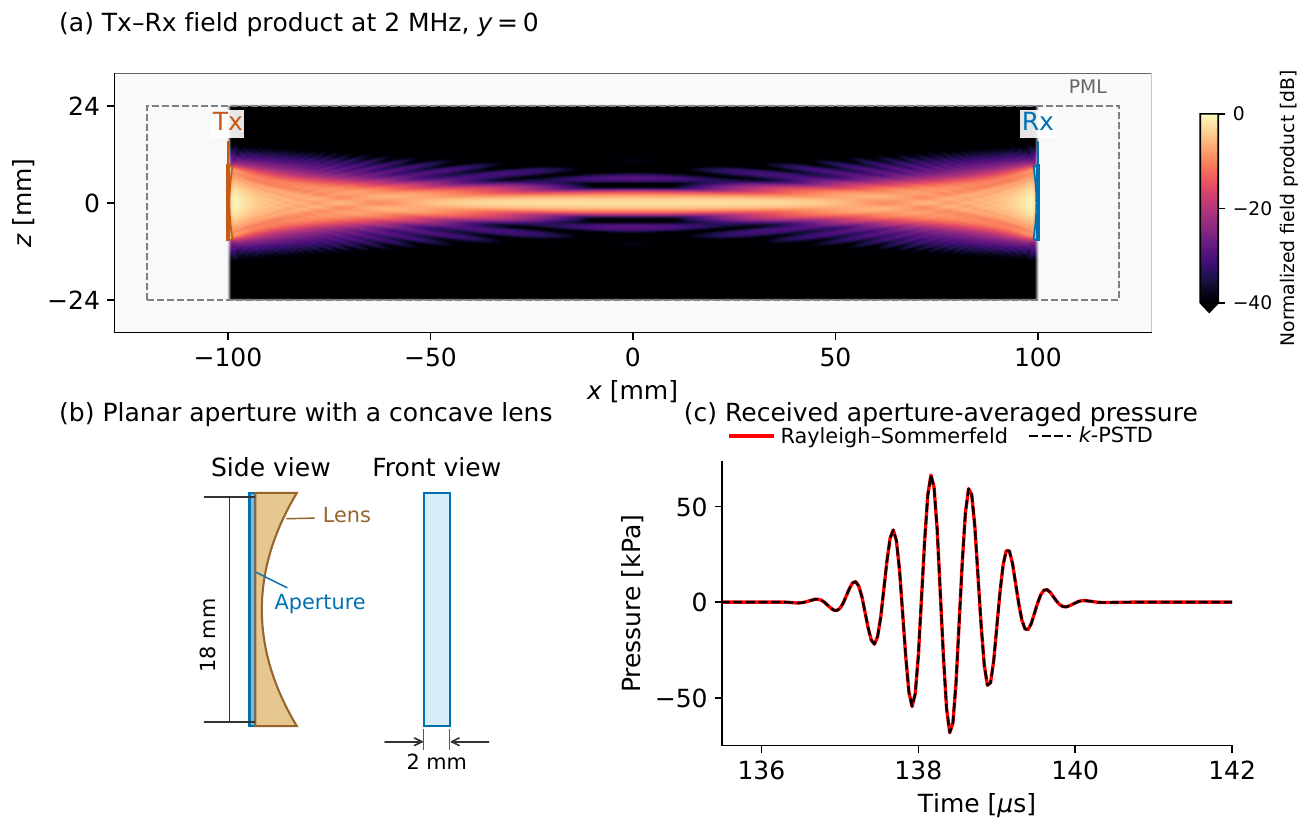}
\caption{Verification with a concave-lens transmitter--receiver pair.
(a) Magnitude of the product of the two radiated complex pressure fields at 2~MHz in the $y=0$ plane, normalized to its maximum between the apertures and displayed using $20\log_{10}$. Dashed lines mark the inner PML boundary; the transmit and receive apertures and their schematic lens profiles are overlaid.
(b) Side and front views of the rectangular aperture, with a schematic lens profile.
(c) Aperture-averaged pressures from the Rayleigh--Sommerfeld reference (red solid) and the \textit{k}-PSTD model (black dashed).}
\label{fig:lens-example}
\end{figure}
}
\FloatBarrier

\section{3D virtual imaging study}\label{sec:applications}

The 3D virtual imaging study demonstrates a USCT-augmented PACT reconstruction workflow using the forward and gradient routines, with finite-aperture transducer responses.
The SOS was first reconstructed from USCT data and then used for IP estimation in PACT to compensate for SOS-induced phase aberration.
Both reconstructions were performed with noisy data and prescribed model mismatch.
Each reconstruction used a linear lossy model to estimate one parameter distribution, with the remaining medium parameters and waveforms held fixed.

\subsection{Data generation}\label{sec:3d-data}
\begin{figure*}[!htbp]
    \centering
    \includegraphics[width=\linewidth]{\arxivdir 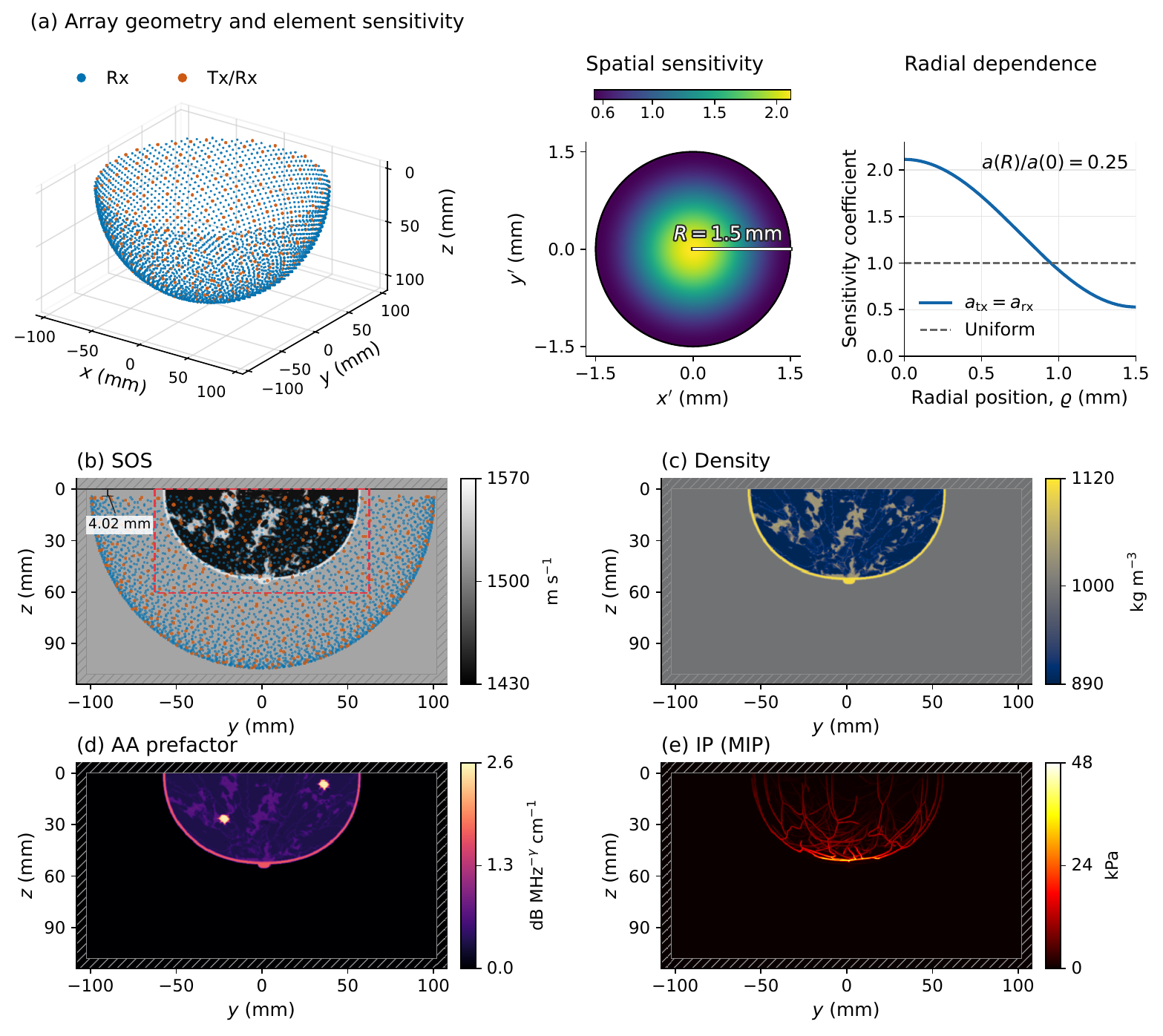}
    \caption{Numerical breast phantom and transducer array for the 3D virtual imaging study. (a) Centers of the 3072 disc elements on a hemisphere of radius 100~mm centered at $(0,0,4)$~mm. Orange marks the 384 elements used for both transmission and reception; blue marks the remaining receivers. Each disc has diameter 3~mm and faces the hemisphere center. The plots on the right show the element apodization and its radial profile; $x'$ and $y'$ are local aperture coordinates. (b)--(d) SOS, density, and AA prefactor at $x=0$. In (b), all element centers are projected along $x$ onto the $y$--$z$ plane using the colors in (a). The marked 4.02-mm distance extends from the chest-wall plane ($z=0$) to the nearest element center. The dashed box marks the final USCT ROI used for SOS-gradient evaluation. The AA prefactor is shown in $\mathrm{dB}\,\mathrm{MHz}^{-\gamma}\mathrm{cm}^{-1}$, with $\gamma=1.1642$. (e) Maximum-intensity projection (MIP) of IP along $x$. Hatched strips mark the 6-mm PML.}
    \label{fig:common-phantom-array}
\end{figure*}

Figure~\ref{fig:common-phantom-array} illustrates the numerical phantom and array of transducer elements used to simulate USCT and PACT data.
An anatomically realistic numerical breast phantom~\cite{liStochasticNumericalBreast2022,parkStochasticThreeDimensional2023,parkVirtualImaging2026} was used to specify the acoustic-property maps and IP distribution.
The USCT and PACT datasets were generated with linear, lossy propagation using the heterogeneous SOS, density, and AA-prefactor maps in Fig.~\ref{fig:common-phantom-array}(b)--(d), with $\gamma=1.1642$.
The $216\times216\times120~\mathrm{mm^3}$ domain was discretized on a $720\times720\times400$ grid with 0.3-mm spacing, including a 20-point (6-mm) PML on each face.
The primary-grid origin was $\br^{[\vec{0}]}=(-108,-108,-6)$~mm, with the chest-wall plane at $z=0$ coinciding with the first non-PML grid plane.
The reference SOS $c_{\mathrm{ref}}$ for the \textit{k}-space correction was set to the maximum of the SOS distribution over the full grid ($1568.90~\mathrm{m\,s^{-1}}$).

The array comprised 3072 disc elements, each 3~mm in diameter, on a hemisphere of radius 100~mm centered at $(0,0,4)$~mm, with normals directed toward its center.
With $z$ increasing downwards, the hemisphere's equatorial plane was therefore 4~mm below the chest-wall plane.
The nonuniform apodization in Fig.~\ref{fig:common-phantom-array}(a), normalized to unit aperture average, was used for both transmission and reception and retained for reconstruction.
Quadrature points were sampled at an upsampling rate of four relative to the 0.3-mm grid. Spatial interpolation used the truncation parameter $\mu=0.075$.

\begin{figure*}[!htbp]
    \centering
    \includegraphics[width=\linewidth]{\arxivdir 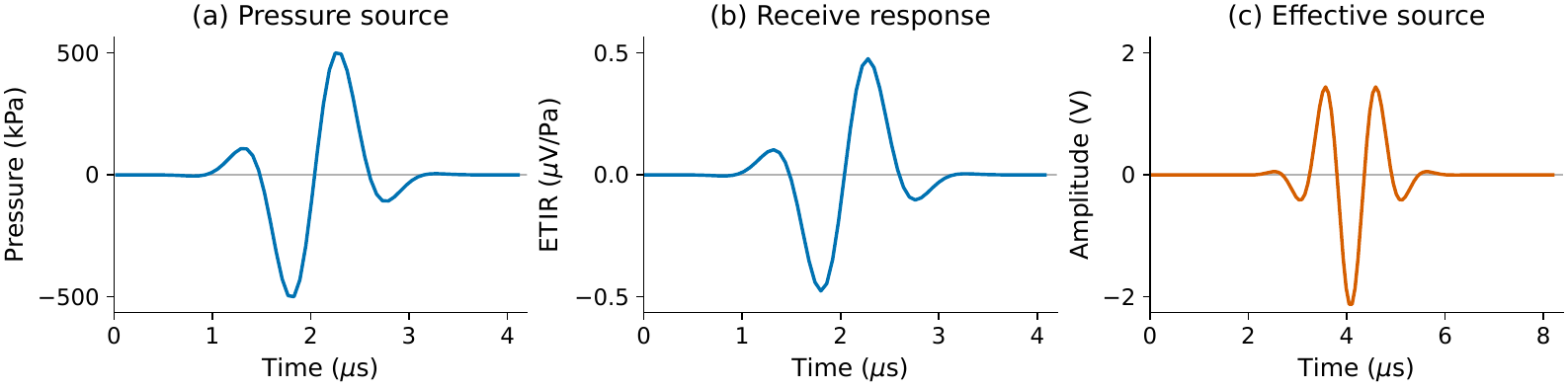}
    \caption{Transmit waveforms and receive response for the 3D virtual imaging study. (a) Pressure waveform shared by the 384 transmitting elements during USCT data generation, with a peak amplitude of 500~kPa. (b) Discrete receive ETIR shared by all 3072 receivers in USCT and PACT, with a maximum frequency-response magnitude of $5~\mu\mathrm{V}\,\mathrm{Pa}^{-1}$. The source waveform and receive ETIR were designed with a center frequency of 0.9~MHz and a nominal 100\% fractional bandwidth at $-6$~dB. (c) Their discrete convolution, used as the effective source in USCT reconstruction. Its center frequency is approximately 0.9~MHz and its $-6$-dB fractional bandwidth is approximately 71\% (0.58--1.22~MHz). The sampling interval is $0.06~\mu\mathrm{s}$.}
    \label{fig:3d-waveforms}
\end{figure*}

The \textit{k}-PSTD time marching was performed using $L=2400$ updates with a time step of $\dt=0.06~\mu\mathrm{s}$ for both USCT and PACT.
No temporal upsampling was applied to the simulated data, so $\Dt=\dt$ and $S=L$; each receiver trace contained 2401 samples over $144~\mu\mathrm{s}$.
For USCT, 384 spatially distributed elements were selected from the array by farthest-point sampling and indexed by $i_{\mathrm{tx}}=0,\dots,383$.
For each acquisition, the transmit weights in Eq.~\eqref{eq:scaled-pressure-source} were specified by a one-hot vector $\vec{w}^{[i_{\mathrm{tx}}]}$, with $w_{\mathrm{tx}}^{[i'_{\mathrm{tx}}]}=[\vec{w}^{[i_{\mathrm{tx}}]}]_{i'_{\mathrm{tx}}}=\delta_{i'_{\mathrm{tx}},i_{\mathrm{tx}}}$ for $i'_{\mathrm{tx}}=0,\dots,383$.
One shot gather was thus generated per transmitting element.
All selected elements shared the same discrete pressure waveform, $\vec{s}_p^{[i_{\mathrm{tx}}]}=\vec{s}_p$, shown in Fig.~\ref{fig:3d-waveforms}(a), with a peak amplitude of 500~kPa.
For PACT, the IP distribution in Fig.~\ref{fig:common-phantom-array}(e) was used.
All 3072 elements were used for reception in both USCT and PACT, with the same ETIR applied to every channel.
The common discrete coefficients $[\vH]_{r,i_{\mathrm{rx}}}=\Dt\,h^{[i_{\mathrm{rx}}]}(r\Dt)$, $r=0,\dots,S_r$, are shown in Fig.~\ref{fig:3d-waveforms}(b).
The source waveform and receive ETIR were designed with a center frequency of 0.9~MHz and a nominal 100\% fractional bandwidth at $-6$~dB.
Each contained 69 samples ($L_s=S_r=68$) at a sampling interval of $0.06~\mu\mathrm{s}$.
The maximum frequency-response magnitude of the ETIR coefficients was $5~\mu\mathrm{V}\,\mathrm{Pa}^{-1}$.
No additional zero-phase filtering was applied.

Zero-mean Gaussian noise with a standard deviation of $46.76~\mu\mathrm{V}$ was added to every USCT shot gather and the PACT dataset.
Independent noise realizations were generated for the 384 USCT shots and the single PACT acquisition using fixed random seeds.
The signal-to-noise ratio (SNR) was defined as $20\log_{10}$ of the signal RMS divided by the noise standard deviation.
For USCT, the signal RMS was evaluated over all shots, excluding only the trace recorded by the transmitting element in each shot.
The resulting SNRs were approximately 20~dB for USCT and 7.8~dB for PACT.
The noisy USCT shot gathers are denoted by $\vG_{\mathrm{USCT}}^{[i_{\mathrm{tx}}]}$, $i_{\mathrm{tx}}=0,\dots,383$, and the noisy PACT dataset by $\vG_{\mathrm{PACT}}$.

\FloatBarrier
\subsection{Image reconstruction}\label{sec:3d-reconstruction}
Both reconstructions were performed on coarser grids than those used for data generation.
For USCT, the successive grid spacings were 1.2, 0.9, and 0.6~mm, with corresponding time steps of 0.24, 0.18, and 0.12~$\mu\mathrm{s}$.
For PACT, the spacings were 1.2, 0.8, and 0.5~mm, with time steps of 0.24, 0.16, and 0.10~$\mu\mathrm{s}$.
The same aperture quadrature and $\mu=0.075$ were retained on all reconstruction grids, with the spatial interpolation coefficients recomputed for each grid.

At each stage, the upsampling operator $\mathsf Z$ was configured to map receiver-pressure samples from the corresponding time step $\dt$ to the measurement interval $\Dt=0.06~\mu\mathrm{s}$.
The zero-phase filter $\sfB$ was constructed at this sampling interval with the two-PPW cutoff for the current grid and $c_{\min}=1400~\mathrm{m\,s^{-1}}$; the same filter was applied to the measured data.
Because the cutoff scales inversely with the grid spacing, the grid schedule also provides frequency continuation: the cutoffs were approximately 0.58, 0.78, and 1.17~MHz for USCT and 0.58, 0.88, and 1.40~MHz for PACT, compared with the nominal $-6$-dB band set by the USCT transmit waveform and receive ETIR (0.58--1.22~MHz).
Each subsequent stage was initialized by interpolating the preceding reconstruction in physical coordinates.

\begin{figure*}[!htbp]
    \centering
    \includegraphics[width=\linewidth]{\arxivdir 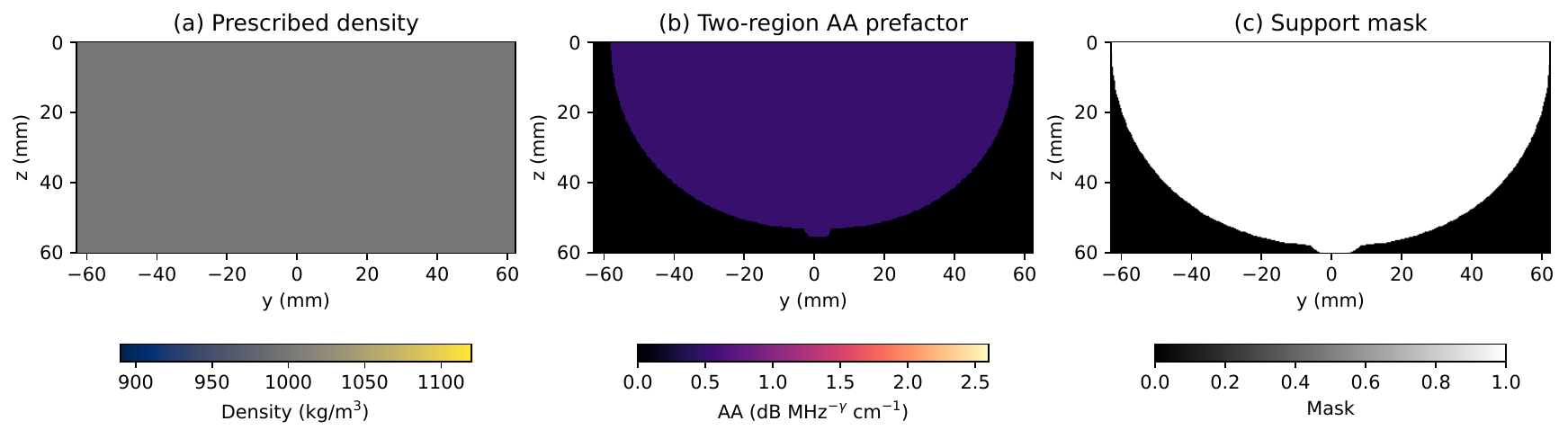}
    \caption{Prescribed medium properties for USCT and PACT reconstruction and support region on the 0.3-mm grid, shown at $x=0$ within the enclosing ROI. (a) Homogeneous density of $1000~\mathrm{kg\,m^{-3}}$, (b) two-region AA prefactor before resampling and smoothing, and (c) support mask, with one indicating the allowed reconstruction region. The AA prefactor is shown in $\mathrm{dB}\,\mathrm{MHz}^{-\gamma}\mathrm{cm}^{-1}$, with $\gamma=1.3$.}
    \label{fig:reconstruction-setup}
\end{figure*}

To represent imperfect knowledge of density and AA, simplified maps were prescribed for both reconstructions.
Figure~\ref{fig:reconstruction-setup}(a) and (b) show these maps on the 0.3-mm grid.
The heterogeneous density used for data generation was not assumed known during reconstruction; a uniform density of $1000~\mathrm{kg\,m^{-3}}$ was prescribed, differing from the true water-background value of $993.36~\mathrm{kg\,m^{-3}}$.
For the two-region AA map, the prefactor was set to the mean breast value, 0.5051~$\mathrm{dB}\,\mathrm{MHz}^{-\gamma}\mathrm{cm}^{-1}$, within the breast and the water value, 0.0022 in the same units, elsewhere.
The power-law exponent was fixed at $\gamma=1.3$ rather than 1.1642.
For each reconstruction grid, the 0.3-mm two-region AA map was resampled in physical coordinates by trilinear interpolation, then smoothed with a Gaussian kernel of standard deviation 0.3~mm.

The support mask used for the reconstruction constraints (Fig.~\ref{fig:reconstruction-setup}(c)) was obtained by dilating the phantom's breast mask by 5~mm and excluding the PML.
This support mask was resampled to each reconstruction grid by nearest-neighbor interpolation.
The constraints defined below allowed SOS or IP to vary only within this support, with the respective background values fixed outside it.
The unknowns were represented within a rectangular ROI enclosing the support on each grid.

\FloatBarrier
\subsubsection{SOS estimation in USCT}\label{sec:usct-sos}
Here, the dependence of $\sfX$ in Eq.~\eqref{eq:dd-full} on the transmit-weight vector $\vec w$ is made explicit as $\sfX_{\vec w}$ to describe source encoding.
The individual USCT acquisitions in Section~\ref{sec:3d-data} were simulated using $\sfX_{\vec w^{[i_{\mathrm{tx}}]}}$, $i_{\mathrm{tx}}=0,\dots,383$.
The effective waveform $\vec s_{p,\mathrm{eff}}$ in Fig.~\ref{fig:3d-waveforms}(c) was obtained by discrete convolution of $\vec s_p$ with the common ETIR coefficients $\Dt\,h^{[i_{\mathrm{rx}}]}(r\Dt)$.
Let $\cS_{\mathrm{eff}}$ denote $\cS$ with every pressure waveform replaced by $\vec s_{p,\mathrm{eff}}$ and all other entries unchanged.
Let $[\vH_{\mathrm{id}}]_{r,i_{\mathrm{rx}}}$ equal one for $r=0$ and zero otherwise, so that $\mathsf C_{\vH_{\mathrm{id}}}$ is the identity.
On a common time grid ($\dt=\Dt$), linear, time-invariant propagation from a zero initial state gives
\begin{equation}\label{eq:usct-effective-source}
    \sfhG_{\vH}\!\left(\sfX_{\vec w}(\cS)\right)
    =\sfhG_{\vH_{\mathrm{id}}}\!\left(\sfX_{\vec w}(\cS_{\mathrm{eff}})\right).
\end{equation}
For reconstruction, the effective waveform was resampled from the $0.06~\mu\mathrm{s}$ sampling grid to the coarser half-step source times by cubic interpolation, which does not generally preserve this discrete equivalence.
The reference SOS $c_{\mathrm{ref}}$ was fixed at $1600~\mathrm{m\,s^{-1}}$ (the upper SOS bound) for the \textit{k}-space correction throughout reconstruction, as assumed in the SOS-gradient formula (Section~\ref{sec:grad-spatial}).

Waveform inversion with source encoding (WISE)~\cite{wangWaveformInversionSource2015,matthewsRegularizedDualAveraging2017} was used to combine the individual acquisitions into one encoded problem per iteration.
A random encoding vector $\boldsymbol\zeta$ was drawn at each iteration, with independent Rademacher entries taking values $-1$ and $+1$ with equal probability.
Encoded propagation was computed using $\sfX_{\boldsymbol\zeta}$, with $\vec w=\boldsymbol\zeta$.
With an isotropic total variation (TV) penalty, the reconstruction was formulated as
\begin{equation}\label{eq:usct-reconstruction}
    \widehat{\vCo}
    =\argmin_{\vCo\in\mathcal{C}_c}
    \left\{
    \frac{1}{2}\mathbb{E}_{\boldsymbol{\zeta}}\!\left[
    \left\|
    \sfhG_{\vH_{\mathrm{id}}}\!\left(\sfX_{\boldsymbol\zeta}(\cS_{\mathrm{eff}})\right)
    -\sfB\!\left(\vG_{\boldsymbol\zeta}\right)
    \right\|_{\mathrm{F}}^2
    \right]
    +\lambda_c\,\mathrm{TV}(\vCo)
    \right\},
\end{equation}
where $\vG_{\boldsymbol\zeta}=\sum_{i_{\mathrm{tx}}=0}^{383}[\boldsymbol\zeta]_{i_{\mathrm{tx}}}\vG_{\mathrm{USCT}}^{[i_{\mathrm{tx}}]}$ denotes the encoded data.
The convex constraint set was
\begin{equation*}
    \mathcal{C}_c
    =\left\{\vCo\in\cR\;\middle|\;
    \begin{array}{ll}
    1400\leq[\vCo]_{\vn}\leq1600,
        & \vn\in\mathbb{N}_{\mathrm{supp}},\\
    {[\vCo]_{\vn}}=1520.6,
        & \vn\in\mathbb{N}_{\Omega}\setminus\mathbb{N}_{\mathrm{supp}}
    \end{array}
    \right\},
\end{equation*}
where $\mathbb{N}_{\mathrm{supp}}\subseteq\mathbb{N}_{\mathrm{ROI}}$ denotes the index set of the support region and SOS values are expressed in $\mathrm{m\,s^{-1}}$.
On the isotropic 3D grids, where $d_{x_1}=d_{x_2}=d_{x_3}$, the TV penalty for a spatial array $\vV\in\cR$ was discretized as
\begin{equation}\label{eq:reconstruction-tv}
    \mathrm{TV}(\vV)
    =d_{x_1}^2\sum_{\vn\in\mathbb{N}_{\mathrm{ROI}}}
    \left(\sum_{m=1}^{3}
    \left|[\vV]_{\vn+\vec e_m}-[\vV]_{\vn}\right|^2\right)^{1/2},
\end{equation}
with each difference set to zero when the neighboring index lies outside the ROI.

Weighted regularized dual averaging (RDA) was applied following Algorithm~2 of Matthews \textit{et al.}~\cite{matthewsRegularizedDualAveraging2017}, with gradients computed by DTA.
On the coarsest 1.2-mm grid, the SOS was initialized by resampling the true map and applying Gaussian smoothing with a standard deviation of 8~mm, followed by enforcement of the support and bounds.
Weighted stochastic gradients were accumulated, and the estimate was updated with a constrained TV proximal map, evaluated by fast gradient projection~\cite{beckFastGradientBasedAlgorithms2009} with support and bounds imposed at each proximal iteration.
Following Algorithm~3 in the appendix of Matthews \textit{et al.}~\cite{matthewsRegularizedDualAveraging2017}, a trial averaging weight of one was halved until the objective, including the TV penalty, decreased for the fixed encoding.
The TV weights $\lambda_c$ were 0.15, 0.15, and 0.7 on the successive 1.2-, 0.9-, and 0.6-mm grids; the RDA step scales were $1.6\times10^5$, $7\times10^4$, and $1.6\times10^4$.
The step scale multiplies the cumulative weighted gradient; its product with the cumulative averaging weight sets the proximal step.
Each stage comprised 200 iterations, with the accumulators reset at grid transitions.

\subsubsection{IP estimation in PACT}\label{sec:pact-ip}
For the linear PACT model, $\cS$ contained the prescribed medium parameters and the IP $\vPo$, which was the only unknown and specified the initial condition for $\sfX$.
The ETIR from data generation was retained in $\sfhG_{\vH}$.
With the TV penalty in Eq.~\eqref{eq:reconstruction-tv}, the reconstruction was formulated as
\begin{equation}\label{eq:pact-reconstruction}
    \widehat{\vPo}
    =\argmin_{\vPo\in\mathcal{C}_p}
    \left\{
    \frac{1}{2}\left\|
    \sfhG_{\vH}\!\left(\sfX(\cS)\right)
    -\sfB\!\left(\vG_{\mathrm{PACT}}\right)
    \right\|_{\mathrm{F}}^2
    +\lambda_p\,\mathrm{TV}(\vPo)
    \right\}.
\end{equation}
The convex constraint set was
\begin{equation*}
    \mathcal{C}_p
    =\left\{\vPo\in\cR\;\middle|\;
    \begin{array}{ll}
    {[\vPo]_{\vn}}\geq0,
        & \vn\in\mathbb{N}_{\mathrm{supp}},\\
    {[\vPo]_{\vn}}=0,
        & \vn\in\mathbb{N}_{\Omega}\setminus\mathbb{N}_{\mathrm{supp}}
    \end{array}
    \right\}.
\end{equation*}
On the coarsest 1.2-mm grid, the IP was initialized to zero.

Two PACT reconstructions were performed: one using the final USCT SOS estimate interpolated to each PACT grid, and the other using a homogeneous SOS of $1520.6~\mathrm{m\,s^{-1}}$.
In both cases, the SOS was held fixed and $c_{\mathrm{ref}}$ was set to its maximum on the current grid.
The noisy data and all other reconstruction settings were identical.

Barzilai--Borwein (BB) step sizes were used with the constrained TV proximal map to form update candidates.
A nonmonotone line search adapted from the spectral projected-gradient method~\cite{birginNonmonotoneSpectralProjected2000} selected the step toward each fixed candidate.
The step was halved until the objective, including the TV penalty, satisfied a sufficient-decrease condition relative to the largest of the previous ten objective values.
Because the PACT forward map is linear in IP for a fixed medium, only the first trial estimate in each line search required forward propagation.
Predicted data for subsequent halved steps were obtained by linear combinations of the forward predictions already computed at the current and first-trial estimates.
The prediction at the accepted estimate was reused in the next gradient evaluation, avoiding another forward computation.
The TV weights $\lambda_p$ were $8\times10^{-6}$, $8\times10^{-6}$, and $4\times10^{-5}$, and the initial BB steps were $6\times10^{11}$, $6\times10^{11}$, and $2\times10^{12}$.
Each stage was terminated when the maximum absolute change in IP divided by the maximum absolute IP was at most $2\times10^{-4}$, or after 100 iterations.

\subsection{Results}\label{sec:3d-results}

\begin{figure*}[!htb]
    \centering
    \includegraphics[width=\linewidth]{\arxivdir 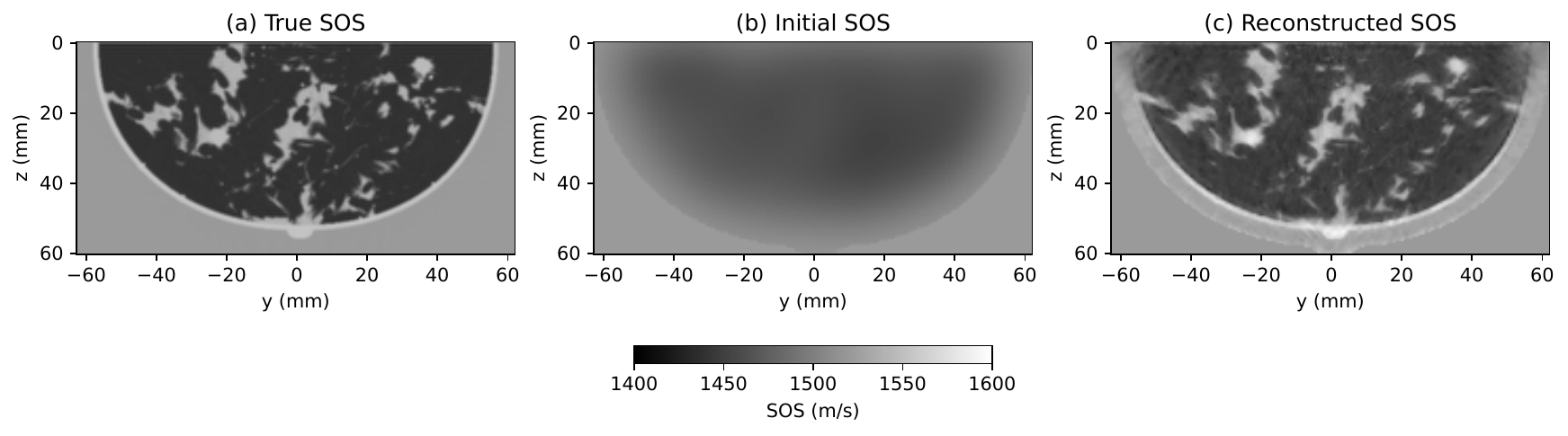}
    \caption{Comparison of true, initial, and reconstructed SOS within the ROI on the final 0.6-mm USCT grid. (a) Spatially filtered and resampled true SOS, (b) initial SOS obtained by Gaussian smoothing with an 8-mm standard deviation on the 1.2-mm grid, then interpolated to the displayed grid, and (c) reconstructed SOS. All panels show the cross section at $x=0$ with a common display range of 1400--1600~$\mathrm{m\,s^{-1}}$, matching the reconstruction bounds.}
    \label{fig:usct-reconstruction}
\end{figure*}

\begin{figure*}[!htb]
    \centering
    \includegraphics[width=\linewidth]{\arxivdir 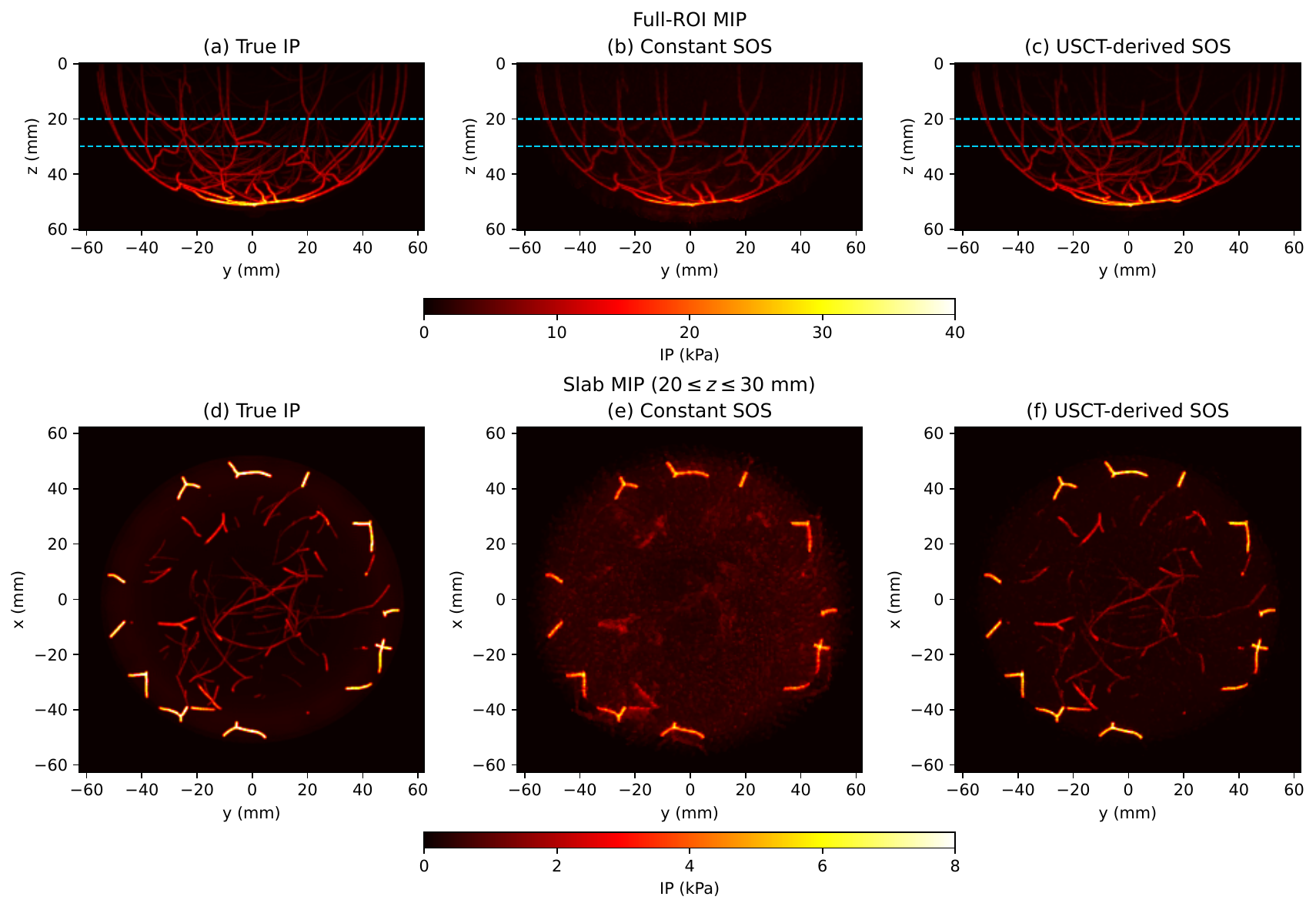}
    \caption{Comparison of PACT reconstructions using homogeneous and USCT-derived SOS. From left to right: spatially filtered and resampled true IP; IP reconstructed with a homogeneous SOS of $1520.6~\mathrm{m\,s^{-1}}$; and IP reconstructed with the USCT-derived SOS. (a)--(c) Full-ROI maximum-intensity projections (MIPs) along $x$, with a common display range of 0--40~kPa. Dashed lines mark the slab $20\leq z\leq30$~mm. (d)--(f) Slab MIPs along $z$, with a common display range of 0--8~kPa. Values above the display limits are saturated for visualization only. All panels retain the full ROI extent in $y$; this extent also defines the $x$ range in the slab views. All maps are shown on the final 0.5-mm grid without amplitude normalization. Both reconstructions used the same noisy data and all other reconstruction settings.}
    \label{fig:pact-preview}
\end{figure*}

Figures~\ref{fig:usct-reconstruction} and~\ref{fig:pact-preview} show the USCT and PACT reconstruction results on their final 0.6- and 0.5-mm grids, respectively.
For comparison, the true SOS and IP maps were low-pass filtered in the spatial Fourier domain to the corresponding Nyquist wavenumbers, downsampled onto the 0.6-mm USCT and 0.5-mm PACT grids, respectively, and then cropped to the reconstruction ROI in each case.

The main SOS heterogeneities in Fig.~\ref{fig:usct-reconstruction} were recovered despite the added noise, the coarser reconstruction grids, and the simplified AA and ambient-density maps.
Features near the chest wall were less clearly resolved; the limited angular coverage near this boundary may contribute to this loss of detail.
The root-mean-square error (RMSE) within the support decreased from 34.16~$\mathrm{m\,s^{-1}}$ for the initial estimate to 16.21~$\mathrm{m\,s^{-1}}$ for the reconstruction.

The PACT images reconstructed with the USCT-derived SOS and a homogeneous SOS are compared in Fig.~\ref{fig:pact-preview}.
With a homogeneous SOS, fine vessels became less distinct toward the breast interior, as shown in the slab projections in Fig.~\ref{fig:pact-preview}(d)--(f).
With the USCT-derived SOS, phase aberration was partly compensated and these internal vascular structures were more clearly resolved.
Within the support, the RMS error divided by the RMS of the true IP was 60.30\% with the homogeneous SOS and 29.06\% with the USCT-derived SOS, without amplitude normalization.

\FloatBarrier
Table~\ref{tab:reconstruction-time} reports computation time and peak memory for evaluating the gradient together with the data-fidelity value, and computation time for evaluating the value alone.
The gradient evaluations in Table~\ref{tab:reconstruction-time} included both forward and adjoint computations.
During PACT reconstruction, the accepted forward prediction was reused in the next gradient evaluation, which required only the adjoint computation (Section~\ref{sec:pact-ip}).
Measurements for the linear model with power-law AA and dispersion disabled are included as a computational reference.

All gradient evaluations fit within GPU memory without checkpointing.
On the finest grids, peak memory was 3.53~GiB for the PACT IP gradient and 22.06~GiB for the USCT SOS gradient, despite the finer PACT grid.
For the linear model, the IP gradient requires no stored forward history, whereas the SOS gradient retains $\vV_c$ within the ROI (Eqs.~\eqref{eq:grad-p0} and~\eqref{eq:grad-sos}).
An additional memory check on the final USCT grid showed that reverse-mode AD of the \pausct{} forward computation failed GPU memory allocation despite ROI-restricted SOS inputs, whereas DTA completed; neither used checkpointing.
This check and comparisons under matched recomputation schedules are detailed in~\ref{app:ad-dta-comparison}.

Each USCT stage comprised 200 iterations; the PACT stages comprised 49, 84, and 59 iterations with the USCT-derived SOS and 86, 83, and 63 with the homogeneous SOS.
The complete reconstructions took approximately 4~h 5~min for USCT, 1~h 41~min for PACT with the USCT-derived SOS, and 2~h 33~min for PACT with the homogeneous SOS, each on one GPU.
These times include input loading, setup, compilation, data encoding for USCT, iterative TV proximal updates on the GPU, and file output.
All reconstructions used the linear lossy model.

\begin{table}[!htbp]
\centering
\caption{Computation time and peak memory for SOS-gradient evaluation in USCT and IP-gradient evaluation in PACT.
Measurements for the lossless model, with AA and its associated dispersion disabled, are shown in parentheses.
Gradient evaluations include both forward (without reuse) and adjoint computations without checkpointing.
Timings include host--device transfers but exclude spatial-coefficient precomputation.
The PACT measurements used the USCT-derived SOS.}
\label{tab:reconstruction-time}
\small
\begin{tabular}{ccccc}
\hline
& \shortstack{Spacing\\(mm)} & \multicolumn{2}{c}{\shortstack{Gradient and\\data-fidelity value}} & \shortstack{Data-fidelity\\value alone} \\
& & Time (s) & Peak memory (GiB) & Time (s) \\
\hline
\multirow{3}{*}{USCT} & 1.2 & 1.463 (1.095) & 1.645 (1.581) & 0.776 (0.566) \\
& 0.9 & 6.613 (4.985) & 4.757 (4.603) & 3.414 (2.505) \\
& 0.6 & 38.821 (28.850) & 22.057 (21.436) & 19.795 (14.638) \\
\hline
\multirow{3}{*}{PACT} & 1.2 & 1.363 (1.021) & 0.351 (0.348) & 0.647 (0.484) \\
& 0.8 & 8.454 (6.327) & 0.950 (0.732) & 3.991 (2.941) \\
& 0.5 & 83.974 (61.490) & 3.530 (2.644) & 40.458 (29.295) \\
\hline
\end{tabular}
\end{table}

\FloatBarrier

\section{Discussion}\label{sec:discussion}
\paragraph{Gradient evaluation with DTA and AD}
The storage savings depend on the requested gradients and the propagation model.
For nonlinear propagation, the full-grid histories required by the adjoint increments remain even when the unknowns are restricted to an ROI; checkpointing reduces the temporal extent of these histories rather than their spatial extent.
For joint estimation, sharing the adjoint solution avoids repeated adjoint solves for the requested DTA-based medium and source gradients, but the required forward histories depend on the selected parameters (Section~\ref{sec:parameter-gradients}).
The comparisons in~\ref{app:ad-dta-comparison} (Fig.~\ref{fig:ad-dta-memory}) show how DTA and JAX reverse-mode AD differ in time and GPU memory when applied to the same \pausct{} forward models with matched recomputation schedules.

We also compared \pausct{} with \texttt{j-Wave} (0.2.1), which relies on JAX AD for gradient evaluation (\ref{app:jwave-comparison}).
Its time-domain model is linear and lossless, so we disabled power-law AA and cumulative nonlinearity in \pausct{} for this USCT comparison.
With JAX reverse-mode AD in both packages, \pausct{} used about half the peak GPU memory and took slightly less time to evaluate the SOS gradient.
The difference under a common AD method illustrates how the forward operations and the intermediate arrays they produce affect the cost of gradient evaluation.

Extending the physical model or changing its discretization may require revising and independently verifying the adjoint and gradient expressions, whereas AD can differentiate the modified forward computation without a separate derivation.
\pausct{} can combine both approaches: the derivative of the data-fidelity function with respect to the predicted traces can be obtained by AD or supplied algebraically, whereas the discrete adjoint recursion uses the explicit DTA formulas.

\paragraph{Limitations}
Time-fractional absorption models can accommodate a spatially varying power-law exponent.
For example, the static-memory method of King \textit{et al.}~\cite{king2025modelling} replaces full-history storage with recursively updated auxiliary fields, keeping memory use independent of the number of time steps.
The auxiliary fields can nevertheless require substantial memory on large 3D grids.
The fractional-Laplacian absorption formulation adopted here avoids this additional storage but assumes a spatially constant exponent $\gamma$, while permitting a spatially varying AA prefactor.
The wave model is intended for media that can be treated as acoustic fluids; elastic and shear-wave propagation are not included.

In the current implementation, each wave simulation is performed on one GPU.
Distributed spatial FFTs and high-bandwidth GPU interconnects could support an extension to simulations spanning multiple GPUs.

\paragraph{Outlook}
Virtual imaging studies can compare reconstruction with prescribed medium properties against joint estimation of IP and SOS or of medium and receive-response parameters.
Gradients with respect to transmit waveforms and receive ETIR coefficients also support transducer calibration.
The $B/A$ gradient supports studies of nonlinear multiparameter inversion~\cite{shultzman2022nonlinear}.
Gradient accuracy alone does not establish whether the selected parameters can be recovered; acquisition design, parameter coupling, and prior information must also be considered~\cite{matthewsParameterizedJointReconstruction2018}.

\section{Conclusion}\label{sec:conclusion}
This work presented a framework to compute gradients for optimization-based reconstruction in PACT and USCT.
The gradients are consistent with the complete discrete forward model, which accounts for heterogeneous, lossy, and nonlinear acoustic propagation and spatial or spatiotemporal responses of finite-aperture transmitters and receivers.
Requested adjoint-based gradients share one adjoint solution per acquisition and objective, with forward quantities stored or recomputed according to the parameters being estimated.
Numerical tests verified agreement with independently computed derivatives in 2D, and a 3D virtual imaging study demonstrated USCT-augmented PACT with noisy data and prescribed model mismatch.

\appendix
\counterwithin*{figure}{section}
\section{Spatial apodization in transmission and reception}\label{app:transducer-operators}
\subsection{Spatial interpolation coefficients}
To present the separable transmit and receive formulations together, let $i$ denote the index of an arbitrary transmitting or receiving element.
For a finite aperture $\Gamma^{[i]}$, let $\{\hbr^{[i,j]}\}_{j=0}^{M^{[i]}-1}$ denote the quadrature points and $\{\varpi_{j}^{[i]}\}_{j=0}^{M^{[i]}-1}$ the corresponding normalized weights, with $\sum_{j=0}^{M^{[i]}-1}\varpi_{j}^{[i]}=1$.
The quadrature weights are $\lvert\Gamma^{[i]}\rvert\varpi_{j}^{[i]}$ and may differ between points.
The hat notation indicates continuous coordinates in $\Gamma^{[i]}\subset\mathbb{R}^d$ that are not constrained to the Cartesian grid.
To evaluate the pressure values $p(\hbr^{[i,j]},t^{(\ell)})$, the on-grid discrete pressure $\vP^{(\ell)}$ is interpolated using a truncated sinc kernel~\cite{wiseRepresentingArbitraryAcoustic2019}:
\begin{equation}\label{eq:bl-interp}
    p(\hbr^{[i,j]},t^{(\ell)}) \approx \sum_{\vn\in\mathbb{N}_{\Omega}} \varphi(\br^{[\vn]}; \hbr^{[i,j]},\mu)\,[\vP^{(\ell)}]_{\vn}.
\end{equation}
Here, the interpolation kernel $\varphi$ centered at $\hbr^{[i,j]}$ is defined as
\begin{equation}\label{eq:interp-kernel}
    \varphi(\br^{[\vn]};\,\hbr^{[i,j]},\mu) = \mathbbm{1}_{\vn\in\mathcal{V}^{[i,j]}(\mu)}\,\prod_{m=1}^{d} \sinc\!\left(\pi \xi_m^{[\vn]}\right),
\end{equation}
where $\xi_m^{[\vn]} \coloneqq (x_m^{[\vn]}-\hat{x}_m^{[i,j]})/d_{x_m}$.
The indicator function $\mathbbm{1}_{\vn\in\mathcal{V}^{[i,j]}(\mu)}$ restricts the evaluation to a sparse local neighborhood.
For every quadrature point, \pausct{} uses the integer-offset set
\begin{equation*}
    \mathcal{O}(\mu)\coloneqq
    \left\{\boldsymbol{o}\in\mathbb{Z}^{d}\middle|
    \prod_{m=1}^{d}\max\bigl(|o_m|,1\bigr)\leq n_{\mu}
    \right\},
\end{equation*}
where $n_{\mu} \coloneqq \lceil 1/(\pi\mu) \rceil$ sets the maximum index offset along each axis for a truncation parameter $\mu>0$.
Let $\bar{\vn}^{[i,j]}$ denote the nearest grid index to $\hbr^{[i,j]}$. The grid indices included in the interpolation form the set
\begin{equation} \label{eq:R-def}
    \mathcal{V}^{[i,j]}(\mu)
    =
    \left\{\bar{\vn}^{[i,j]}+\boldsymbol{o}\in\mathbb{N}_{\Omega}\setminus\mathbb{N}_{\mathrm{PML}}\middle|
    \boldsymbol{o}\in\mathcal{O}(\mu)
    \right\}.
\end{equation}
Although these grid indices are selected from integer offsets relative to the nearest grid index, the sinc values in Eq.~\eqref{eq:interp-kernel} use the exact fractional offsets $\xi_m^{[\vn]}$.
The grid index set for element $i$ is the union over its quadrature points:
\begin{equation} \label{eq:R-def-total}
    \widetilde{\mathcal{V}}^{[i]}(\mu)
    \coloneqq
    \bigcup_{j=0}^{M^{[i]}-1}\mathcal{V}^{[i,j]}(\mu).
\end{equation}

\subsection{Discrete spatial transmit and receive coefficients}
For a spatial amplitude weighting $a^{[i]}$, the element's coefficient array $\vec{S}^{[i]}\in\cR$ is defined by
\begin{equation} \label{eq:sensitivity-sir}
    [\vec{S}^{[i]}]_{\vn} = \sum_{j=0}^{M^{[i]}-1} \varpi_{j}^{[i]}a^{[i]}(\hbr^{[i,j]})\,\varphi(\br^{[\vn]}; \hbr^{[i,j]},\mu).
\end{equation}

When the element is modeled as a point at $\hbr^{[i]}$, aperture quadrature is omitted. For an off-grid point, the coefficient array reduces to
\begin{equation}\label{eq:sensitivity-point-off}
    [\vec{S}^{[i]}]_{\vn} = \varphi(\br^{[\vn]}; \hbr^{[i]},\mu).
\end{equation}
The coefficient arrays in Eqs.~\eqref{eq:sensitivity-sir} and~\eqref{eq:sensitivity-point-off} are sparse: nonzero entries are restricted to $\widetilde{\mathcal{V}}^{[i]}(\mu)$, with the union reducing to a single neighborhood for an off-grid point.
If $\hbr^{[i]}$ coincides with a primary grid point in the non-PML region ($\hbr^{[i]} = \br^{[\vn']}$), Eq.~\eqref{eq:sensitivity-point-off} reduces to a Kronecker delta:
\begin{equation}\label{eq:sensitivity-point-on}
    [\vec{S}^{[i]}]_{\vn} = \delta_{\vn'\vn},
\end{equation}
where $\delta_{\vn'\vn} = 1$ if $\vn = \vn'$ and $0$ otherwise.
The receive and transmit arrays $\vec{S}_{\mathrm{rx}}^{[i_{\mathrm{rx}}]}$ and $\vec{S}_{\mathrm{tx}}^{[i_{\mathrm{tx}}]}$ follow these definitions with $(i,a^{[i]})=(i_{\mathrm{rx}},a_{\mathrm{rx}}^{[i_{\mathrm{rx}}]})$ and $(i,a^{[i]})=(i_{\mathrm{tx}},a_{\mathrm{tx}}^{[i_{\mathrm{tx}}]})$, respectively.
For coincident apertures with identical apodization, quadrature points and weights, computational grid, and truncation parameters, $\vec{S}_{\mathrm{rx}}^{[i_{\mathrm{rx}}]}=\vec{S}_{\mathrm{tx}}^{[i_{\mathrm{tx}}]}$; the additional transmit scaling is defined in~\ref{app:source-convention}.

The aperture-averaged pressure in Eq.~\eqref{eq:sir-continuous} is approximated using the quadrature points and weights as
\begin{equation}\label{eq:sir-discrete}
    \bar{p}^{[i_{\mathrm{rx}}]}(t^{(\ell)}) \approx \sum_{j=0}^{M^{[i_{\mathrm{rx}}]}-1} \varpi_{j}^{[i_{\mathrm{rx}}]}a_{\mathrm{rx}}^{[i_{\mathrm{rx}}]}(\hbr^{[i_{\mathrm{rx}},j]})\,p(\hbr^{[i_{\mathrm{rx}},j]},t^{(\ell)}).
\end{equation}
Substituting Eq.~\eqref{eq:bl-interp} into Eq.~\eqref{eq:sir-discrete} and using Eq.~\eqref{eq:sensitivity-sir} yields the discrete receive action
\begin{equation}\label{eq:sir-innerproduct}
    [\bar{\vP}]_{\ell,i_{\mathrm{rx}}} \coloneqq \langle\vec{S}_{\mathrm{rx}}^{[i_{\mathrm{rx}}]}, \vP^{(\ell)}\rangle_{\cR},
    \qquad i_{\mathrm{rx}}=0,\dots,\Nr-1,
\end{equation}
which is the componentwise form of Eq.~\eqref{eq:dd-sir-step}.

\subsection{Discrete source increments}\label{app:source-convention}
The sampled waveforms $\vec{s}_p^{[i_{\mathrm{tx}}]},\vec{s}_u^{[i_{\mathrm{tx}}]}\in\mathbb{R}^{L_s+1}$ corresponding to Eq.~\eqref{eq:continuous-source-waveforms} are defined by
\begin{equation*}
[\vec{s}_p^{[i_{\mathrm{tx}}]}]_{\ell}\coloneqq s_p^{[i_{\mathrm{tx}}]}(t^{\slp}),
\qquad
[\vec{s}_u^{[i_{\mathrm{tx}}]}]_{\ell}\coloneqq s_u^{[i_{\mathrm{tx}}]}(t^{(\ell)}),
\qquad \ell=0,\dots,L_s.
\end{equation*}
Both sequences vanish for $\ell>L_s$.
A uniform time step $\dt$ is used throughout time marching, and the prescribed source waveforms must be adequately resolved at this interval.
When the source \textit{k}-space correction is used, define~\cite{coxAccurateTimeVaryingSources2018}
\begin{equation}
\sfKsrc(\vV)
\coloneqq
\sfFri\!\left(
\boldsymbol{\kappa}_{\mathrm{src}}\had\sfFr(\vV)
\right),
\qquad
\boldsymbol{\kappa}_{\mathrm{src}}
\coloneqq
\cos\!\left(\frac{c_{\mathrm{ref}}\dt}{2}\vK\right).
\label{eq:source-kspace-correction}
\end{equation}
Without this correction, $\sfKsrc$ is the identity.
For particle-velocity sources, let $\nu_m^{[i_{\mathrm{tx}}]}$, $m=1,\dots,d$, denote the components of the constant unit normal to transmitting element $i_{\mathrm{tx}}$.
With the transmitters in water (Section~\ref{back:cd:wave}), $c_0(\br')=c_{\mathrm{water}}$ over their apertures, and the total source increments are
\begin{align}
\vtMm^{\slp}
&\coloneqq
\frac{2\dt}{d\,c_{\mathrm{water}}\bar d_x}
\sum_{i_{\mathrm{tx}}=0}^{\Ntx-1}w_{\mathrm{tx}}^{[i_{\mathrm{tx}}]}
\frac{\lvert\Gamma^{[i_{\mathrm{tx}}]}\rvert}{\Delta_{\Gamma}^{[i_{\mathrm{tx}}]}}
[\vec{s}_p^{[i_{\mathrm{tx}}]}]_{\ell}\sfKsrc(\vec{S}_{\mathrm{tx}}^{[i_{\mathrm{tx}}]}), \label{eq:scaled-pressure-source}\\
\vtFm^{(\ell)}
&\coloneqq
\frac{2c_{\mathrm{water}}\dt}{\bar d_x}
\sum_{i_{\mathrm{tx}}=0}^{\Ntx-1}w_{\mathrm{tx}}^{[i_{\mathrm{tx}}]}
\frac{\lvert\Gamma^{[i_{\mathrm{tx}}]}\rvert}{\Delta_{\Gamma}^{[i_{\mathrm{tx}}]}}
\nu_m^{[i_{\mathrm{tx}}]}[\vec{s}_u^{[i_{\mathrm{tx}}]}]_{\ell}\sfKsrc(\vec{S}_{\mathrm{tx}}^{[i_{\mathrm{tx}}]}),
\qquad m=1,\dots,d.
\label{eq:scaled-velocity-source}
\end{align}
Here, $\bar d_x\coloneqq(\prod_{q=1}^{d}d_{x_q})^{1/d}$ denotes the geometric-mean grid spacing.
For an aperture of geometric dimension $d_{\Gamma}^{[i_{\mathrm{tx}}]}$ (1 for a line and 2 for a surface), define the grid measure $\Delta_{\Gamma}^{[i_{\mathrm{tx}}]}\coloneqq(\bar d_x)^{d_{\Gamma}^{[i_{\mathrm{tx}}]}}$.

The spatial source coefficients are $\big(\lvert\Gamma^{[i_{\mathrm{tx}}]}\rvert/\Delta_{\Gamma}^{[i_{\mathrm{tx}}]}\big)\vec{S}_{\mathrm{tx}}^{[i_{\mathrm{tx}}]}$, so transmission retains the aperture measure.
For a line in 2D or a surface in 3D, $\bar d_x\Delta_{\Gamma}^{[i_{\mathrm{tx}}]}=\prod_{q=1}^{d}d_{x_q}$, so that the aperture coefficients reduce to $\lvert\Gamma^{[i_{\mathrm{tx}}]}\rvert/\prod_{q=1}^{d}d_{x_q}$, as required by the normalization of the spatial Dirac delta in Eq.~\eqref{eq:continuous-source-waveforms}.
For point sources, $\lvert\Gamma^{[i_{\mathrm{tx}}]}\rvert/\Delta_{\Gamma}^{[i_{\mathrm{tx}}]}$ is replaced by 1 in the source formulas.
The remaining factors $2\dt/(d\,c_{\mathrm{water}}\bar d_x)$ and $2c_{\mathrm{water}}\dt/\bar d_x$ in Eqs.~\eqref{eq:scaled-pressure-source} and~\eqref{eq:scaled-velocity-source} have the form of the additive pressure- and velocity-source scalings in \texttt{k-Wave}~\cite{treebyWaveMatlabToolbox2010}, with $\bar d_x$ equal to the grid spacing on isotropic grids.

The same entries of $\sfKsrc(\vec{S}_{\mathrm{tx}}^{[i_{\mathrm{tx}}]})$ are used at corresponding array indices on the primary and staggered grids. These indices represent the physical locations $\br^{[\vn]}$ and $\br^{[\vn]}+\tfrac{d_{x_m}}{2}\bm{e}_m$, respectively; no additional spatial interpolation is applied to shift a velocity source onto the staggered grids.
The same $\vtMm^{\slp}$ is added to every split-density component; the factor $1/d$ in Eq.~\eqref{eq:scaled-pressure-source} accounts for this split.
The increments are linear in the waveform samples and vanish for $\ell>L_s$.
The increments in Eqs.~\eqref{eq:scaled-pressure-source} and~\eqref{eq:scaled-velocity-source} already include $\dt$ and have units of density and particle velocity, respectively; they are not samples of the continuous source rates $m$ and $\bm{f}$.

\section{Spatiotemporal aperture responses}\label{app:aperture-temporal}
{
Sections~\ref{back} and~\ref{sec:dta} develop the forward and discrete-adjoint formulations for the separable aperture models in Eqs.~\eqref{eq:continuous-source-waveforms} and~\eqref{eq:sir-continuous}, in which spatial apodization is independent of time~\cite{jensenCalculationPressureFields1992}.
This appendix gives the more general spatiotemporal transmit and receive formulation supported by \pausct{}, allowing the temporal response to vary over each aperture~\cite{cuetoSpatialResponseIdentification2022}.
Its discretization and discrete adjoints use the aperture quadrature and spatial interpolation developed in~\ref{app:transducer-operators}.
An example with concave acoustic lenses is presented in Section~\ref{sec:lens-example}.

\subsection{Continuous model}
Let $k_{\mathrm{tx}}^{[i_{\mathrm{tx}}]}(\br',\tau)$ and $k_{\mathrm{rx}}^{[i_{\mathrm{rx}}]}(\br',\tau)$ denote effective spatiotemporal transmit and receive responses defined over the corresponding apertures.
These prescribed responses have units of $\mathrm{s}^{-1}$ and depend on temporal lag $\tau\in\mathbb{R}$, not on absolute time; the resulting operators are linear and time-invariant.
By convention, temporal transmit effects common to all points of an element are included in its prescribed waveform.
For reception, the common receive-chain and temporal sampling-aperture response is included in the receiving element's ETIR.
These effects are not included again in the corresponding aperture responses.

For $s^{[i_{\mathrm{tx}}]}\in\{s_p^{[i_{\mathrm{tx}}]},s_u^{[i_{\mathrm{tx}}]}\}$, the factor $a_{\mathrm{tx}}^{[i_{\mathrm{tx}}]}(\br')s^{[i_{\mathrm{tx}}]}(t)$ in Eq.~\eqref{eq:continuous-source-waveforms} is replaced by
\begin{equation}\label{eq:spatiotemporal-source}
 \int_{\mathbb{R}}k_{\mathrm{tx}}^{[i_{\mathrm{tx}}]}(\br',\tau)
 s^{[i_{\mathrm{tx}}]}(t-\tau)\,d\tau.
\end{equation}

Equation~\eqref{eq:sir-continuous} becomes
\begin{equation}\label{eq:spatiotemporal-receive}
 \bar{p}^{[i_{\mathrm{rx}}]}(t)
 =\frac{1}{\lvert\Gamma^{[i_{\mathrm{rx}}]}\rvert}
 \int_{\Gamma^{[i_{\mathrm{rx}}]}}\int_{\mathbb{R}}
 k_{\mathrm{rx}}^{[i_{\mathrm{rx}}]}(\br',\tau)\,p(\br',t-\tau)\,d\tau\,d\Gamma^{[i_{\mathrm{rx}}]}(\br'),
\end{equation}
followed by ETIR convolution as in Eq.~\eqref{eq:meas}.

When $k_{\mathrm{tx}}(\br',\tau)=a_{\mathrm{tx}}(\br')\delta(\tau)$ and $k_{\mathrm{rx}}(\br',\tau)=a_{\mathrm{rx}}(\br')\delta(\tau)$, with $\delta(\tau)$ the one-dimensional Dirac delta, these expressions reduce to Eqs.~\eqref{eq:continuous-source-waveforms} and~\eqref{eq:sir-continuous}.

\subsection{Discrete model and adjoints}\label{app:aperture-temporal-discrete}
Let $\sigma\in\{\mathrm{tx},\mathrm{rx}\}$ and let $i$ index the corresponding elements.
For time-independent apodization, the quadrature contributions are combined into one array per element in Eq.~\eqref{eq:sensitivity-sir}.
For a general spatiotemporal response $k_{\sigma}^{[i]}(\br',\tau)$, temporal convolution is applied separately at each quadrature point before summation.
For quadrature point $\hbr^{[i,j]}$ of element $i$, denote the single-point array in Eq.~\eqref{eq:sensitivity-point-off} by $\vec{S}_{\sigma}^{[i,j]}\in\cR$:
\begin{equation*}
 [\vec{S}_{\sigma}^{[i,j]}]_{\vn}
 =\varphi(\br^{[\vn]};\hbr^{[i,j]},\mu).
\end{equation*}

Let $\widehat{\vec{k}}_{\sigma}^{[i,j]}$ denote the Fourier multiplier for temporal convolution at quadrature point $\hbr^{[i,j]}$.
It is formed from $k_{\sigma}^{[i]}(\hbr^{[i,j]},\tau)$, including temporal integration weights and lag alignment.
The multiplier is evaluated at the FFT frequencies of the input trace, whose sampling interval is $\dt$ for transmission and $\Dt$ for reception.
For an input trace $\vec{v}$ of length $n$, define the convolution operator and its adjoint by
\begin{align*}
 \mathsf{C}_{\sigma}^{[i,j]}(\vec{v})
 &=\sfT_n\sfFti\!\left[
 \widehat{\vec{k}}_{\sigma}^{[i,j]}\had\sfFt(\vec{v})\right],\\
 \bigl(\mathsf{C}_{\sigma}^{[i,j]}\bigr)^*(\vec{v})
 &=\sfT_n\sfFti\!\left[
 \overline{\widehat{\vec{k}}_{\sigma}^{[i,j]}}\had\sfFt(\vec{v})\right].
\end{align*}
Here, $\sfFt$ denotes the columnwise temporal FFT and $\sfFti$ its inverse, with inputs zero-padded to a common transform length; $\sfT_n$ retains the first $n$ samples.
An overbar denotes elementwise complex conjugation.

For either source waveform $\vec{s}^{[i]}\in\{\vec{s}_p^{[i]},\vec{s}_u^{[i]}\}$, write
$\hat{\vec{s}}^{[i,j]}\coloneqq\mathsf{C}_{\mathrm{tx}}^{[i,j]}(\vec{s}^{[i]})$.
For $m=1,\dots,d$, the source increments in Eqs.~\eqref{eq:scaled-pressure-source} and~\eqref{eq:scaled-velocity-source} become
\begin{align}
 \vtMm^{\slp}
 &=\frac{2\dt}{d\,c_{\mathrm{water}}\bar d_x}
 \sum_{i=0}^{\Ntx-1}w_{\mathrm{tx}}^{[i]}
 \frac{\lvert\Gamma^{[i]}\rvert}{\Delta_{\Gamma}^{[i]}}
 \sum_{j=0}^{M^{[i]}-1}\varpi_j^{[i]}
 [\hat{\vec{s}}_p^{[i,j]}]_{\ell}
 \sfKsrc(\vec{S}_{\mathrm{tx}}^{[i,j]}),
 \label{eq:pointwise-aperture-transmit}\\
 \vtFm^{(\ell)}
 &=\frac{2c_{\mathrm{water}}\dt}{\bar d_x}
 \sum_{i=0}^{\Ntx-1}w_{\mathrm{tx}}^{[i]}
 \frac{\lvert\Gamma^{[i]}\rvert}{\Delta_{\Gamma}^{[i]}}
 \nu_m^{[i]}
 \sum_{j=0}^{M^{[i]}-1}\varpi_j^{[i]}[\hat{\vec{s}}_u^{[i,j]}]_{\ell}
 \sfKsrc(\vec{S}_{\mathrm{tx}}^{[i,j]}).
 \label{eq:pointwise-aperture-transmit-velocity}
\end{align}
The chain rule gives
\begin{equation*}
 \dd{\sfJ}{\vec{s}^{[i]}}
 =\sum_{j=0}^{M^{[i]}-1}
 \bigl(\mathsf{C}_{\mathrm{tx}}^{[i,j]}\bigr)^*
 \dd{\sfJ}{\hat{\vec{s}}^{[i,j]}}.
\end{equation*}
The derivatives on the right are evaluated using Eq.~\eqref{eq:grad-source-pressure} for pressure sources or Eq.~\eqref{eq:grad-source-velocity} for particle-velocity sources, with $\vec{S}_{\mathrm{tx}}^{[i]}$ replaced by $\varpi_j^{[i]}\vec{S}_{\mathrm{tx}}^{[i,j]}$.

For reception, pressure is interpolated at each quadrature point, resampled to interval $\Dt$, and convolved with the corresponding aperture response before summation:
\begin{equation}\label{eq:pointwise-aperture-receive}
 [\bar{\vP}_{\Dt}]_{s,i}
 =\sum_{j=0}^{M^{[i]}-1}\varpi_j^{[i]}
 \left[\mathsf{C}_{\mathrm{rx}}^{[i,j]}
 \left(\mathsf{Z}\!\left(
 \bigl(\langle\vec{S}_{\mathrm{rx}}^{[i,j]},\vP^{(\ell)}\rangle_{\cR}\bigr)_{\ell=0}^{L}
 \right)\right)\right]_s,
\end{equation}
where $i=i_{\mathrm{rx}}$ and $s=0,\dots,S$.

For Eq.~\eqref{eq:pointwise-aperture-receive}, Eqs.~\eqref{eq:adjoint-source} and~\eqref{eq:dd-sir-adjoint} give the adjoint source as
\begin{equation}\label{eq:spatiotemporal-adjoint-source}
 \adsrc^{(k)}
 =\sum_{i=0}^{\Nr-1}\sum_{j=0}^{M^{[i]}-1}\varpi_j^{[i]}
 \left[\mathsf{Z}^{*}\bigl(\mathsf{C}_{\mathrm{rx}}^{[i,j]}\bigr)^{*}
 \left(\left[\mathsf{C}_{\vH}^{*}\sfB^{*}(\nabla_{\vV}\sfJ)\right]_{\cdot,i}\right)\right]_k
 \vec{S}_{\mathrm{rx}}^{[i,j]},
\end{equation}
where $k=0,\dots,L$ and $\nabla_{\vV}\sfJ$ is evaluated at the predicted data.
}

\section{Temporal interpolation, ETIR, and filtering}\label{app:temporal-operators}
The three operations in Eq.~\eqref{eq:dd-eir-full} and their discrete adjoints are evaluated channelwise in parallel.

\subsection{Temporal interpolation}
When $\Dt<\dt$, $\mathsf{Z}$ upsamples the pressure traces to the measurement time grid before ETIR convolution using cubic convolution interpolation~\cite{keysCubicConvolutionInterpolation1981}.
The adjoint $\mathsf{Z}^*$ accumulates contributions on the \textit{k}-PSTD time grid using the interpolation indices and weights.
For $\Dt=\dt$, $\mathsf{Z}$ is the identity.

\subsection{ETIR convolution}\label{app:erir-convolution}
The channel responses are applied through the causal convolution defined in Section~\ref{back:dd:receive}.
The discrete coefficients $\vH$ already include the temporal integration factor $\Dt$.
For $\vec{A},\vec{E}\in\cG$, the convolution and its adjoint are evaluated as
\begin{align*}
    \mathsf{C}_{\vH}(\vec{A})
    &=\sfT_{S+1}\sfFti\!\left[
    \sfFt(\vec{A})\had\sfFt(\vH)\right],\\
    \mathsf{C}_{\vH}^*(\vec{E})
    &=\sfT_{S+1}\sfFti\!\left[
    \sfFt(\vec{E})\had\overline{\sfFt(\vH)}\right].
\end{align*}
The FFT inputs are zero-padded to a common transform length of at least $S+S_r+1$ samples.
If the ETIR is incorporated into the spatiotemporal receive response in~\ref{app:aperture-temporal}, the separate ETIR convolution and its adjoint must be omitted to avoid applying the ETIR twice.

\subsection{Temporal filtering}
{
The optional filter $\sfB$ is specified by a frequency response $\widehat{\vec b}$ on the temporal FFT grid for sampling interval $\Dt$, with conjugate symmetry to give real output.
The response may be shared across receiving elements or specified separately for each element.
For $\vec{A},\vec{E}\in\cG$, the filter and its adjoint are
\begin{align*}
    \sfB(\vec{A})
    &=\sfT_{S+1}\sfFti\!\left[
    \widehat{\vec b}\had\sfFt(\vec{A})\right],\\
    \sfB^{*}(\vec{E})
    &=\sfT_{S+1}\sfFti\!\left[
    \overline{\widehat{\vec b}}\had\sfFt(\vec{E})\right].
\end{align*}
Inputs are zero-padded to the transform length at which $\widehat{\vec b}$ is specified.
For a real-valued frequency response, $\sfB^{*}=\sfB$.
For low-pass filtering based on spatial resolution, the cutoff can be chosen as $c_{\min}/(n_{\mathrm{PPW}}\max_m d_{x_m})$, where $c_{\min}$ is a prescribed lower bound on SOS.
At least two points per wavelength ($n_{\mathrm{PPW}}\ge2$) are recommended.
}

\section{Forward-model comparison with \texttt{k-Wave}}\label{app:kwave-validation}
The nonlinear USCT data-generation setup in Section~\ref{sec:verification} was used to compare forward pressure traces with \texttt{k-Wave}~\cite{treebyWaveMatlabToolbox2010,treebyModelingNonlinearUltrasound2012}, using MATLAB toolbox version 1.4 and CUDA solver version 1.3, compiled with CUDA 12.9.
The \texttt{k-Wave} simulations were also performed in single precision.
Two cases were considered for both transmission and reception: (i) on-grid points obtained by snapping the element centers to the nearest grid nodes and (ii) the line apertures used in that section.
Both packages used identical grid indices or aperture quadrature points, with source \textit{k}-space correction enabled.
For this comparison, uniform unit apodization was used for both transmission and reception instead of the tapered profile in Section~\ref{sec:verification}.
Temporal receive operations were omitted so that pressure traces were compared directly.

Table~\ref{tab:kwave-forward} reports mean execution times over ten runs after one warm-up and relative RMS differences evaluated over all receiver samples, normalized by the RMS of the \texttt{k-Wave} traces.
For line reception, \pausct{} evaluates the aperture integrals on the GPU during time marching, whereas \texttt{k-Wave} combines the recorded grid pressures into element traces on the CPU afterward.
The \texttt{k-Wave} simulation and CPU receive postprocessing were therefore timed separately.

The \pausct{} computations were timed after JIT compilation, with spatial coefficients computed in advance and inputs already on the GPU; GPU execution was synchronized.
The \texttt{k-Wave} simulation times include GPU data transfers and recording pressure at sensor grid points, but exclude input loading and FFT planning.
The CPU reception times include calculating the spatial interpolation weights and integrating the recorded grid pressures over each aperture.\footnote{These CPU measurements used MATLAB R2024a with eight computational threads on a host with two AMD EPYC 9124 processors.}

\begin{table}[htbp]
\centering
\caption{Computation time and relative RMS differences for nonlinear USCT pressure traces.}
\label{tab:kwave-forward}
\small
\setlength{\tabcolsep}{4pt}
\begin{tabular}{lrrrr}
\hline
& \pausct{} & \multicolumn{2}{c}{\texttt{k-Wave}} & Relative RMS \\
Element model & \shortstack{\textit{k}-PSTD + reception\\(GPU, s)} & \shortstack{\textit{k}-PSTD\\(GPU, s)} & \shortstack{Reception\\(CPU, s)} & difference (\%) \\
\hline
On-grid points & 17.62 & 18.66 & 0.001 & $1.57\times10^{-3}$ \\
Line apertures & 17.89 & 18.86 & 2.913 & $2.59\times10^{-2}$ \\
\hline
\end{tabular}
\end{table}

For the 96 receiving line elements at $\mu=0.05$, \texttt{k-Wave} returned 76,384 grid traces with 8640 samples each, requiring 2.46~GiB of host memory for this single-precision pressure array alone.
These grid traces were then combined on the CPU into 96 element traces.
The number of recorded grid traces generally increases with the number of receiving elements and with decreasing $\mu$.
For polygonal or elliptical apertures in 3D, the quadrature points lie on a surface, but their interpolation supports extend into the surrounding volume.
In contrast, \pausct{} records only the 96 element traces during propagation.

The comparisons covered lossy PACT and linear and nonlinear USCT with pressure or particle-velocity sources, each with on-grid points, off-grid points, and line apertures.
Across all 15 two-dimensional configurations, relative RMS differences were below $3.4\times10^{-2}\%$.

\section{Direct differentiation with respect to the power-law exponent}\label{app:gamma-recursion}
Throughout this appendix, a dot denotes $\mathrm{d}/\mathrm{d}\gamma$ (not a time derivative), with all other inputs, including the SI coefficient $\alpha_0$, held fixed (Section~\ref{sec:grad-gamma}).
The initial conditions in Eq.~\eqref{eq:fwd_ic} give
\begin{equation*}
\vdP^{(0)}=\vec{0},\qquad
\vdUm^{(-\frac{1}{2})}=\vec{0},\qquad
\vdRm^{(0)}=\vec{0}.
\end{equation*}
Let $\vdU^{\slpm}=(\vdUm^{\slpm})_{m=1}^{d}$, $\vdRs^{(\ell)}=\sum_m\vdRm^{(\ell)}$, and $\vdbR^{(\ell)}=2\vdRs^{(\ell)}$ for nonlinear propagation.
Differentiating Eqs.~\eqref{eq:fwd_u}--\eqref{eq:fwd_p} gives the following updates for $\ell=0,\dots,L-1$:
\begin{subequations}\label{eq:gamma-recursion}
\begin{align}
\vdUm^{\slp}
&=\pmlp\had\Big[\pmlp\had\vdUm^{\slm}
-\dt\,\sfDp\big(\vdP^{(\ell)}\big)\hdiv\sfIp\big(\vRo\big)\Big],
\label{eq:velocity-gamma}\\
\vdRm^{(\ell+1)}
&=\pml\had\Big[\pml\had\vdRm^{(\ell)}
-\dt\Big(\vbR^{(\ell)}\had\sfDm\big(\vdUm^{\slp}\big)
+\vdbR^{(\ell)}\had\sfDm\big(\vUm^{\slp}\big)\Big)\Big],
\label{eq:density-gamma}\\
\vdP^{(\ell+1)}
&=\vCo^2\had\Big[\vdRs^{(\ell+1)}
+\vTo\had\sfQ_{\mathrm{A}}\big(\vdU^{\slp}\big)
-\vEo\had\sfQ_{\mathrm{D}}\big(\vdRs^{(\ell+1)}\big)\notag\\
&\qquad\quad+2\sfN\big(\vRs^{(\ell+1)}\had\vdRs^{(\ell+1)}\big)
+\vV_{\gamma}^{(\ell+1)}\Big].
\label{eq:pressure-gamma}
\end{align}
\end{subequations}
The source increments have zero derivatives because the prescribed excitation, $c_{\mathrm{water}}$, and the source \textit{k}-space correction are independent of $\gamma$.
For linear propagation, $\vbR^{(\ell)}=\vRo$, $\vdbR^{(\ell)}=\vec{0}$, and $\sfN(\cdot)\equiv\vec{0}$.
The quantity $\vV_{\gamma}^{(\ell+1)}$ defined in Section~\ref{sec:grad-gamma} is evaluated at each step without retaining its history.
At each step, the updated pressure and its derivative are projected by $\sftS$ and recorded.
Applying $\sfH$ to the derivative traces then gives the second argument of the inner product in Eq.~\eqref{eq:grad-gamma-forward}.

\section{Comparison of reverse-mode AD and DTA within \pausct{}}\label{app:ad-dta-comparison}
Reverse-mode AD and DTA were compared using the nonlinear USCT observations and acquisition setup of Section~\ref{sec:verification}.
The evaluation models were linear lossless (AA, dispersion, and nonlinear terms omitted), linear lossy (nonlinear terms omitted), and nonlinear lossy; identical forward models were differentiated by both methods.
Gradients were evaluated on a $1280\times1280$ grid with 0.2-mm spacing, a 40-point PML on each side, and $L=4320$ updates at $\dt=0.04~\mu\mathrm{s}$.
The medium maps were subsampled and Gaussian-smoothed with a 4-mm standard deviation, and $\gamma$ was set to 1.3, as in Section~\ref{sec:verification}.
Transmit and receive SIRs were retained, and $\sfH$ combined upsampling to $\Dt=0.02~\mu\mathrm{s}$, ETIR convolution, and zero-phase filtering at the two-PPW cutoff of 3.5~MHz.

The SOS inputs and gradients covered either the full grid including the PML or the $128\times112$~mm ROI ($640\times560$ points).
ROI inputs were padded with background values within the JIT-compiled computation, giving identical full-grid maps.
The reference SOS and PML coefficients were held fixed during differentiation.

For $N_c\in\{1,2,3,5,7,9,11,15\}$ internal checkpoints, the updates were divided into $N_c+1$ equal segments as in Section~\ref{sec:check}; both methods were also tested without checkpointing.
Reverse-mode AD used \texttt{jax.\allowbreak value\_and\_grad} on the complete objective and \texttt{jax.\allowbreak checkpoint} with \texttt{prevent\_cse=False} for the first $N_c$ segments in \texttt{jax.lax.scan}.
The final segment's intermediates were retained, matching DTA's recomputation schedule.
The temporal receive operators were applied after the receiver traces were assembled, outside the time-stepping loop.
Each configuration used a fresh process, the same cuFFT spatial-FFT backend for both methods, and the measurement protocol in Section~\ref{sec:implementation}.
The checkpointing comparisons are limited to the SOS gradient on one 2D grid size and one acquisition duration.

Table~\ref{tab:ad-dta-checkpoint} reports the 108 configurations. 93 completed and 15 failed GPU memory allocation.
Within each method, model, and gradient region, gradients were compared across checkpoint settings against $N_c=0$ where feasible or the smallest feasible $N_c$ otherwise.
Relative RMS differences, normalized by the RMS of the corresponding reference gradient, were below $8.5\times10^{-5}\%$ for DTA and $9.5\times10^{-5}\%$ for reverse-mode AD.
The peak-memory values are plotted in Fig.~\ref{fig:ad-dta-memory}, and the memory and timing comparisons are summarized in Section~\ref{sec:check}.
Checkpointing made all three models feasible with reverse-mode AD.
Without checkpointing, DTA completed both linear models, whereas reverse-mode AD completed only the lossless model.
For full-grid lossless evaluations, peak memory was slightly lower with reverse-mode AD at larger checkpoint counts; at $N_c=1$, however, its peak did not decrease relative to $N_c=0$.

\begin{table*}[!htb]
\centering
\caption{Time and peak memory for evaluating the SOS gradient and data-fidelity value with reverse-mode AD and DTA on the $1280\times1280$ grid with 4320 updates. $N_c$ counts internal checkpoints ($N_c=0$: none). An out-of-memory (OOM) error indicates a failed GPU memory allocation.}
\label{tab:ad-dta-checkpoint}
\small
\setlength{\tabcolsep}{4pt}
\begin{tabular}{@{}cc*{8}{c}@{}}
\hline
& & \multicolumn{4}{c}{Full grid, including PML} & \multicolumn{4}{c}{ROI} \\
\cline{3-6}\cline{7-10}
& & \multicolumn{2}{c}{\shortstack{Reverse-mode\\AD}} & \multicolumn{2}{c}{DTA} & \multicolumn{2}{c}{\shortstack{Reverse-mode\\AD}} & \multicolumn{2}{c}{DTA} \\
Model & $N_c$ & s & GiB & s & GiB & s & GiB & s & GiB \\
\hline
\multirow{9}{*}{\shortstack{Linear\\lossless}} & 0 & 3.23 & 26.54 & 2.81 & 26.50 & 3.24 & 26.53 & 2.52 & 5.90 \\
 & 1 & 4.02 & 26.58 & 3.56 & 13.43 & 4.01 & 26.57 & 3.33 & 3.12 \\
 & 2 & 4.18 & 9.08 & 3.94 & 9.01 & 4.17 & 9.08 & 3.58 & 2.13 \\
 & 3 & 4.30 & 6.91 & 3.93 & 6.84 & 4.31 & 6.91 & 3.68 & 1.68 \\
 & 5 & 4.42 & 4.78 & 4.04 & 4.76 & 4.40 & 4.78 & 3.80 & 1.32 \\
 & 7 & 4.47 & 3.74 & 4.14 & 3.77 & 4.48 & 3.74 & 3.82 & 1.19 \\
 & 9 & 4.50 & 3.14 & 4.17 & 3.22 & 4.50 & 3.14 & 3.89 & 1.15 \\
 & 11 & 4.53 & 2.76 & 4.17 & 2.89 & 4.53 & 2.76 & 3.91 & 1.17 \\
 & 15 & 4.54 & 2.33 & 4.18 & 2.56 & 4.57 & 2.34 & 3.56 & 1.02 \\
\hline
\multirow{9}{*}{\shortstack{Linear\\lossy}} & 0 & \multicolumn{2}{c}{OOM} & 4.18 & 26.54 & \multicolumn{2}{c}{OOM} & 4.07 & 5.93 \\
 & 1 & \multicolumn{2}{c}{OOM} & 5.16 & 13.46 & \multicolumn{2}{c}{OOM} & 5.07 & 3.15 \\
 & 2 & 6.56 & 26.73 & 5.52 & 9.05 & 6.52 & 26.73 & 5.38 & 2.16 \\
 & 3 & 6.69 & 20.17 & 5.66 & 6.88 & 6.68 & 20.16 & 5.55 & 1.71 \\
 & 5 & 6.86 & 13.64 & 5.83 & 4.79 & 6.88 & 13.63 & 5.74 & 1.34 \\
 & 7 & 6.97 & 10.40 & 5.90 & 3.80 & 6.97 & 10.40 & 5.81 & 1.21 \\
 & 9 & 7.01 & 8.49 & 5.95 & 3.25 & 7.01 & 8.48 & 5.87 & 1.17 \\
 & 11 & 7.03 & 7.23 & 5.98 & 2.92 & 7.05 & 7.23 & 5.88 & 1.19 \\
 & 15 & 7.11 & 5.70 & 6.03 & 2.59 & 7.10 & 5.70 & 5.83 & 1.03 \\
\hline
\multirow{9}{*}{\shortstack{Nonlinear\\lossy}} & 0 & \multicolumn{2}{c}{OOM} & \multicolumn{2}{c}{OOM} & \multicolumn{2}{c}{OOM} & \multicolumn{2}{c}{OOM} \\
 & 1 & \multicolumn{2}{c}{OOM} & \multicolumn{2}{c}{OOM} & \multicolumn{2}{c}{OOM} & 5.93 & 42.73 \\
 & 2 & \multicolumn{2}{c}{OOM} & 6.35 & 35.45 & \multicolumn{2}{c}{OOM} & 6.25 & 28.56 \\
 & 3 & \multicolumn{2}{c}{OOM} & 6.55 & 26.69 & \multicolumn{2}{c}{OOM} & 6.40 & 21.52 \\
 & 5 & 7.65 & 31.24 & 6.67 & 18.00 & 7.67 & 31.22 & 6.53 & 14.55 \\
 & 7 & 7.74 & 23.61 & 6.84 & 13.72 & 7.74 & 23.60 & 6.65 & 11.13 \\
 & 9 & 7.79 & 19.05 & 6.85 & 11.19 & 7.82 & 19.04 & 6.72 & 9.11 \\
 & 11 & 7.83 & 16.06 & 6.92 & 9.54 & 7.83 & 16.03 & 6.75 & 7.81 \\
 & 15 & 7.89 & 12.31 & 6.95 & 7.56 & 7.87 & 12.31 & 6.84 & 5.99 \\
\hline
\end{tabular}
\end{table*}
\FloatBarrier

A separate memory check used the final 0.6-mm USCT configuration of Section~\ref{sec:3d-reconstruction}, with a $360\times360\times200$ grid, a $199\times208\times101$ SOS ROI, and 1200 updates.
Only SOS was differentiated, using the same saved SOS estimate and a zero-valued observation array of the original dimensions for both methods.
Without checkpointing, DTA completed, whereas reverse-mode AD of the \pausct{} forward computation failed GPU memory allocation.

\section{Comparison with \texttt{j-Wave}}\label{app:jwave-comparison}
The pressure traces and computational costs of \pausct{} and \texttt{j-Wave}~\cite{stanziolaJWaveOpensourceDifferentiable2023} were compared under common modeling assumptions.
SOS gradients computed by reverse-mode AD were also compared on a coarser grid.

The comparison used \texttt{j-Wave} 0.2.1 with JaxDF 0.2.8 and JAX/\allowbreak JAXlib 0.4.38 using CUDA 12.9.
Its time-domain \textit{k}-PSTD solver supports heterogeneous SOS and ambient density but does not implement power-law AA, cumulative nonlinear effects, or time-varying particle-velocity sources.

Its acquisition utilities include full-grid periodic band-limited interpolation for off-grid point reception~\cite{wiseRepresentingArbitraryAcoustic2019} and grid-aligned binary masks for line-aperture transmission and reception.
Grid-aligned masks can introduce staircasing errors for apertures not aligned with the grid~\cite{wiseRepresentingArbitraryAcoustic2019}.
Periodic band-limited interpolation is not implemented for off-grid transmission, and temporal receive operations are not provided.

To compare computational performance under matched propagation and acquisition assumptions, both packages were configured for linear, lossless USCT with on-grid point pressure sources and receivers, without temporal receive operations.
Source \textit{k}-space correction in Eq.~\eqref{eq:source-kspace-correction} was also omitted to match the evaluated \texttt{j-Wave} solver, which does not implement it.

The forward comparison used the $2560\times2560$ data-generation grid, time discretization, and heterogeneous SOS and ambient-density maps from Section~\ref{sec:verification}.
Both packages used identical grid indices for the 96 receivers and four simultaneous pressure sources, with the same source waveform, the maximum SOS as the reference SOS, and matching quartic PML profiles with a strength parameter of 2.
The relative RMS difference between the pressure traces was $2.49\times10^{-3}\%$, normalized by the RMS of the \texttt{j-Wave} traces.
Mean execution times were 26.58~s for \texttt{j-Wave} and 12.43~s for \pausct{}, corresponding to a 2.14-fold speedup for \pausct{}.
The respective peak-memory values were 0.625 and 0.350~GiB.
For spatial differentiation, \pausct{} uses real-input FFTs, whereas the evaluated \texttt{j-Wave} implementation uses full complex FFTs.
The contribution of FFT choice to the timing and memory differences was not measured separately.

The reverse-mode AD comparison used the $640\times640$ gradient-evaluation grid, time discretization, and Gaussian-smoothing kernel of Section~\ref{sec:verification}, retaining the linear, lossless, on-grid configuration above.
Only SOS within a $96\times80$~mm ROI was smoothed and differentiated; density and SOS outside this ROI were unchanged.
Observations were every fourth time sample of the fine-grid \texttt{j-Wave} data.
Both packages evaluated the half-squared-sum data-fidelity value and SOS gradient using \texttt{jax.\allowbreak value\_\allowbreak and\_\allowbreak grad} without checkpointing, with the existing per-time-step checkpointing option in \texttt{j-Wave} disabled.
The relative RMS difference between the gradients was $1.01\times10^{-2}\%$, normalized by the RMS of the \texttt{j-Wave} gradient.
The evaluation required 0.539~s and 6.647~GiB in \texttt{j-Wave}, compared with 0.492~s and 3.332~GiB in \pausct{}, giving about half the peak memory of \texttt{j-Wave} and a modest reduction in time.

The additional benefit of explicitly organizing DTA histories, including combination of gradient terms and ROI restriction, is assessed in~\ref{app:ad-dta-comparison} using the same time-segment boundaries and recomputed updates.

\section*{Code and data availability}

The measurements used in this work were generated numerically.
The \pausct{} source code and scripts for synthetic data generation and reconstruction are available at \url{https://github.com/gangwonee/pausct}.

\section*{Acknowledgements}
This work was supported in part by the National Institutes of Health (United States) under grants EB031585, EB034261, and UG3DA065155.

\section*{Declaration of generative AI and AI-assisted technologies in the manuscript preparation process}
OpenAI ChatGPT was used during manuscript preparation to support structural review and language refinement.
All AI-assisted suggestions were critically evaluated, verified, and revised by the authors.
Full responsibility for the content of the manuscript is retained by the authors.

\bibliographystyle{elsarticle-num}
\bibliography{\arxivdir References/pausct,\arxivdir References/MA-GJ-additions}

\end{document}